\documentclass[preprint2]{aastex}
\usepackage{graphicx}
\usepackage{lscape} 
\usepackage{amssymb}
\usepackage[figuresright]{rotating}
\usepackage{xcolor}
\definecolor{mag}{rgb}{1,0,1}

\usepackage{ulem}
\usepackage{xcolor}

\newcommand{\gapprox}{$\stackrel {>}{_{\sim}}$}  
\newcommand{\lapprox}{$\stackrel {<}{_{\sim}}$}

\newcommand{\htwo}{H$_2$}
\newcommand{\hi}{\ion{H}{1}}

\newcommand{\fei}{\ion{Fe}{1}}
\newcommand{\feii}{\ion{Fe}{2}}
\newcommand{\sii}{[\ion{S}{2}]}
\newcommand{\ti}{\ion{Ti}{1}}
\newcommand{\tii}{\ion{Ti}{2}}
\newcommand{\cri}{\ion{Cr}{1}}
\newcommand{\crii}{\ion{Cr}{2}}
\newcommand{\hei}{\ion{He}{1}}
\newcommand{\caii}{\ion{Ca}{2}}
\newcommand{\oi}{[\ion{O}{1}]}
\newcommand{\nii}{[\ion{N}{2}]}

\newcommand{\nic}{[\ion{Ni}{2}]}
\newcommand{\av}{A$_V$}

\newcommand{\um}{$\mu$m}

\newcommand{\lsun}{L$_{\odot}$}
\newcommand{\msun}{M$_{\odot}$}

\newcommand{\msunyr}{M$_{\odot}$\,yr$^{-1}$}
\newcommand{\macc}{$\dot{M}_{\mathrm{acc}}$}

\newcommand{\lacc}{$L_{\mathrm{acc}}$}
\newcommand{\lacci}{$L_{\mathrm{acc(i)}}$}
\newcommand{\lumi}{$L_{\mathrm{i}}$}
\newcommand{\rstar}{$R_{\mathrm{*}}$}

\newcommand{\lstar}{$L_{\mathrm{*}}$}
\newcommand{\mstar}{$M_{\mathrm{*}}$}
\newcommand{\teff}{$T_\mathrm{eff}$}
\newcommand{\lbol}{$L_{\mathrm{bol}}$}

\newcommand{\nH}{$n_{\mathrm{H}}$}
\newcommand{\f}{\scriptsize}

\received{June 1, 2019}
\revised{January 10, 2019}
\accepted{\today}

\shorttitle{A spectroscopic survey of young eruptive variables}
\shortauthors{Giannini et al.}

\begin{document}

\title{EXORCISM: a spectroscopic survey of young eruptive variables (EXor and candidates)}
\author{
T. Giannini\altaffilmark{1}, A. Giunta\altaffilmark{2}, M. Gangi\altaffilmark{1}, R. Carini\altaffilmark{1}, D. Lorenzetti\altaffilmark{1}, S. Antoniucci\altaffilmark{1}, A. Caratti o Garatti\altaffilmark{3,4}, L. Cassar{\'a}\altaffilmark{5}, B. Nisini\altaffilmark{1}, A. Rossi\altaffilmark{6}, V. Testa\altaffilmark{1}, F. Vitali\altaffilmark{1} }
\altaffiltext{1}{INAF - Osservatorio Astronomico di Roma, via Frascati 33, 00078, Monte Porzio Catone, Italy}
\altaffiltext{2}{ASI - Agenzia Spaziale Italiana - Via del Politecnico, 00133 Roma, Italy}
\altaffiltext{3}{INAF - Osservatorio Astronomico di Capodimonte, Via Moiariello 16, 80131 Napoli, Italy}
\altaffiltext{4}{Dublin Institute for Advanced Studies, School of Cosmic Physics, Astronomy \& Astrophysics Section, 31 Fitzwilliam Place, Dublin 2, Ireland}
\altaffiltext{5}{INAF-IASF Milano, via Alfonso Corti 12, 20129 Milano, Italy}
\altaffiltext{6}{INAF-Osservatorio di Astrofisica e Scienza dello Spazio, Via Piero Gobetti, 93/3, I-40129 Bologna, Italy}

\begin{abstract} 
We present an  optical/near-IR survey of 11 variable young stars (EXors and EXor candidates) aimed at 
deriving and monitoring their accretion properties.
About 30 optical and near-infrared  spectra ($\Re$ $\sim$
1500-2000) were collected  between 2014-2019 with the Large
Binocular Telescope (LBT). From the
spectral analysis we have derived the
accretion luminosity  (\lacc\,) and
mass accretion rate  (\macc\,), the visual extinction (\av)\,,
the  temperature and density of the
permitted line formation region (T,
\nH), and the signature of the
outflowing matter.

Two sources (ASASSN-13db and iPTF15afq) have been observed in outburst and quiescence, three during a high-level of brightness (XZ Tau, PV Cep, and NY Ori), and the others in quiescence. 
These latter have \lacc\, and  \macc\,  in line with the values measured in classical T Tauri stars of similar mass. 
All sources observed more than once present \lacc\, and \macc\, variability. The most extreme case is  ASASSN-13db, for which \macc\, decreases by two orders of magnitude from the outburst peak in 2015 to quiescence in 2017. Also, in NY Ori \lacc\, decreases by a factor 25 in one year.\\
In 80\% of the sample we detect the \oi\,6300\,\AA\, line, a tracer of mass loss. From the variability of the H$\alpha$/\oi\,6300\,\AA\, ratio, we conclude that mass accretion  variations are larger than mass loss variations.\\
From the analysis of the \hi\, recombination lines a correlation is suggested between the density of the line formation region, and the level of accretion activity of the source.\\
\end{abstract}


\section{Introduction\label{sec:sec1}}

EXor objects (named after the prototype EX Lupi, Herbig 1989) are Pre-Main Sequence (PMS) stars showing episodes of eruptive accretion caused by magnetospheric accretion events (Shu et al. 1994).
Since the unsteady mass accretion is a central theme in the star formation process (e.g. Cieza et al. 2018), these objects have been the subject of many
investigations (Kuffmeier et al. 2018; Contreras-Pe\~{n}a et al. 2019; Meng et al. 2019;
MacFarlane et al. 2019a, 2019b; see also Audard et al. 2014 for a comprehensive view of the EXor phenomenon). EXors are characterized by outbursts of short duration (typically months) occurring at different time-scales (months, years) and showing amplitudes of several magnitudes at optical and near-IR wavelengths.
An evolutionary scheme has been proposed in which EXors represent a less energetic and (possibly) later stage than the powerful FU Orionis eruptions 
(FUors, Hartmann \& Kenyon 1985, Hartmann et al. 1993), although EXors present substantial differences, such as
shorter and more frequent outbursts, spectra dominated by emission lines instead of absorption lines, and smaller values of the mass accretion rate during outbursts (10$^{-7}-10^{-6}$ \msunyr\, vs. 10$^{-5}-10^{-4}$ \msunyr).
Yet, at present, a detailed model for the disk structure and its evolution does not exist for EXors and different hypotheses have been
proposed for the possible trigger of their outbursts: thermal instability in the disk (e.g. Audard et al. 2014 and references therein), gravitational perturbations induced by a binary companion (Lodato $\&$ Clarke 2004) or by the migration of a giant planet  (Bonnell \& Bastien 1992), or  sudden changes in the stellar magnetic activity (D'Angelo \& Spruit 2012, Armitage 2016).

Observational constraints to theoretical models are nowadays provided by photometric monitoring programs (e.g. Grankin et al. 2008; Morales-Calder{\'o}n et al. 2011; Guo et al. 2020) and by ongoing all-sky surveys (e.g. {\it Gaia}\footnote{https://sci.esa.int/web/gaia}, {\it All-Sky Automated Survey for Supernovae} (ASAS-SN)\footnote{http://www.astronomy.ohio-state.edu/asassn/index.shtml}, {\it Intermediate Palomar Transient Factory} (iPTF)\footnote{https://www.ptf.caltech.edu/iptf}, {\it Panoramic Survey Telescope \& Rapid Response System} (Pan-STARSS)\footnote{https://panstarrs.stsci.edu/}, and {\it VISTA Variables in the Via Lactea} (VVV)\footnote{https://vvvsurvey.org/}), which sistematically follow (although often at a poor level of sensitivity) the photometric variations of many EXors and EXor candidates, thus allowing us to track the changes of their integrated properties (light-curves, Spectral Energy Distributions -SEDs- and colors). A dramatic improvement in  this context will be provided in the next future by the the Vera Rubin Observatory\footnote{https://www.lsst.org/}, which will explore the southern sky with unprecedented sensitivity and cadence. 

Conversely, the optical and IR spectroscopic monitoring of the EXors is currently not at a comparable level of sky coverage and cadence, despite its crucial importance for understanding how the gas properties (e.g. excitation, ionization, dynamics) vary during different activity phases. 
Indeed, so far only few EXors have been spectroscopically investigated 
(e.g. EX Lupi: Sipos et al. 2009; Sicilia-Aguilar et al. 2012; V2492 Cyg: Hillenbrand et al. 2013; V1118 Ori: Audard et al. 2010;  Lorenzetti et al. 2015; Giannini et al. 2016, 2017, 2020; HBC722: K{\'o}sp{\'a}l et al. 2016). In addition, Guo et al. (2020) recently presented a near-IR spectroscopic monitoring of a few VISTA variables (supposedly eruptive candidates).

Waiting for instruments like SoXS\footnote{https://www.eso.org/sci/facilities/develop/instruments/SoXS.html} at ESO-NTT (Schipani et al. 2018), which will be specifically dedicated to the spectroscopic monitoring of large samples of objects, we have started in 2014 the EXOR OptiCal and Infrared Systematic Monitoring program EXORCISM (Antoniucci et al. 2013). Over the years, new EXor candidates have been discovered and included in our database. As a result, we provide here the first flux-limited deep spectroscopic atlas of EXors (and candidates) collected through state-of-the-art instrumentation (Large Binocular Telescope - LBT) at optical and near-IR wavelengths. 
As outburst phases are expected to be short-lived and rare events, most of our spectra have been taken during quiescence phases. These spectra allow us to infer the baseline values of the physical parameters, such as intrinsic line fluxes, accretion luminosity (\lacc), and mass accretion rate (\macc), which can be used as reference in case of outburst. This is what we did for the few sources that we observed both in quiescence and outburst, for which we were able to study the variation of the parameters between the two phases.

The paper is organized as follows: Section 2 gives general information on the investigated sample while Section 3 provides details on the observations and data reduction procedures. In Section 4 we comment on the light-curves of the individual sources. Sections 5 and 6 are dedicated to the description of the spectra and to the discussion of the results obtained from 
the spectroscopic analysis. Final remarks are given in Section 7.

\section{The investigated sample\label{sec:sec2}}

Table\,\ref{tab:tab1} lists our targets with their general properties. It provides coordinates, distance, location, binarity, and jet or outflow detection. Together with the name given in the Simbad Astronomical Database\footnote{http://simbad.u-strasbg.fr/simbad/}, we report other designation(s) commonly used in the literature. 

Our sample is divided into two groups.
The first is composed of 5 known EXors (XZ Tau, UZ Tau E, VY Tau, NY Ori, V1143 Ori) listed in the compilation by Audard et al. (2014) and confirmed by many works in literature as members of the class. 
In addition, we have also obtained multi-epoch spectroscopy of the classical EXor V1118 Ori at both optical and IR wavelengths. These data are not discussed here since they have been separately presented in a series of dedicated papers (Giannini et al. 2020 and references therein).

The second group, generically named as 'EXor candidates' in Table\,\ref{tab:tab1}, consists of 6 sources that have been reported through different alerts (e.g.  astronomers' telegrams and public surveys alerts) for being young and highly variables objects ($\Delta$V \gapprox 2 mag). Here we briefly comment on each of them. 
Two objects, namely V1647 Ori and PV Cep, are present in the Audard et al. (2014) compilation, but their nature is still controversial. 
V1647 Ori has been previously classified as a candidate FUor, but it is currently considered as a peculiar object, since its spectrum now differs significantly from that of {\it a bona fide} FUor (Connelley \& Reipurth 2018), showing both FUor
and EXor features. A few more objects with similar characteristics have been identified during the VVV spectroscopic survey and dubbed 'MNors' (Contreras Pe{\~n}a et al. 2017). PV Cep and DR Tau were originally included in the first list of EXor variables by Herbig (1989). PV Cep repeatedly undergoes intense outbursts (up to 5 mag) of short duration and brightness between that of FUors and EXors (Andreasyan et al. 2021).  Its high luminosity ($\approx$ 100 L$_{\odot}$) is likely attributable to a young Herbig Ae star (see e.g. Lorenzetti et al. 2011).
DR Tau has progressively increased its brightness by about three magnitudes between 1960 and 1980. Since then  its average brightness has been at the same level but still with remarkable photometric fluctuations with amplitude up to $\sim$ 1 mag in the optical (Banzatti et al. 2014 and references therein). Similarly, the $B$ magnitude of V350 Cep changed from $>$ 21 to $\sim$ 17 between 1954 and 1978. Afterward, the source has remained at roughly the same level of brightness, with the exception of at least two fading events of about 2 mag in $B$ in 1979 and 2004. For these reasons Herbig (2008) did not classify V350 Cep as an EXor. Considering its high and stable brightness level and its spectrum rich in emission lines, it is likely that also V350 Cep is an object with properties in between those of FUors and EXors (Andreasyan et al. 2021).  iPTF15afq is a new candidate EXor, suggested by Miller et al. (2015) on the basis of the amplitude of its outburst of about 2.5 mag (in $R$-band) and its spectrum dominated by lines in emission. ASASSN-13db (discovered by the ASAS-SN survey) has been originally classified as an EXor by Holoien et al. (2014). Later observations of blue-shifted absorption features in several hydrogen and metallic lines suggest an intermediate behavior between EXors and FUors  (Sicilia-Aguilar et al. 2017).

In Table\,\ref{tab:tab2} we summarize the photometric properties of the sources, with the range of variation observed in the indicated band and a short (non-exhaustive) summary of the variability history as retrieved from the literature. 

In Table\,\ref{tab:tab3} we list the relevant stellar parameters as given in the literature, namely the visual extinction (A$_V$), bolometric (\lbol), stellar (\lstar) and accretion (\lacc) luminosity, spectral type (SpT) and effective temperature (\teff), stellar mass (\mstar), mass accretion rate (\macc), and evolutionary class (typically derived from the spectral slope between 2 and 24 $\mu$m).
All sources but PV Cep, have mass \lapprox 1 \msun\, and spectral types M-K. 
V1647 Ori and PV Cep present a high and variable \av\, while all other sources are only slightly extincted. 
Mass accretion rates have been measured for about 70\% of the sample. During quiescence, they are typically of the order of 10$^{-8}-10^{-7}$ \msunyr, but show variations up to several orders of magnitude in bursts.

In summary, our sample is composed of sources whose variability may differ in amplitude, cadence, and duration. In all cases, however, the increase of brightness has been attributed in the literature to accretion events, sometimes accompanied by significant variations of the local extinction.


\section{Observations and data reduction\label{sec:sec3}}

The observations were carried out between 2014 and 2019 with the 8.4m Large Binocular Telescope (LBT) located at Mount Graham (Arizona, USA). Optical and near-IR spectra of the targets were obtained with the Multi-Object Double Spectrograph (MODS, Pogge et al. 2010) and the LBT Utility Camera in the Infrared (LUCI, Seifert et al. 2003), respectively.

Given the remarkable brightness of the sources, our project was executed as a filler program during  unfavorable atmospheric conditions that usually make observations of weaker sources unfeasible.
As a consequence, we were not able to complete the survey with all the needed spectra. However,  the poor photometric quality of the sky (typical seeing \gapprox 1\farcs5) did not represent a significant problem for the validity of our results, since the flux calibration of the spectra was based on photometric observations taken close in time to the spectra.

In total, we acquired 19 spectra with MODS (11 targets) and 10 spectra with LUCI (8 targets), see the journal of observations in Table\,\ref{tab:tab4}.
Optical and near-IR spectra have been taken during the same night only for V1647 Ori, and for DR Tau at a temporal distance of 4 days. For all the other targets, optical and near-IR spectra are spaced by several months or years, and, in some cases, the spectrum of one of the two segments is missing. 

MODS observations were performed with the dual grating mode (Blue + Red channels, spectral range 350$-$950 nm)  by using a 0$\farcs$80 slit ($\Re \sim$ 1500 and 1800 in the Blue and Red channels, respectively). The slit angle matched the parallactic angle to minimize the wavelength dependence on the 
slit transmission. For each source, the adopted integration time is reported in Table\,\ref{tab:tab4}.

LUCI observations were carried out with the G200 low-resolution grating coupled with the 0$\farcs$75 slit. Two datasets were acquired with the standard ABB'A' technique using the $zJ$ and $HK$ grisms. This provides a final spectrum covering the 1.0$-$2.4 $\mu$m wavelength range  at $\Re \sim$1500. The integration time for the individual sources is provided in Table\,\ref{tab:tab4}.

Data reduction was performed at the Italian LBT Spectroscopic Reduction Center\footnote{http://www.iasf-milano.inaf.it/Research/lbt\_rg.html}, by means of scripts optimized for LBT data.
Data reduction steps of each  MODS spectral-image are: correction for dark and bias, bad-pixel mapping, flat-fielding, sky background subtraction, and extraction of the one-dimensional spectrum by integrating the stellar trace along the spatial direction. In the optical range few telluric features are present, being the most prominent that around 7600\,\AA\,, around which, however,
there are no lines used for the subsequent accretion and ejection analysis. Therefore, we have not applied any telluric correction to the optical spectra. Wavelength calibration was obtained from arc lamps.
Intercalibration between Blue and Red spectral segments was verified superposing the spectral range between 5300 and 5900 \AA\, in common between the two channels. In all cases, the Blue and Red spectra resulted optimally aligned without the need for further corrections. 

The raw LUCI spectral images were flat-fielded, sky-subtracted, and corrected for optical distortions
in both the spatial and spectral directions. Telluric absorptions were removed using the normalized spectrum of a telluric standard star, after the removal of its intrinsic spectral features. Wavelength calibration was obtained from arc lamps spectra. In a few cases, we applied an inter-calibration correction to align the $zJ$ and $HK$ segments, which did not exceeded 20\% of the flux density.

 The bad atmospheric conditions prevented us from flux calibrating the spectra by using spectrophotometric standards. Therefore, we used as absolute calibrators the acquisition images obtained immediately before the spectral images in the filter specified in Table\,\ref{tab:tab4}. Photometry was obtained by taking as reference all the visible stars in the field, for which either a PAN-STARRS\footnote{panstarss.stsci.edu} (in the optical) or a 2MASS\footnote{ipach.caltech.edu/project/2mass} (in the near-IR) photometry is available. In few cases, namely when not enough reference sources were present in the acquisition image or our target appeared saturated, we used the photometry retrieved from public databases and obtained a few days before or after the LBT observation (as specified in the note of Table\,\ref{tab:tab4}).

\section{Light-curves\label{sec:sec4}}
\begin{figure*}
\begin{center}
\includegraphics[trim=0 0 0 0,width=2.0\columnwidth, angle=0]{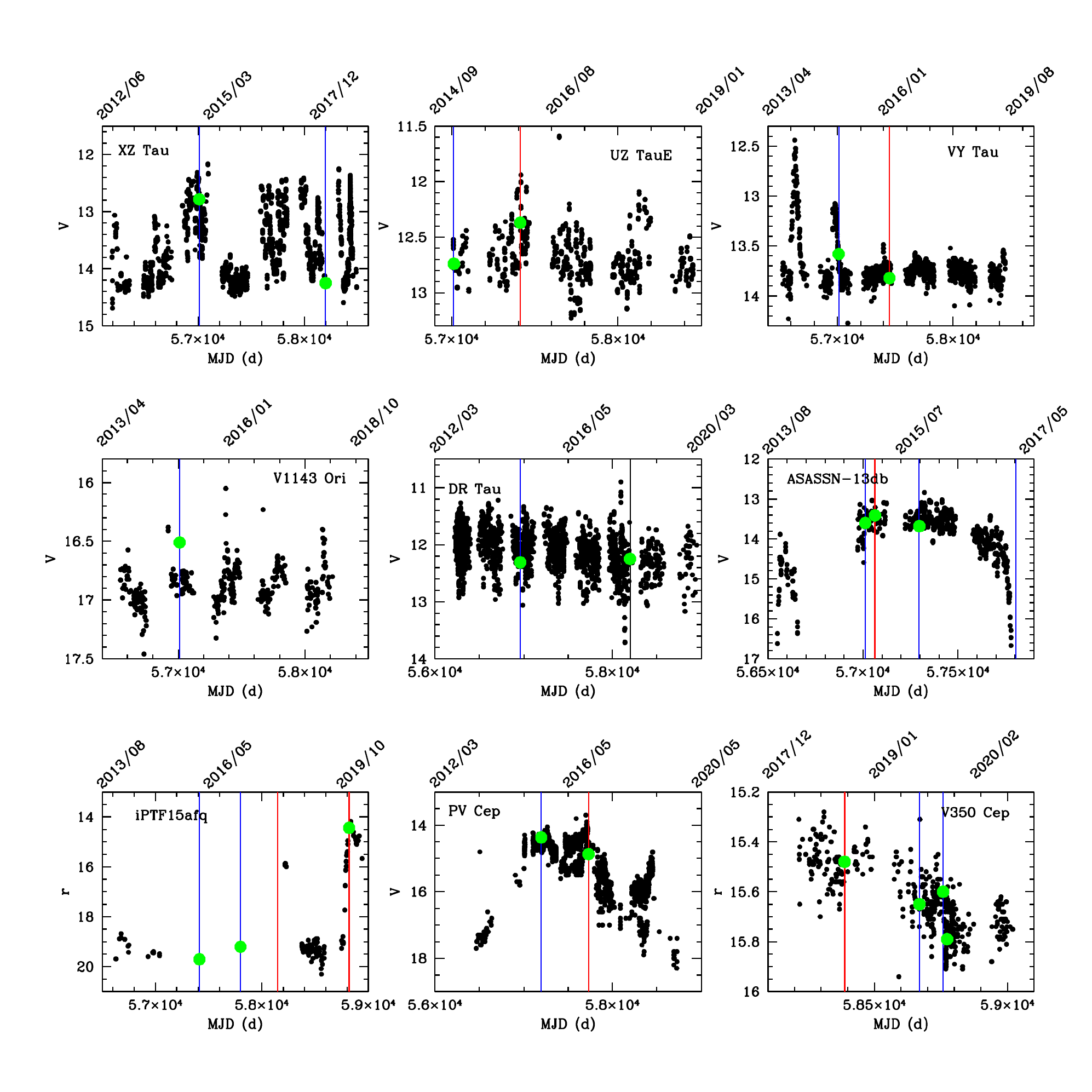}
\caption{Optical ($V-$ or $r-$band) light-curves retrieved from data of public surveys 
(ZTF, ASAS-SN, and AAVSO). The x-axis represents the Modified Julian Date (MJD), while some
calendar dates are given on top as reference. A blue (red) vertical line indicates the date of a MODS (LUCI) observation. The black line corresponds to quasi-simultaneous MODS+LUCI observation of DR Tau. 
Photometry taken closest in time to the spectroscopic observation is marked with a green dot. 
\label{fig:fig1}}
\end{center}
\end{figure*}

To establish the level of accretion activity of each target at the time of the LBT observation we collected data from public surveys (ZTF, ASAS-SN, and AAVSO\footnote{https://www.aavso.org/}) to derive the optical light-curves of the sources over several years, which are displayed in Figure\,\ref{fig:fig1}. 
For NY Ori and V1647 Ori, only the light-curves of the ASAS-SN survey are available. Unfortunately, the ASAS-SN images of these sources are strongly contaminated by the diffuse emission of the ONC and Mc Neil's nebulae, where NY Ori and V1647 Ori are respectively located (Aspin et al. 2009, Herbig 2008). Therefore, these two light-curves are not shown.

In the following, we briefly comment on the light-curves of the remaining 9 objects.
\begin{itemize}
\item[-]{\bf XZ Tau}. The $V$ band magnitude varied between $\sim$ 14.5 and 12 mag within a 6 yr period. Our MODS
spectra cover both a high- and low-level brightness state.

\item[-]{\bf UZ Tau E}. This source shows a very small amplitude level of variability within 5 yr. It has maintained an average brightness level around $V$ = 12.7 mag, with fluctuations of $\sim$\,0.3 mag. The spectroscopic observations were obtained at photometric levels with 0.6 mag of difference.

\item[-]{\bf VY Tau}. A remarkable brightening occurred at the beginning of 2014, unfortunately not covered by our observations. This was followed by a second low-amplitude burst about a year later. The first MODS spectrum refers to the declining phase of this burst. 

\item[-]{\bf V1143 Ori}. Between 2013 and 2018 the source has remained at $V$ $\sim$\,17 mag with typical fluctuations of tenths of magnitude. The MODS spectrum was acquired at V$\sim$ 16.5 mag.

\item[-]{\bf DR Tau}. In the last seven years the DR Tau average photometry has been $V\sim$ 12 mag, with large fluctuations up to 1 mag. All our observations have been performed close to the average level. 

\item[-]{\bf ASASSN-13db}. This source underwent a first, short-lived outburst in 2013, followed by a second
more intense outburst from 2014 to late 2016. We have observed ASASSN-13db four times, the first
three (two with MODS and one with LUCI) during the second burst, and the last with MODS in 2017, when the source was back to quiescence.
  
\item[-]{\bf iPTF15afq}. The source underwent an outburst at the beginning of 2018 (not shown in the $r$ band light-curve, see Hillenbrand 2019), followed by a quiescence period of about 2 years. A second outburst occurred in 2019, covered by our LUCI observation during the rising phase and close to the peak. A slow decay started at the beginninig of 2020 and is still ongoing. The remaining three observations (between 2016 and 2018), have been taken during quiescence ($R\sim$ 18.5-19 mag), according to the light-curve presented by Miller et al. (2015).

\item[-]{\bf PV Cep}. Between 2013 and 2020, PV Cep has experienced a slow increase in brightness, with a peak in 2016 ($V \sim$ 14.5 mag), followed by a decrease down to $V >$ 18 mag at a roughly similar speed. Both our observations were conducted close to the brightness peak.

\item[-]{\bf V350 Cep}. Our first LUCI observation was taken in November 2015, at $R$=15.4 mag (Semkov et al. 2017). The source experienced a deep dimming in spring 2016 followed by a new brightening. Since 2018, V350 Cep is again progressively and slowly fading (from $r \sim$ 15.3 to $r \sim$ 15.8). Our second LUCI observation (2018) was obtained when V350 Cep was still relatively bright ($r\sim$ 15.5 mag), while the next three MODS spectra were taken when $r$ had decreased down to 15.6-15.8 mag.

\end{itemize}

Summarizing, our targets have shown a variety of behaviors during the last decade. Some of them have remained fairly constant, others have undergone strong outbursts or long-term and remarkable brightness variations. As expected, in most cases 
our observations were performed during phases of quiescence. However, some exceptions exist: indeed we have observed XZ Tau
and PV Cep during a high level of their activity and even caught the outbursts of ASASSN-13db and iPTF15afq. In addition, our acquisition images of NY Ori indicate a variation of about two magnitudes in the $r$-band between 2014 ($r \sim$ 12) and 2016 ($r \sim$ 14).

\section{Description of the spectra\label{sec:sec5}}
\begin{figure*}
\includegraphics[trim=0 0 0 0,width=2.5\columnwidth, angle=0]{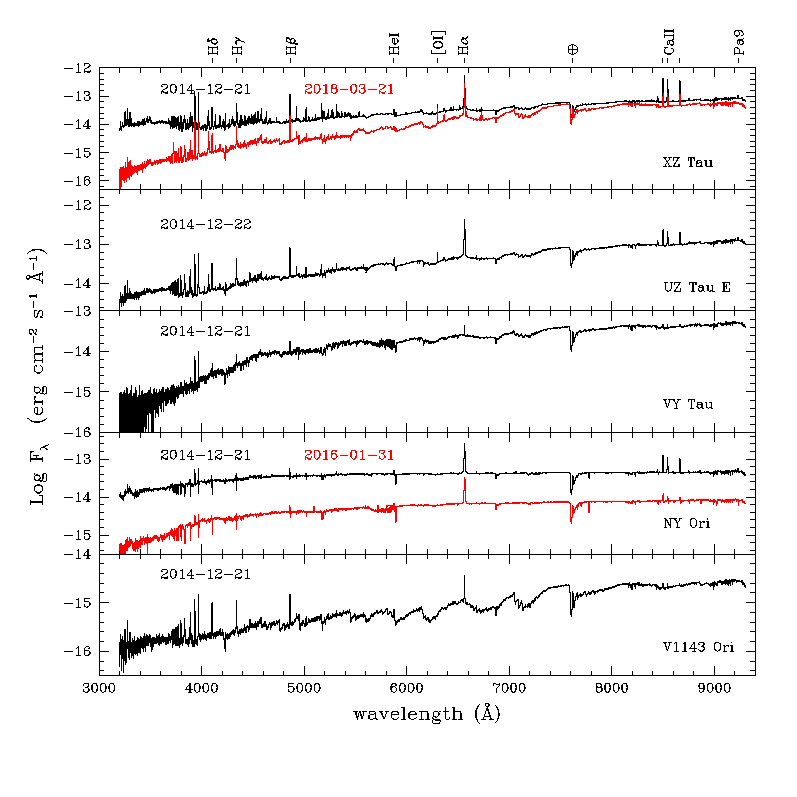}
\caption{MODS spectra. Spectra obtained in different dates are shown with different colors. Brightest lines and telluric features are labelled.\label{fig:fig2}}
\end{figure*}

\begin{figure*}
\includegraphics[trim=0 0 0 0,width=2.5\columnwidth, angle=0]{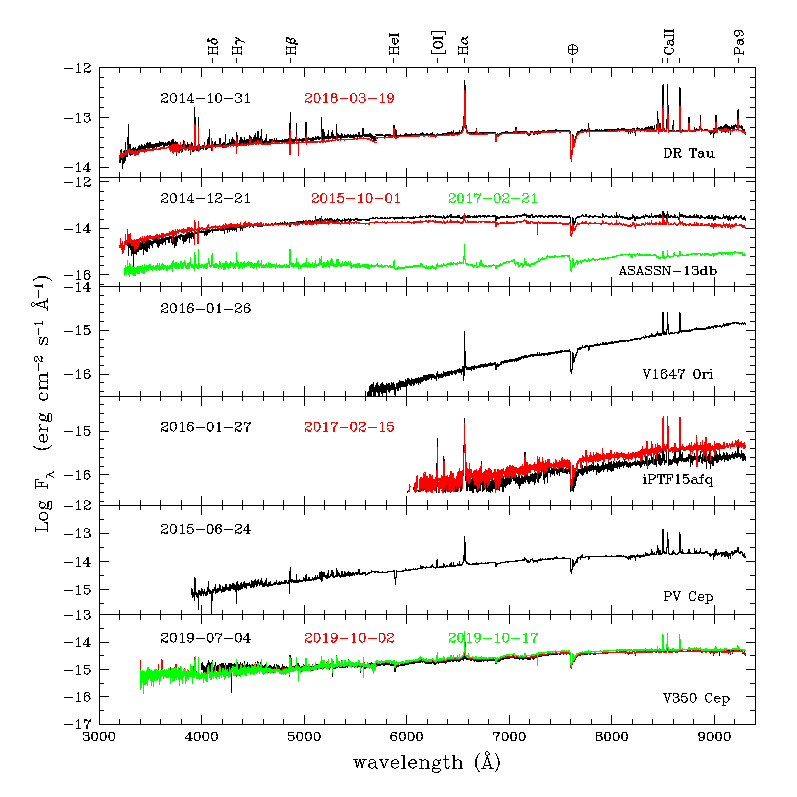}
\caption{As in Figure\,\ref{fig:fig2}..\label{fig:fig3}}
\end{figure*}

\begin{figure*}
\includegraphics[trim=0 0 0 0,width=2.5\columnwidth, angle=0]{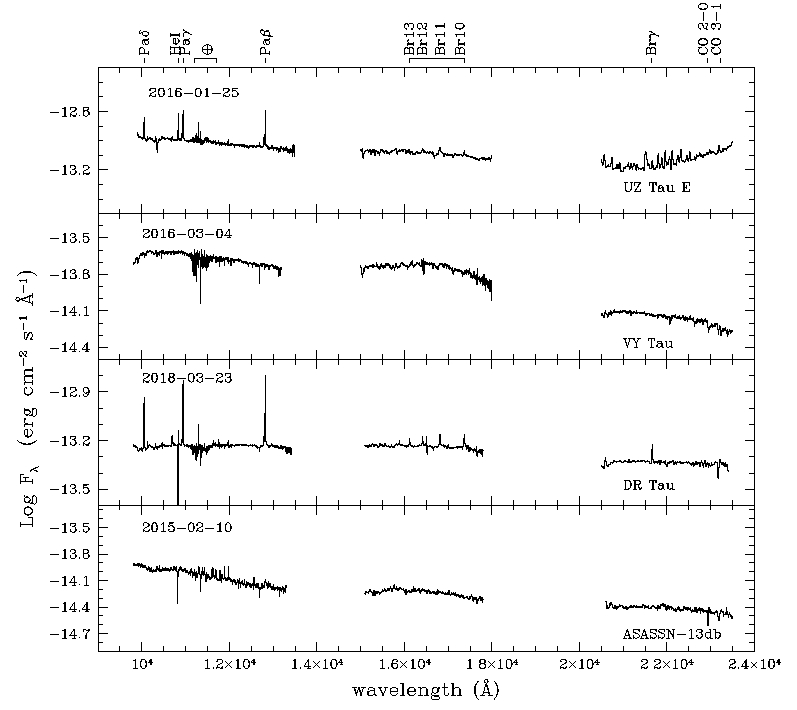}
\caption{LUCI spectra. Spectra obtained in different dates are shown with different colors. Brightest lines are labelled.
\label{fig:fig4}}
\end{figure*}

\begin{figure*}
\includegraphics[trim=0 0 0 0,width=2.5\columnwidth, angle=0]{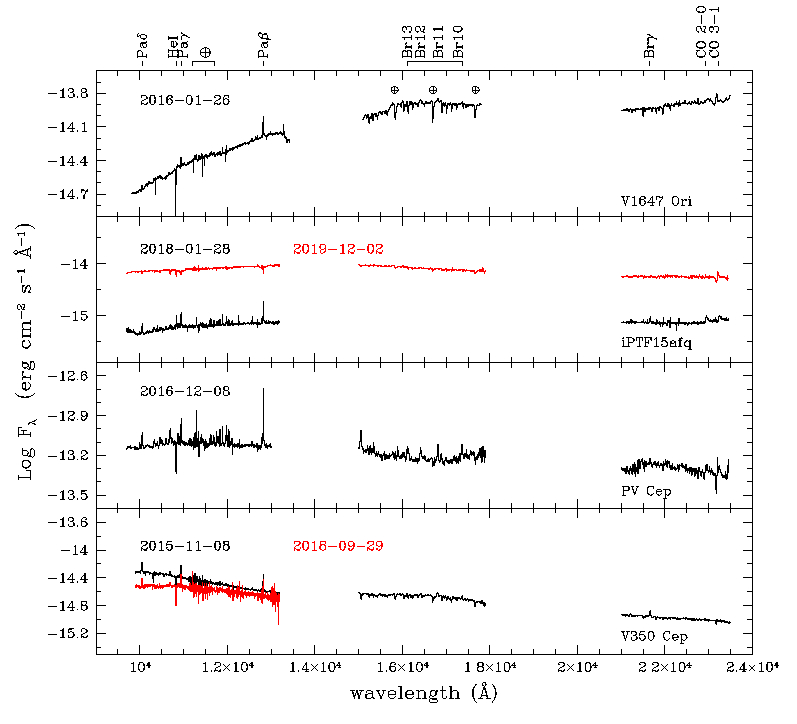}
\caption{As in Figure\,\ref{fig:fig4}.
\label{fig:fig5}}
\end{figure*}

Figures\,\ref{fig:fig2}-\ref{fig:fig3} and Figures\,\ref{fig:fig4}-\ref{fig:fig5} show the MODS and LUCI spectra of our sample, respectively. Line fluxes have been derived by fitting the profiles with a Gaussian function, and the relative uncertainties have been estimated by multiplying the {\it rms} noise of the continuum close to the line times the FWHM. Appendix\, \ref{sec:appendix_a} (Tables\,\ref{tab:A_1}\,-\,\ref{tab:A_11}) report the fluxes of lines used in the analysis.

Lines tracing accretion have been detected in all sources. In particular, the Balmer recombination lines (typically from H$\alpha$ to H$\delta$, but for many sources up to H15) are detected in all MODS spectra. In several  objects (NY Ori, DR Tau, ASASSN-13db, V1647 Ori, PV Cep) some of these lines are partially or totally seen in absorption. Note that, given the late spectral type of the sources (Table\,\ref{tab:tab3}), the photospheric contribution to the lines seen in absorption is negligible, with the possible exceptions of NY Ori and PV Cep.
Paschen and Brackett lines (both in the optical and in near-IR range) are detected in 10 and 5 objects, respectively, and appear all in emission. In all objects observed with LUCI, we detect the \hei\, 1.08\,$\mu$m line, while other
  \hei\, optical lines are seen in the spectra of the brightest objects. Metallic lines of many species (\fei, \feii, \ti, \tii, \caii, \cri, \crii) are also detected.\\
  Signatures of ejection activity, such as the \oi\,6300 \,\AA\, line, are present  in all sources but VY Tau and V1143 Ori.
  In two objects (iPTF15afq and V350 Cep) we also detected \htwo\, emission at 2.12 \um, which is a typical tracer of molecular outflows. In XZ Tau, UZ Tau E, DR Tau, and PV Cep, we detect some \sii\, lines, indicative of gas with a low ionization degree. Lines of  \sii, [\feii], \nic, and \nii\, have been detected in the spectrum of iPTF15afq, which could signal the occurrence of a high-velocity jet (Giannini et al. 2019).
  
In some of the LUCI spectra we have identified at least one of the two CO ro-vibrational bandheads (v=2-0 and v=3-1) in emission  (UZ Tau E, iPTF15afq, and PV Cep), or in absorption (VY Tau), as expected for EXor sources (e.g. Biscaya et al. 1997, Hillenbrand et al. 2013). In the remaining objects the CO bandheads are undetected, probably because the edge of the HK spectra is typically very noisy.
   
In general, we remark that: 1) the spectra well resemble the optical/near-IR spectra of accreting T Tauri stars (e.g. Alcal{\'a} et al. 2017) and no evident differences exist among the spectra of known EXors and EXor candidates;  2) the intensity of the accretion tracers (e.g. \hi\,, \caii) follows the variation of the source brightness (e.g. XZ Tau, Table\,\ref{tab:A_1}) and may even disappear or change from emission to absorption in case of an outburst event (ASASSN-13db and iPTF15afq).

\section{Analysis and discussion\label{sec:sec6}}
\subsection{Extinction\label{sec:sec6.1}}
The primary goal of the present work is to determine the accretion parameters (\lacc\, and \macc\,) of the sample. To get a meaningful estimate of these two quantities we first  derived the visual extinction (\av)\, toward the sources. As listed in Table\,\ref{tab:tab3}, one or more \av\, values are reported in the literature for some of them. Although these values represent an important reference,
several studies have shown that a remarkable reduction (e.g. Hillenbrand et al. 2013), or less often, an increase of the extinction (in edge-on disks, Stock et al. 2020)  often accompany accretion burst episodes. Therefore, we cannot assume \av\,  as a constant, but we need to estimate its value for each observation date.
In the following, we separately discuss the methods used to derive \av\, from both the optical and near-IR spectra.  

\subsubsection{Extinction derived from optical spectra\label{sec:sec6.1.1}}
The method used to derive the extinction from the optical spectra is described in detail in Giannini et al. (2018). This is essentially based on the empirical relationships found by Alcal{\'a} et al. (2014, 2017),
between the accretion luminosity, \lacc, and the luminosities \lumi\, of 
selected emission lines (so-called 'line method'). In the optical range these relationships exist for more than 20 lines, namely the \hi\, recombination lines of the Balmer and 
Paschen series from H$\alpha$ to H15, and from Pa8 to Pa10, along with 
\hei\,, \ion{O}{1}, and \ion{Ca}{2} lines. Therefore, the relationships  
provide up to around 20 independent estimates of \lacc\,, namely one for 
each observed line (\lacci\,). 
 
In our procedure, we fit simultaneously \lacc\, and \av\,, this latter allowed to vary between 0 and 15 mag (in steps of 0.20 mag). The extinction law by Cardelli et al. (1989) and total-to-selective extinction ratio R$_V$\,=\,3.1 are assumed. We iteratively fit \lacc\, to minimize the dispersion among the individual \lacci\,, computed by de-reddening the observed fluxes for the current \av\, and assuming the distance listed in Table\,\ref{tab:tab1}. 
The best estimate of \av\, is the value for which the dispersion between the \lacci\, is minimized and we assume as \lacc\, the average of the \lacci\, corresponding to that \av. 

\begin{figure*}
\begin{center}
\includegraphics[trim=0 0 0 0,width=1.0\columnwidth, angle=0]{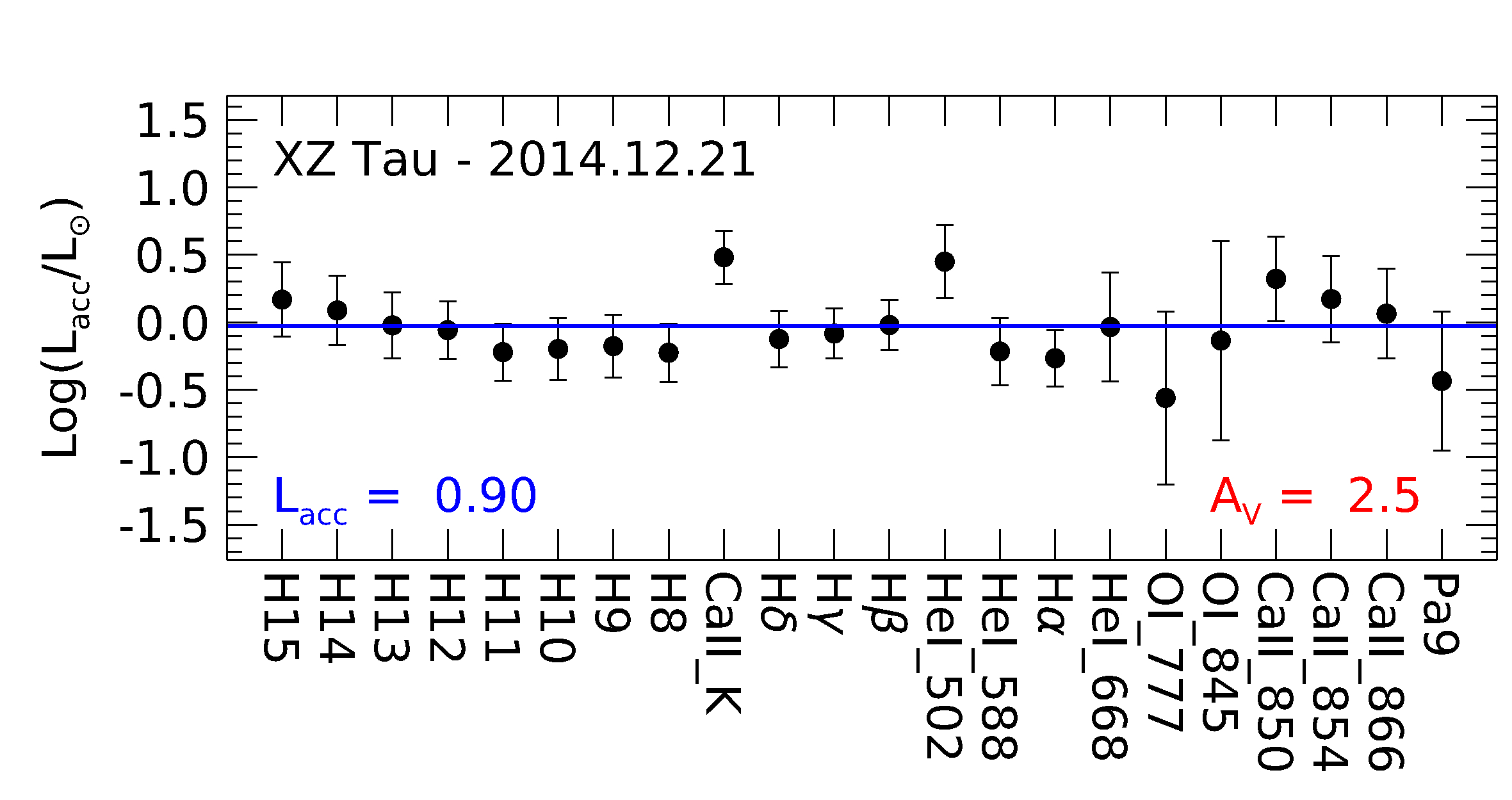}
\includegraphics[trim=0 0 0 0,width=1.0\columnwidth, angle=0]{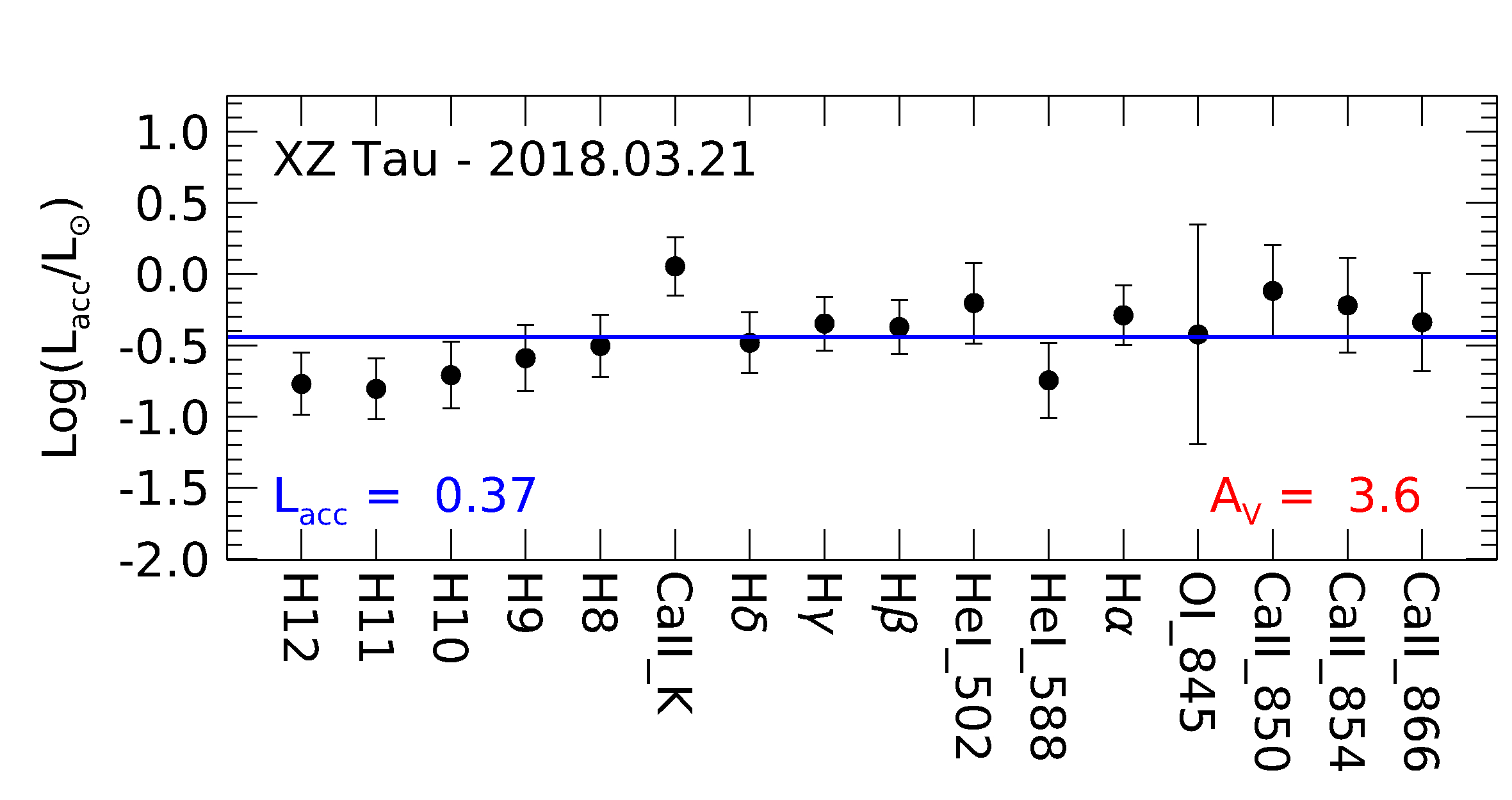}
\includegraphics[trim=0 0 0 0,width=1.0\columnwidth, angle=0]{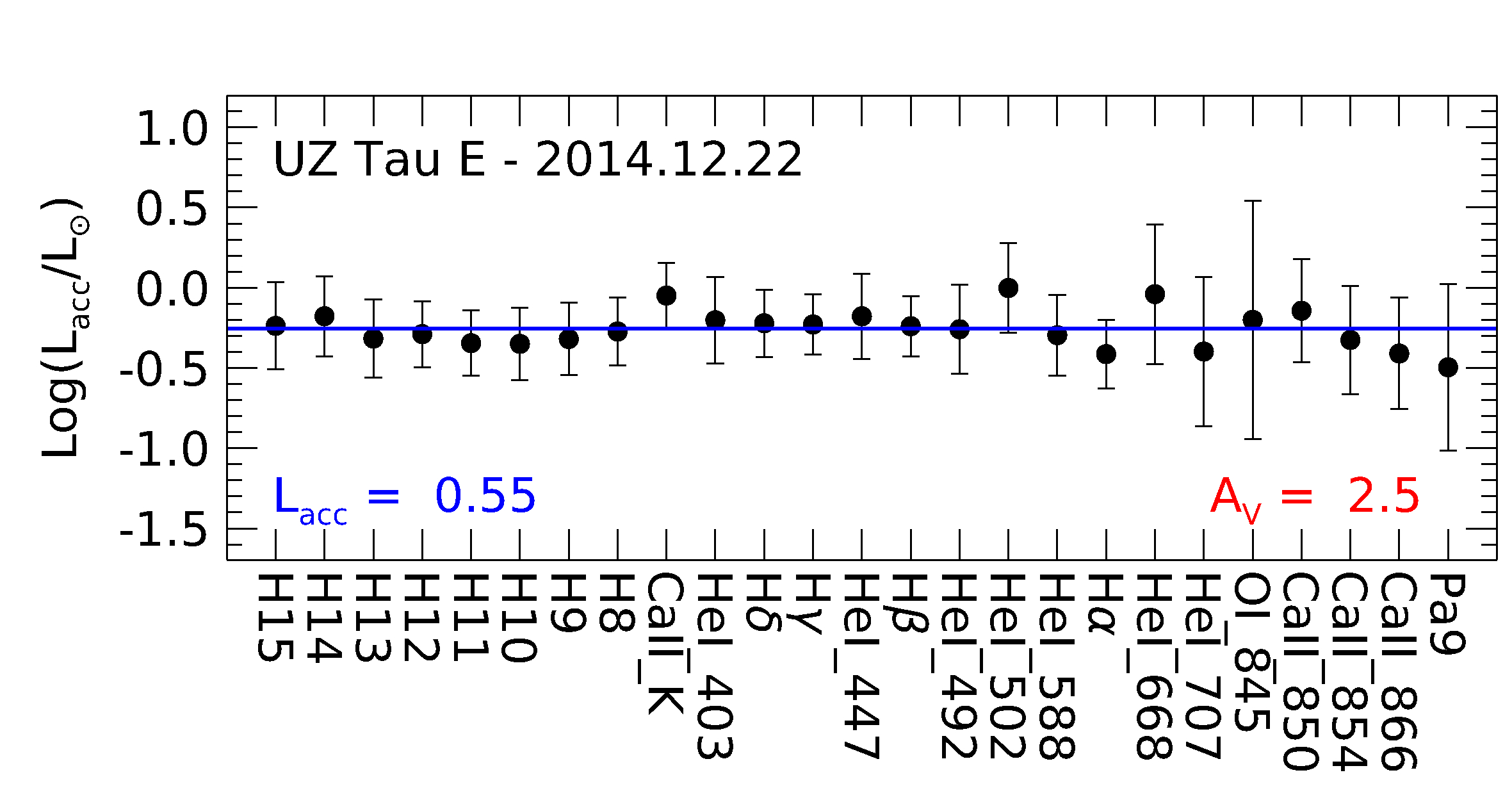}
\includegraphics[trim=0 0 0 0,width=1.0\columnwidth, angle=0]{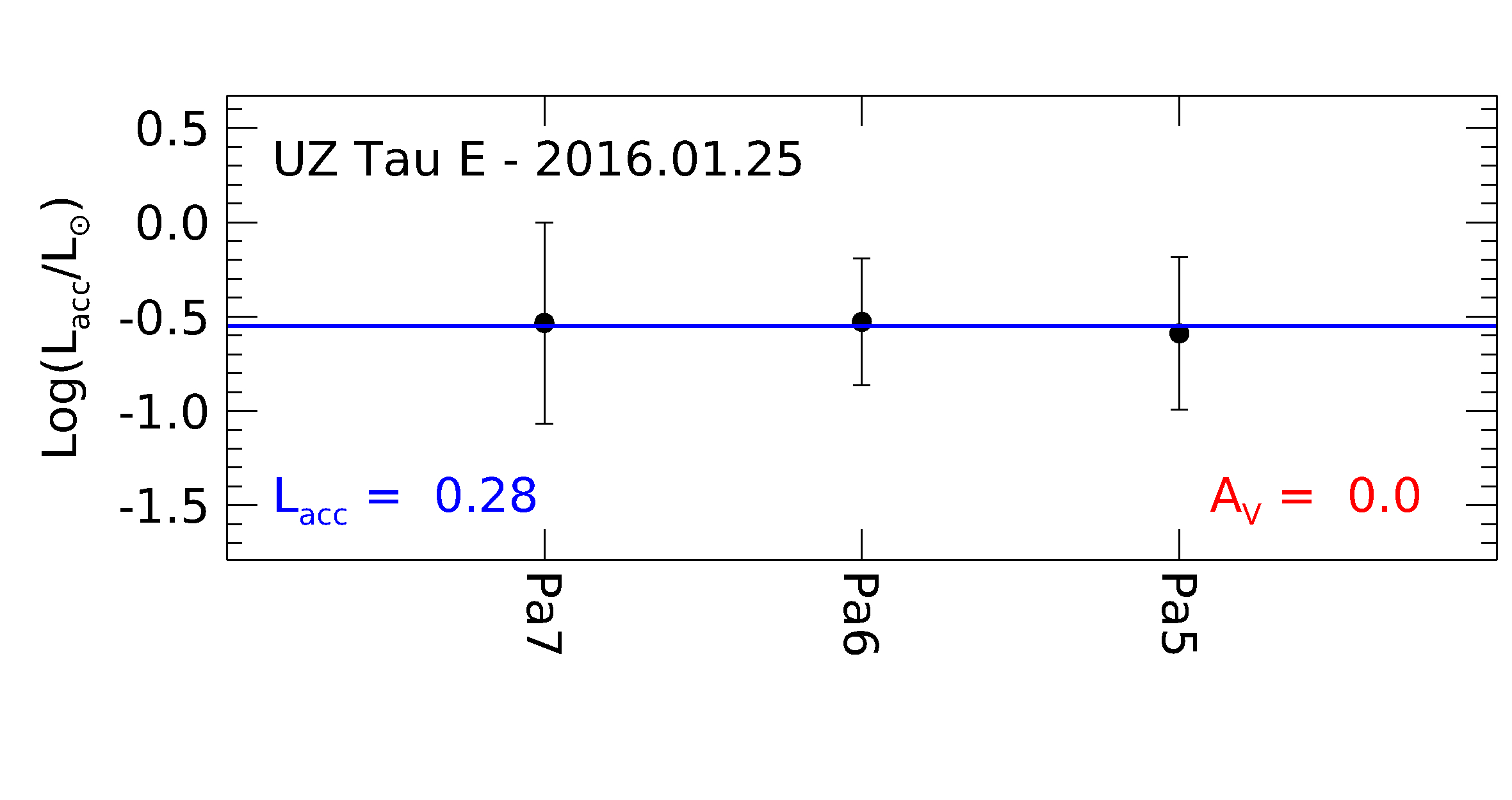}
\includegraphics[trim=0 0 0 0,width=1.0\columnwidth, angle=0]{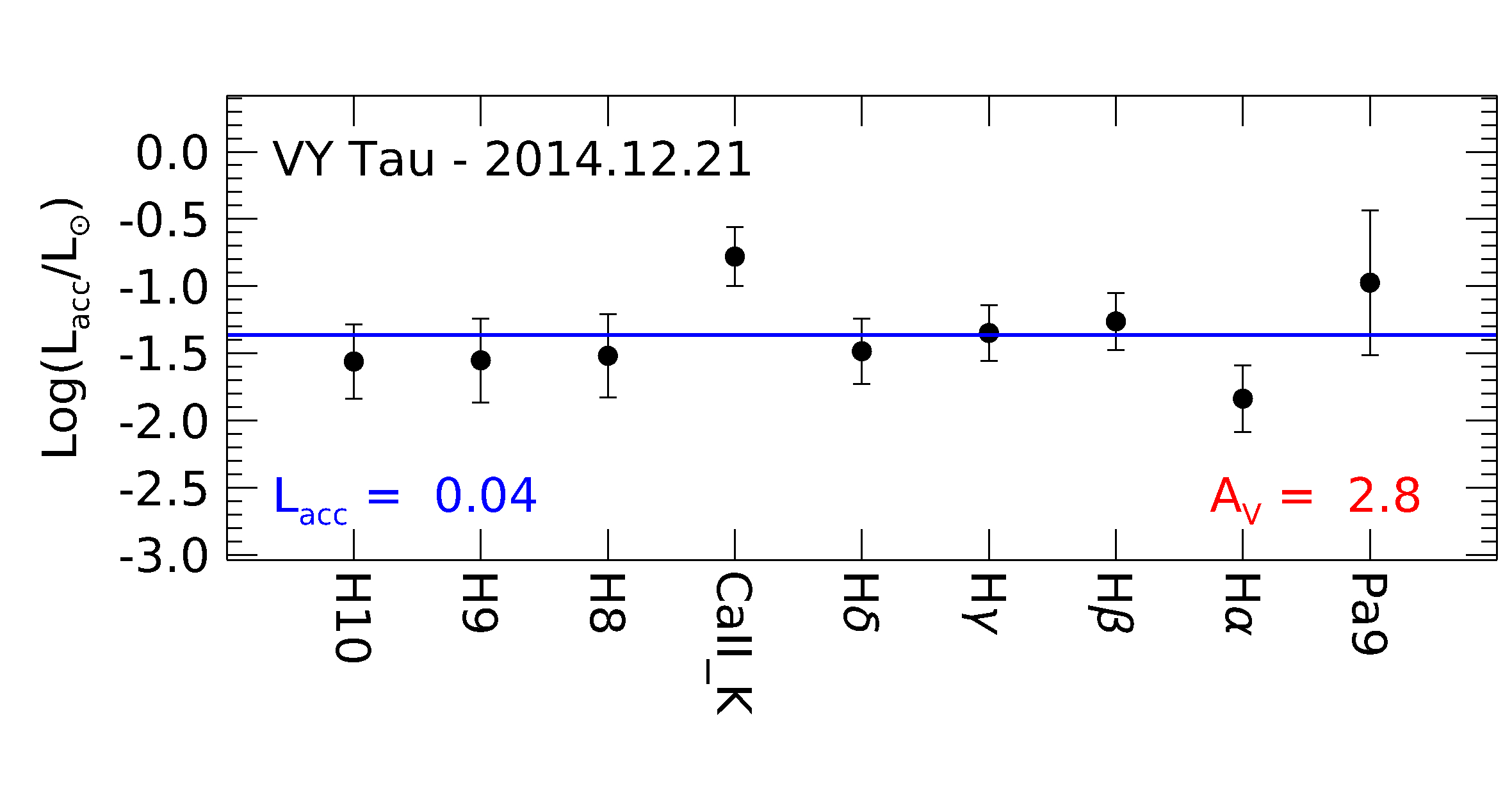}
\includegraphics[trim=0 0 0 0,width=1.0\columnwidth, angle=0]{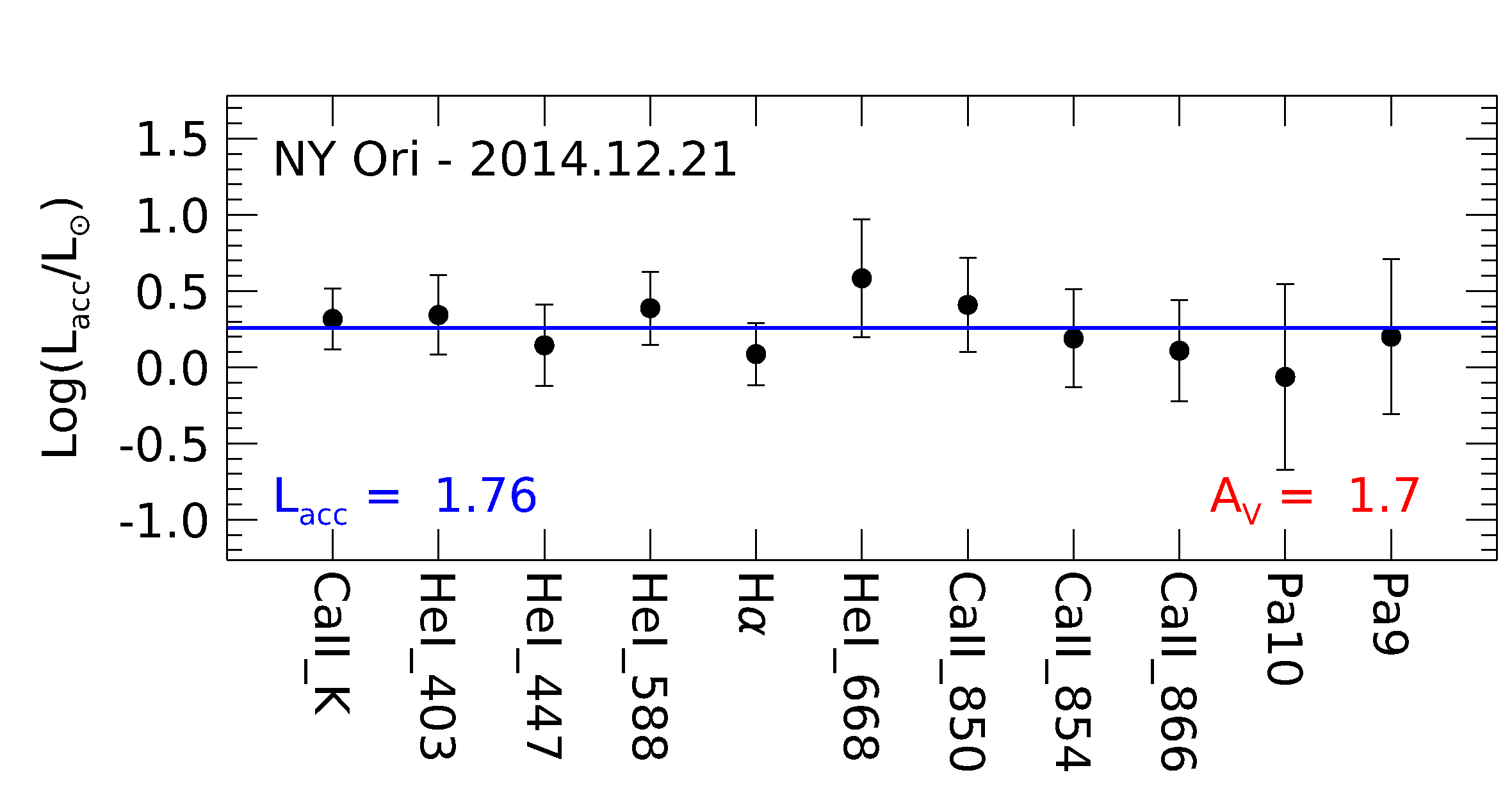}
\includegraphics[trim=0 0 0 0,width=1.0\columnwidth, angle=0]{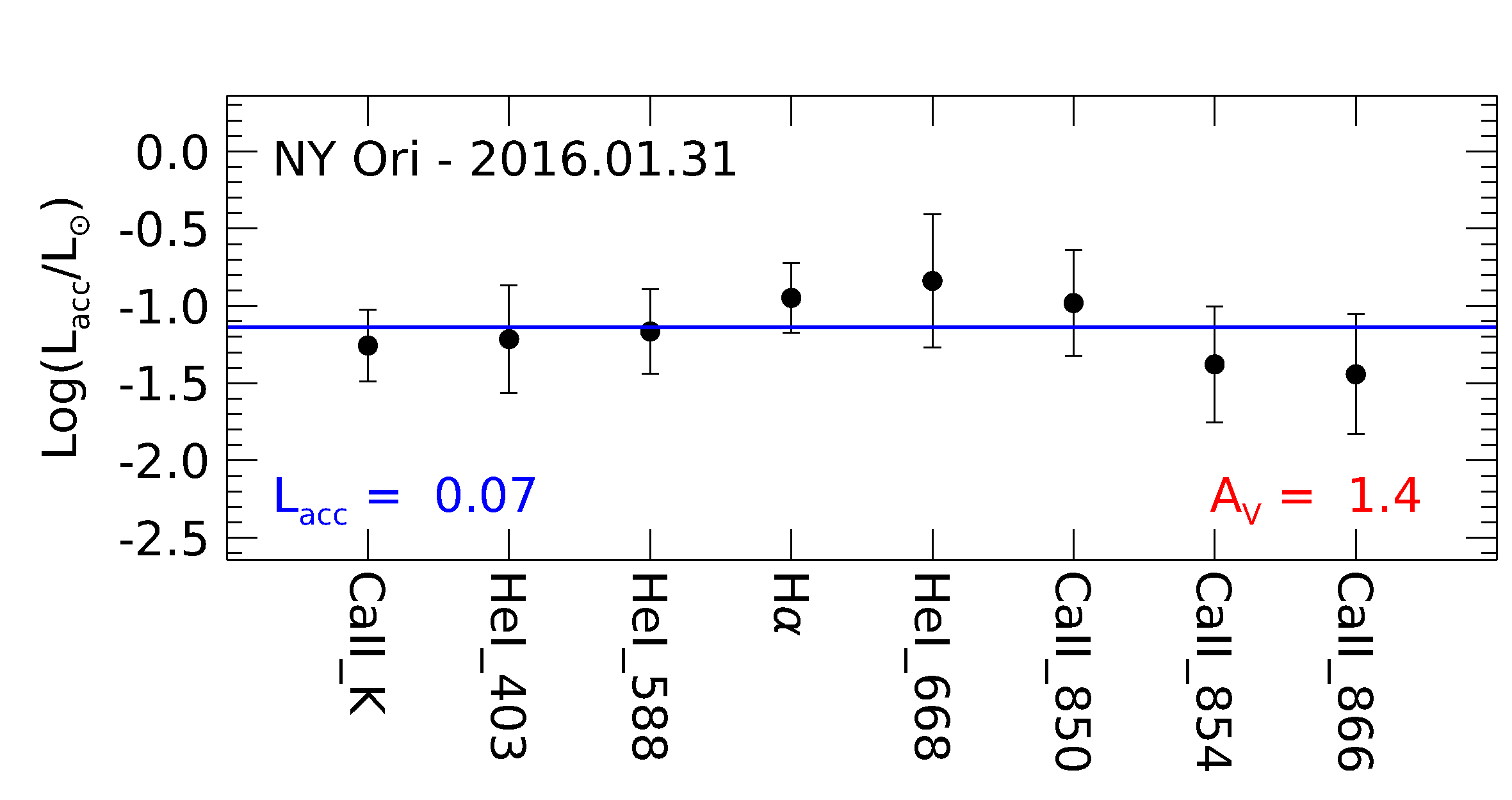}
\includegraphics[trim=0 0 0 0,width=1.0\columnwidth, angle=0]{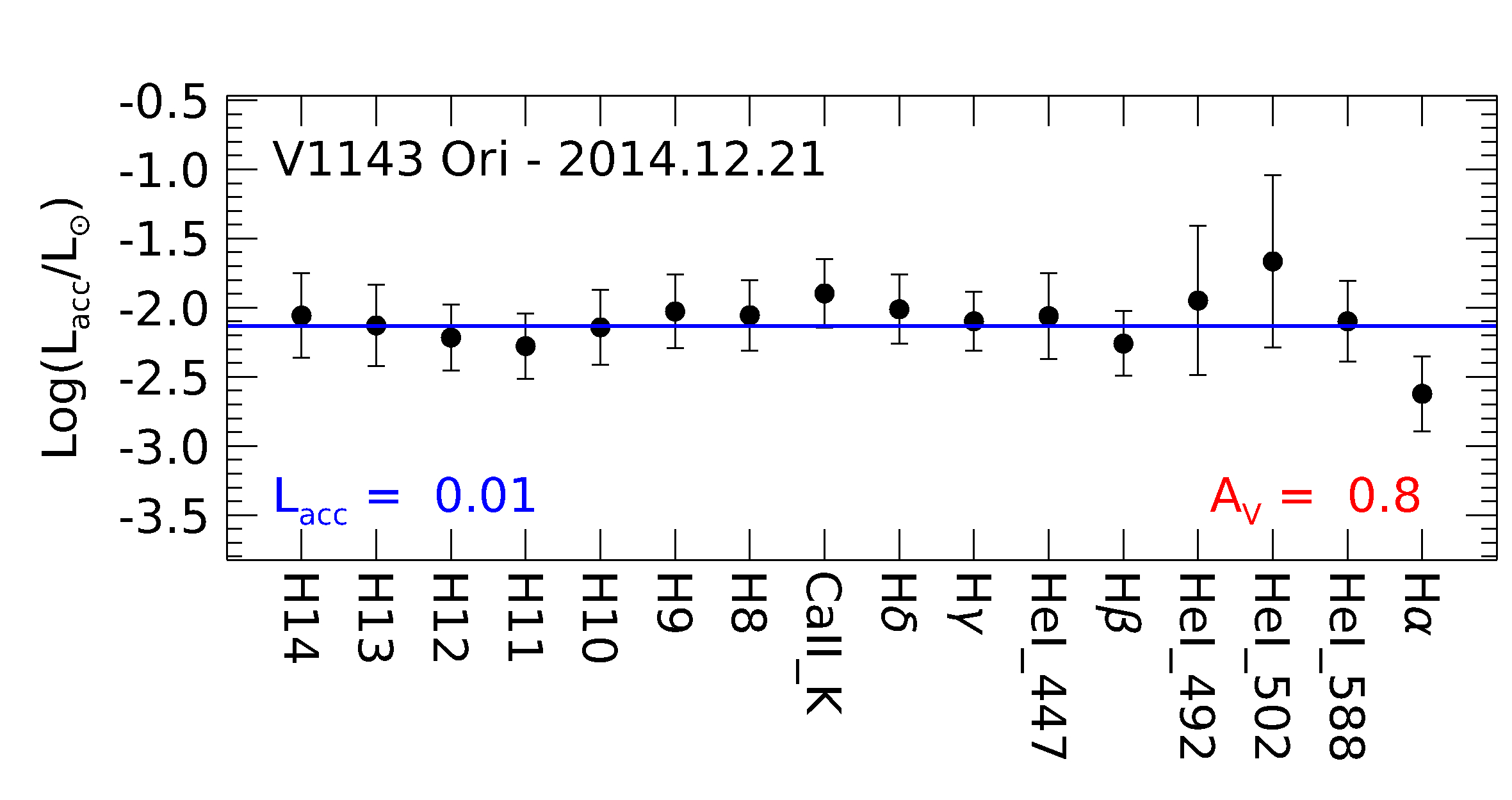}
\end{center}
\caption{Accretion luminosity derived from the flux of the indicated tracers by using the relationships by Alcal{\'a} et al. (2017). The horizontal line shows the average \lacc\,. The source name, the observation date, and the fitted values of \lacc\,, and \av\, are indicated.
\label{fig:fig6}
}
\end{figure*}

\begin{figure*}
\begin{center}
\includegraphics[trim=0 0 0 0,width=1.0\columnwidth, angle=0]{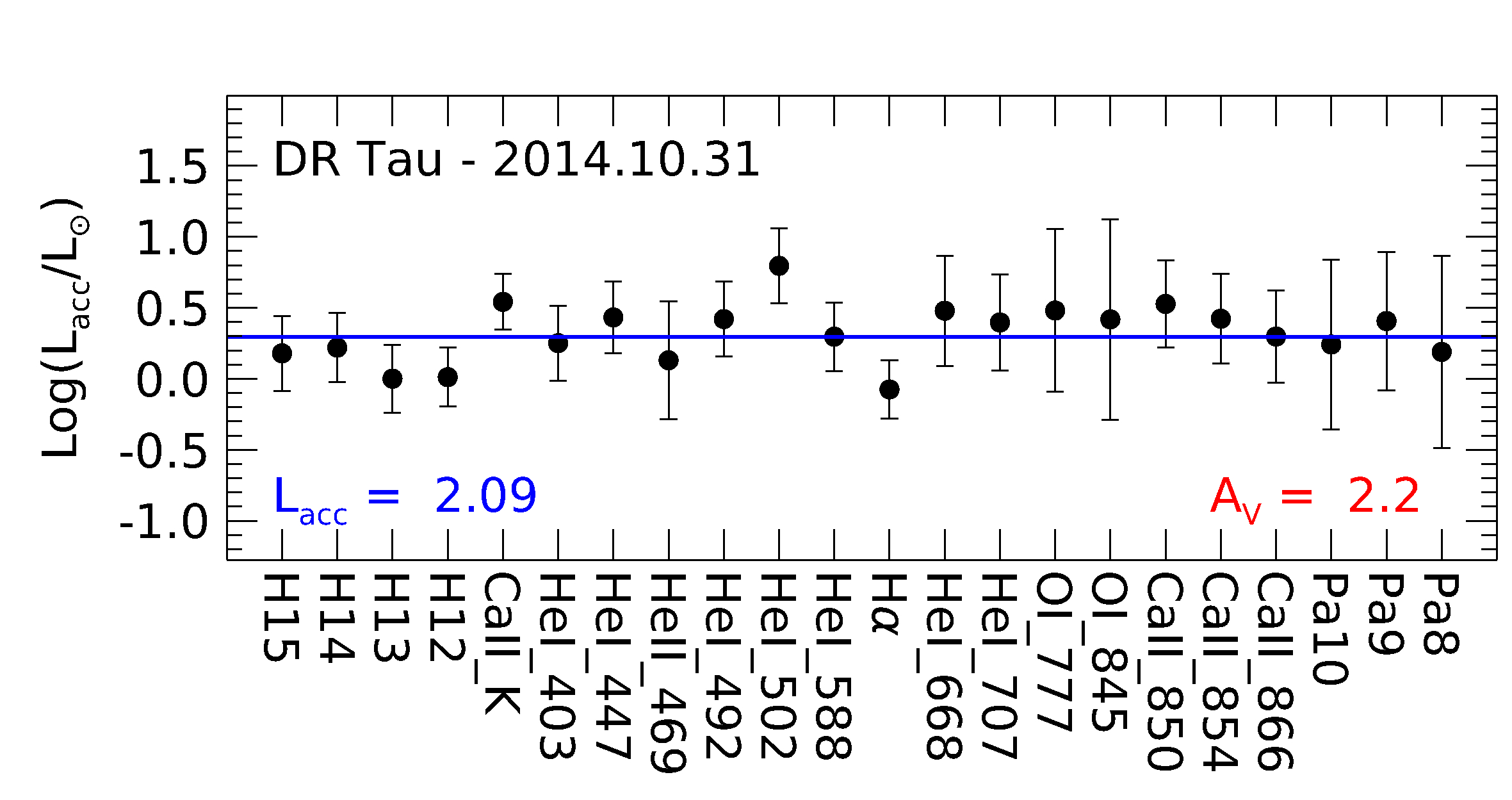}
\includegraphics[trim=0 0 0 0,width=1.0\columnwidth, angle=0]{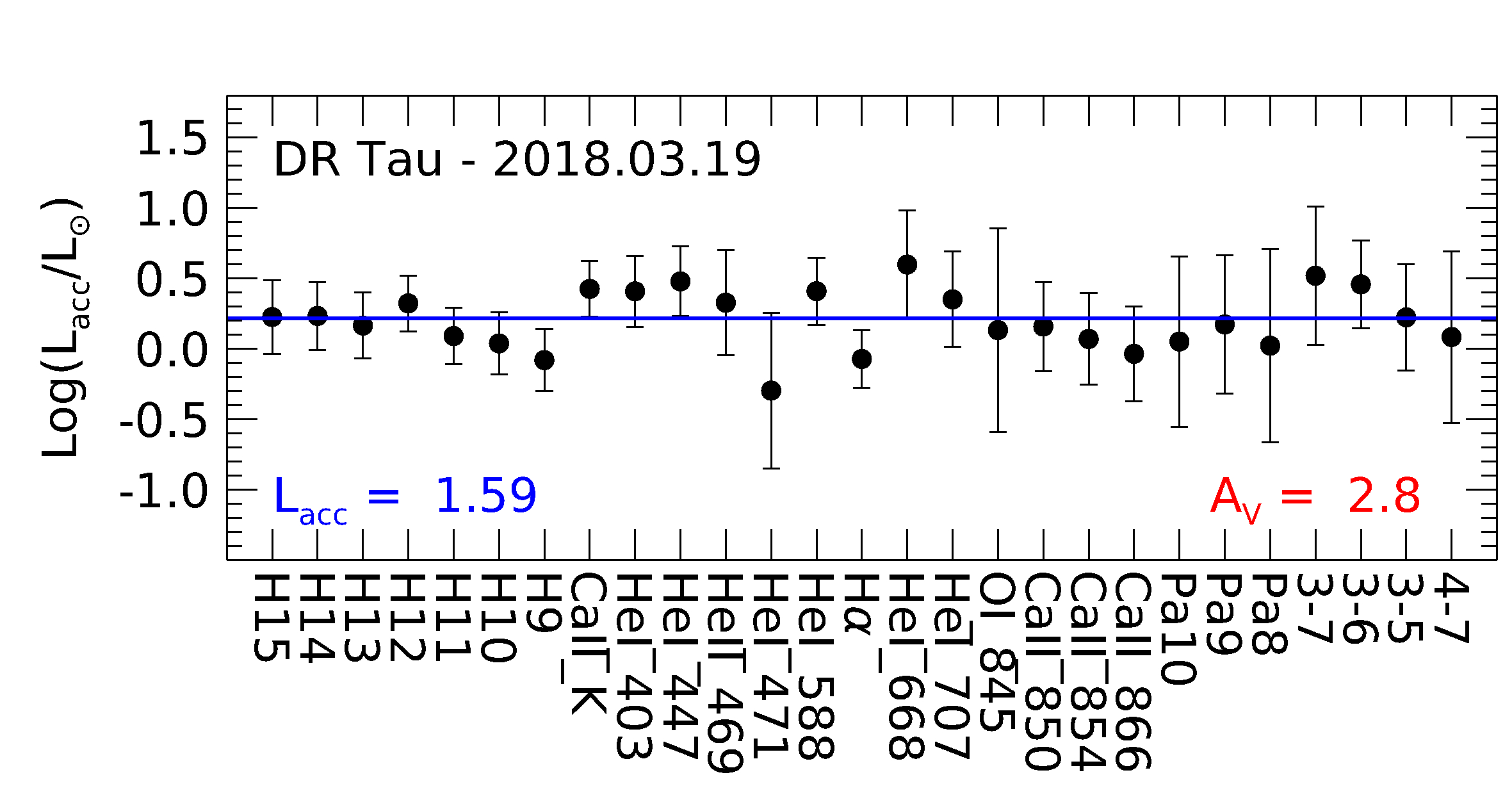}
\includegraphics[trim=0 0 0 0,width=1.0\columnwidth, angle=0]{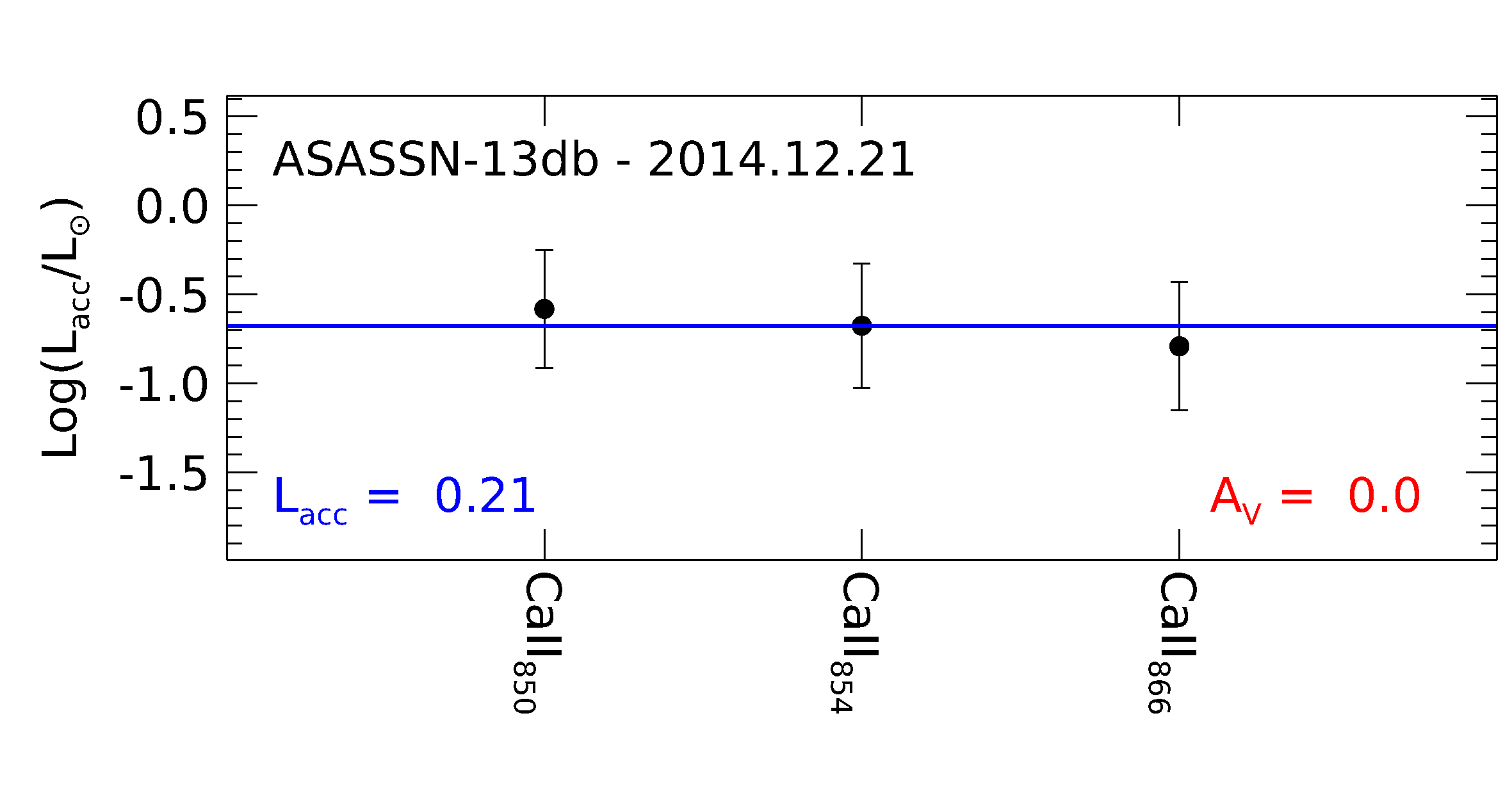}
\includegraphics[trim=0 0 0 0,width=1.0\columnwidth, angle=0]{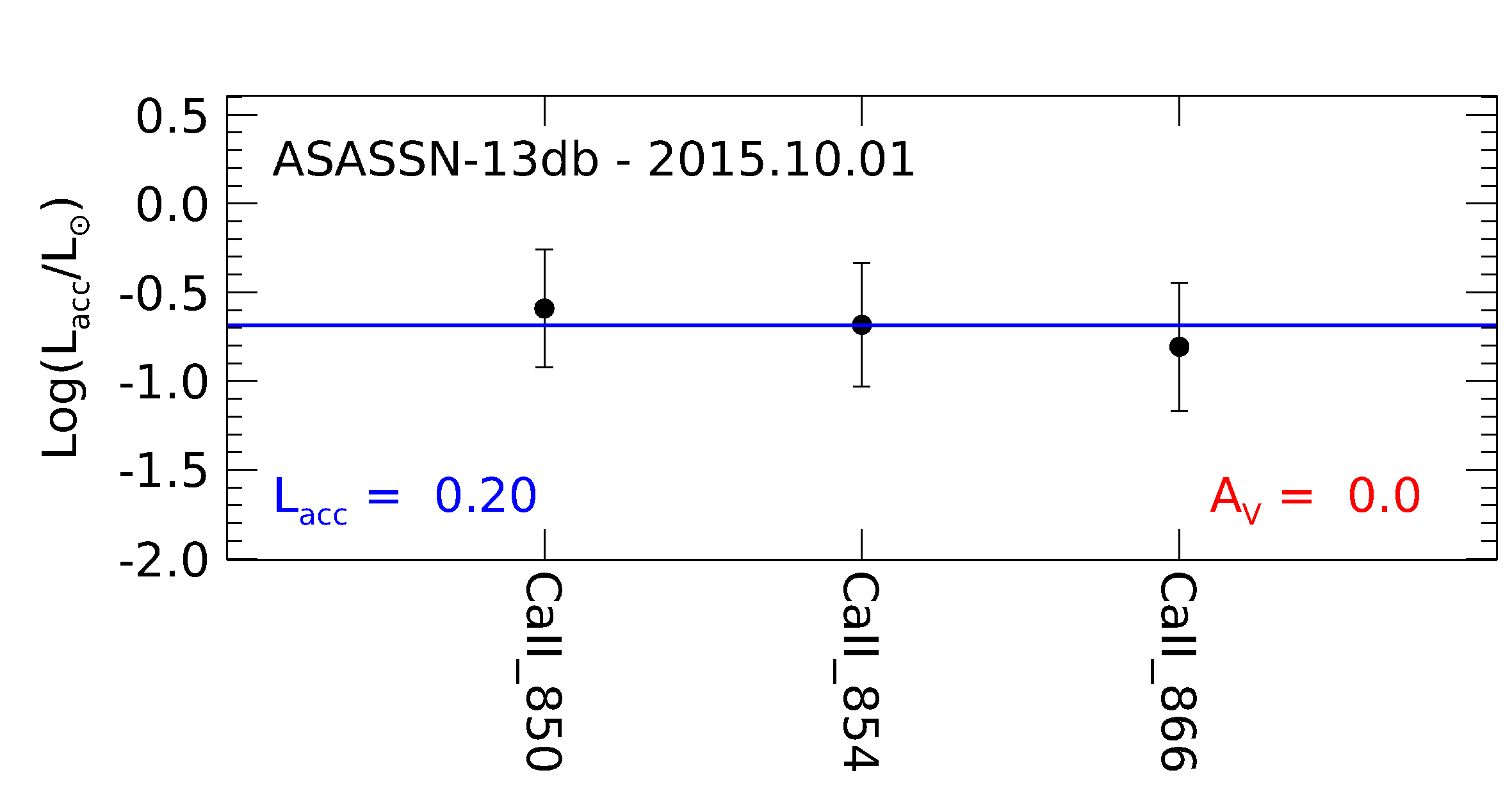}
\includegraphics[trim=0 0 0 0,width=1.0\columnwidth, angle=0]{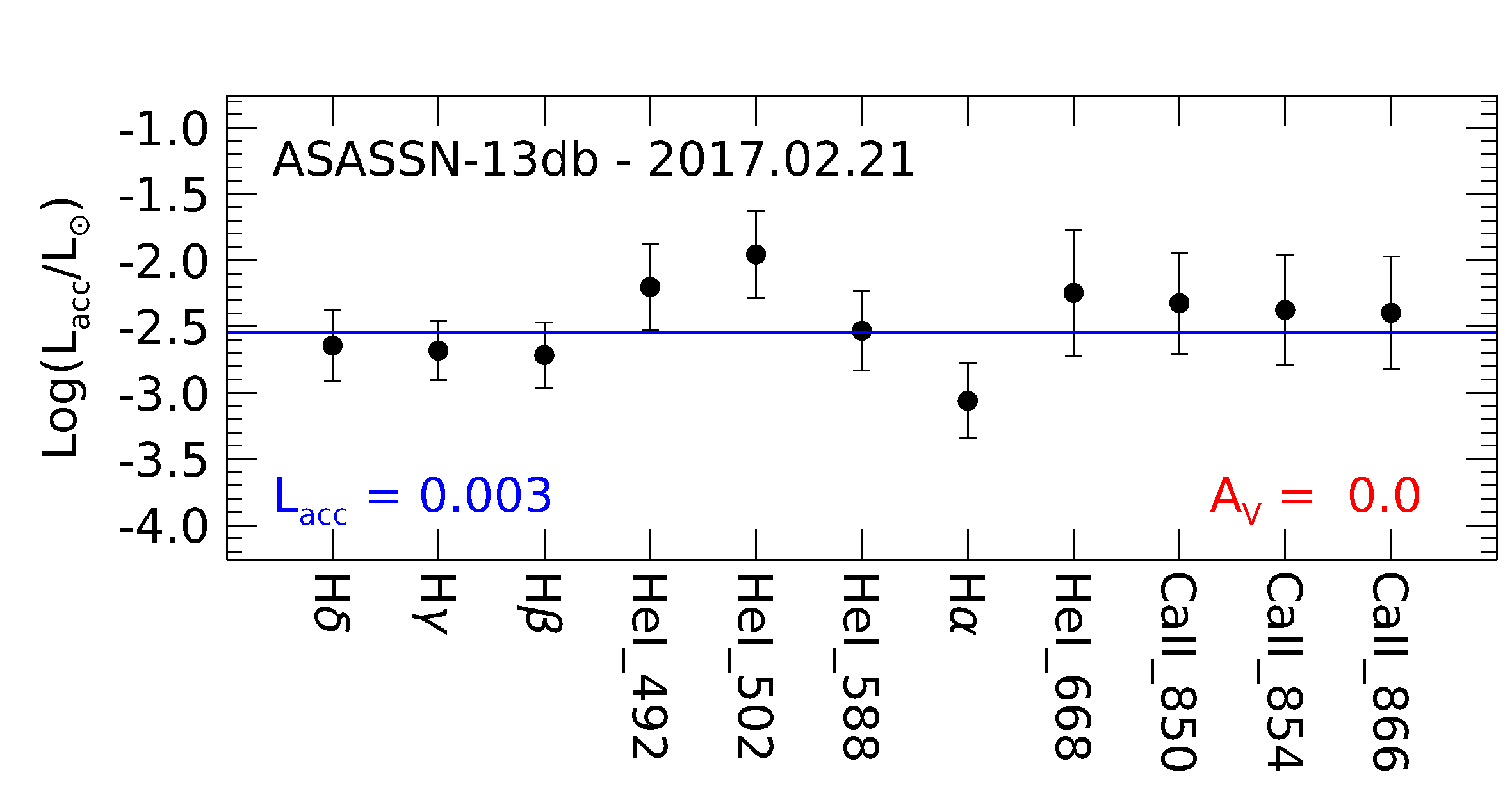}
\includegraphics[trim=0 0 0 0,width=1.0\columnwidth, angle=0]{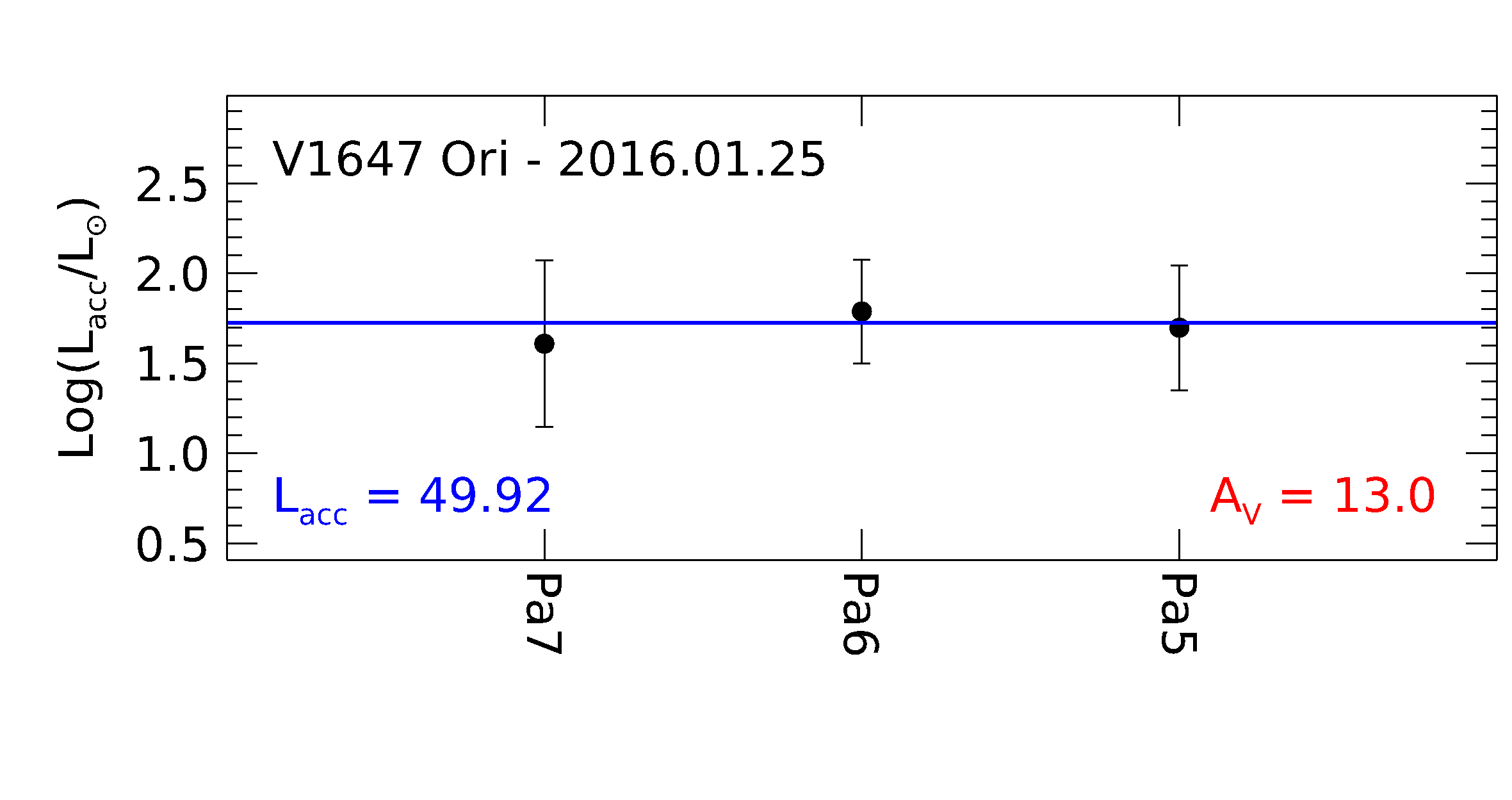}
\includegraphics[trim=0 0 0 0,width=1.0\columnwidth, angle=0]{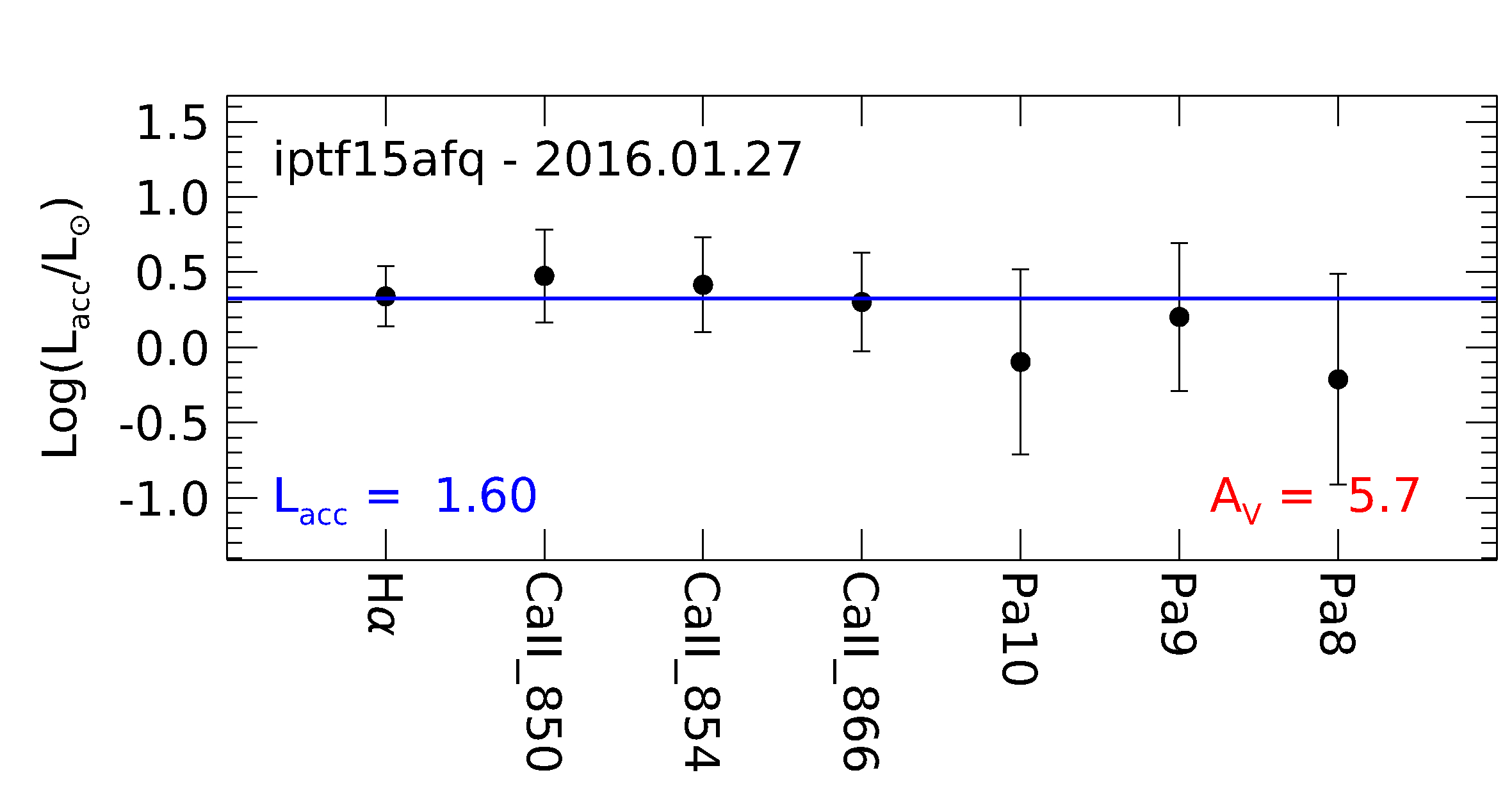}
\includegraphics[trim=0 0 0 0,width=1.0\columnwidth, angle=0]{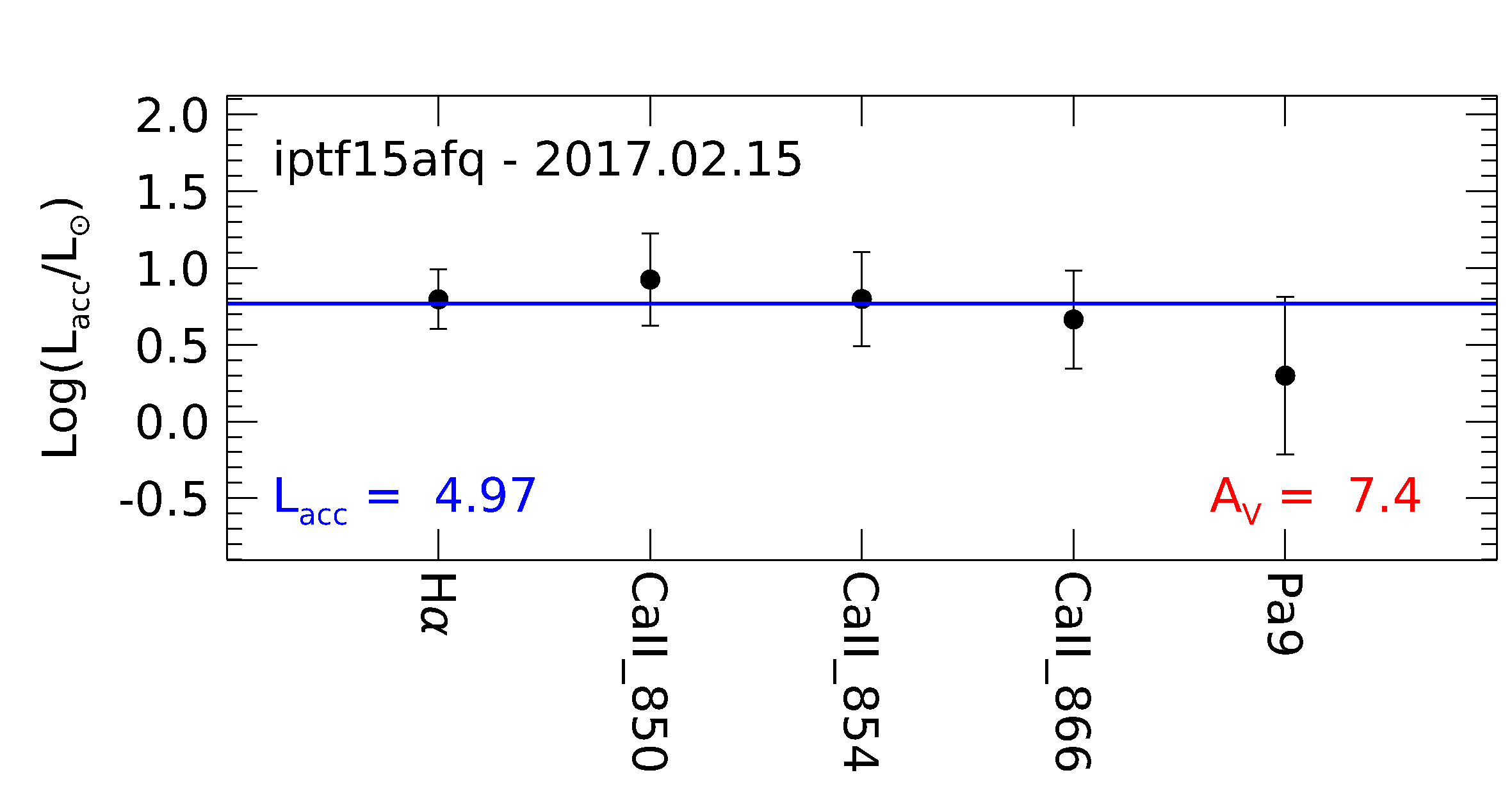}
\end{center}
\caption{As in Fig.\,\ref{fig:fig6}.
\label{fig:fig7}
}
\end{figure*}

\begin{figure*}
\begin{center}
\includegraphics[trim=0 0 0 0,width=1.0\columnwidth, angle=0]{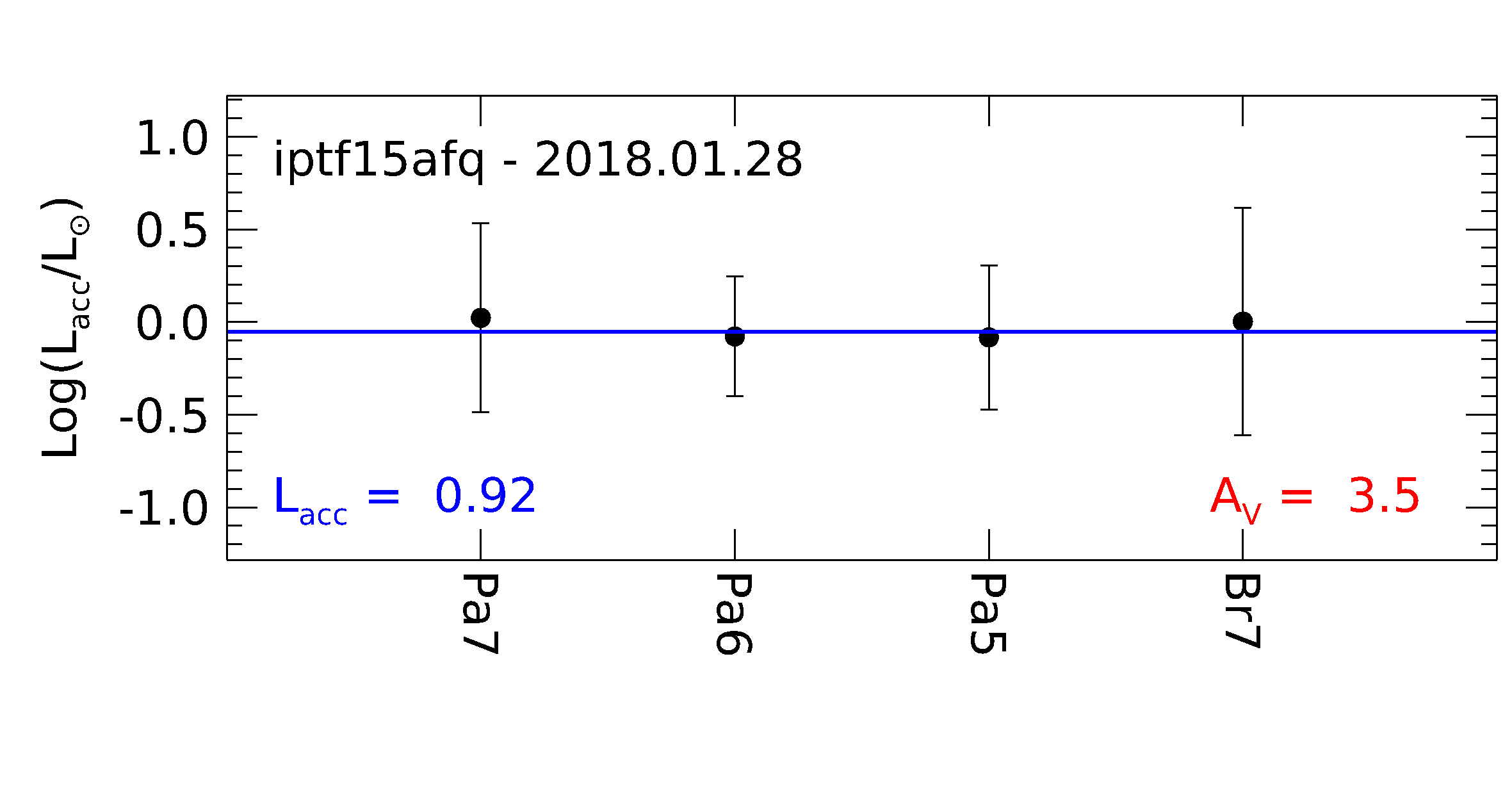}
\includegraphics[trim=0 0 0 0,width=1.0\columnwidth, angle=0]{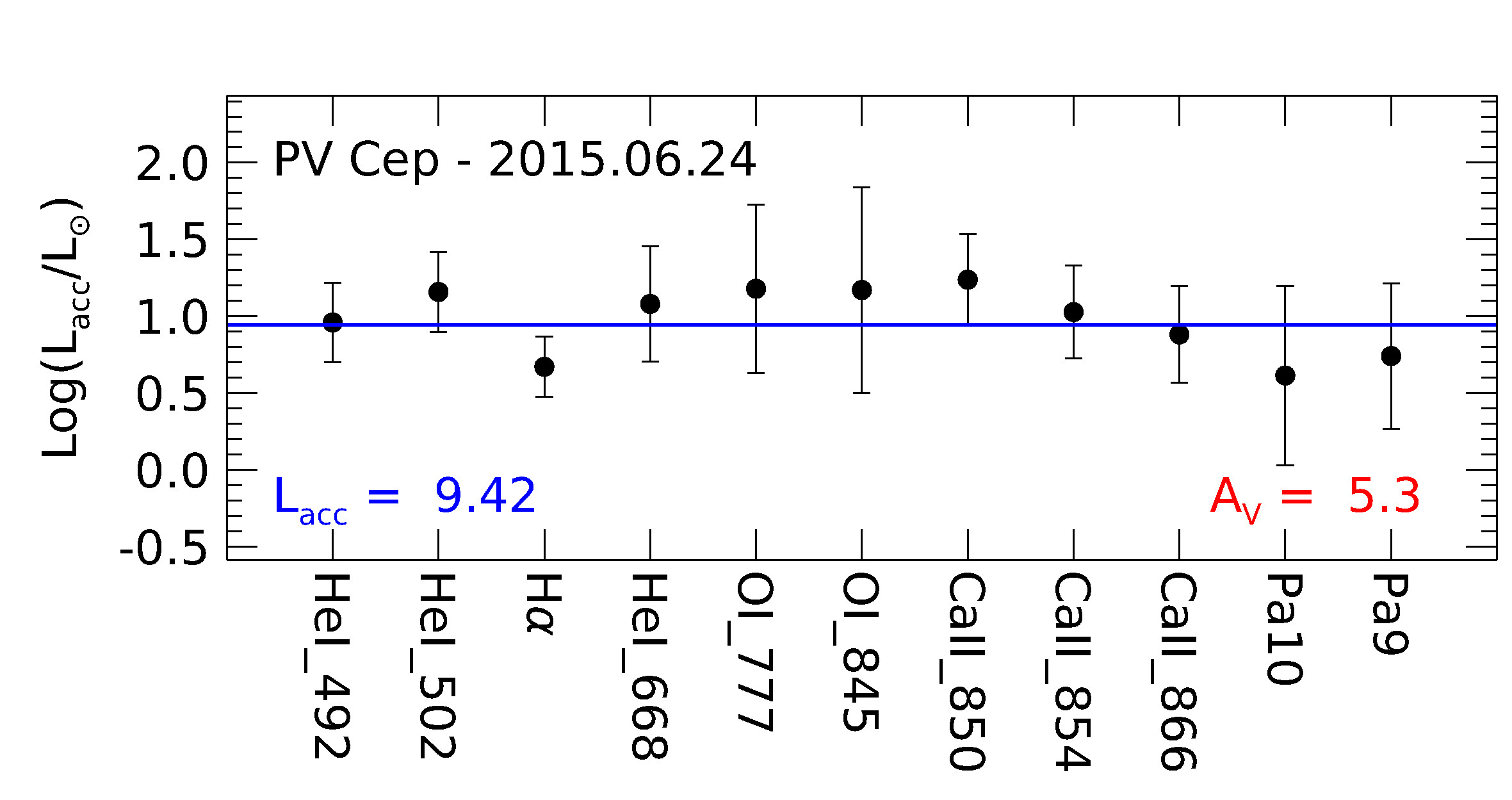}
\includegraphics[trim=0 0 0 0,width=1.0\columnwidth, angle=0]{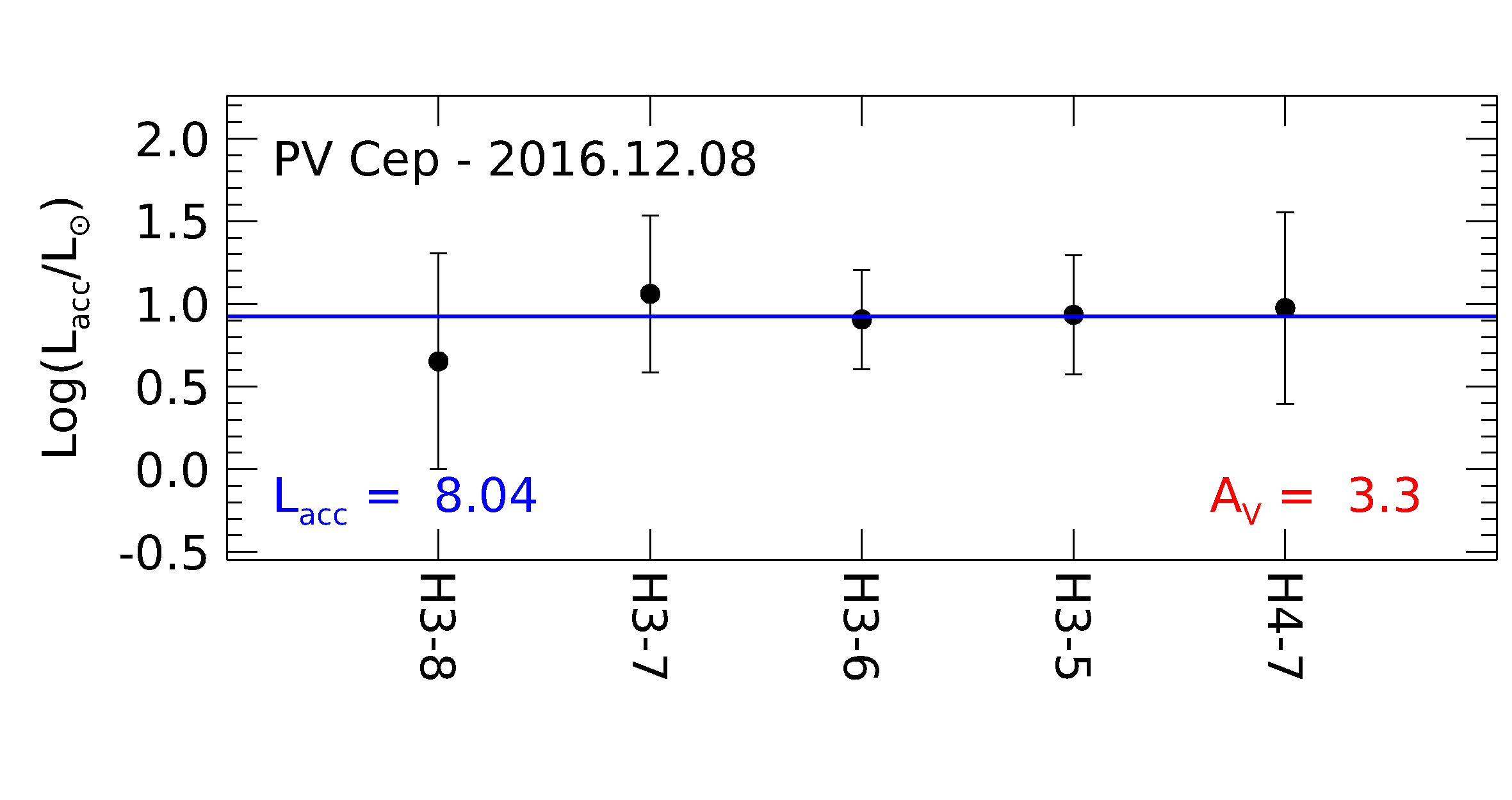}
\includegraphics[trim=0 0 0 0,width=1.0\columnwidth, angle=0]{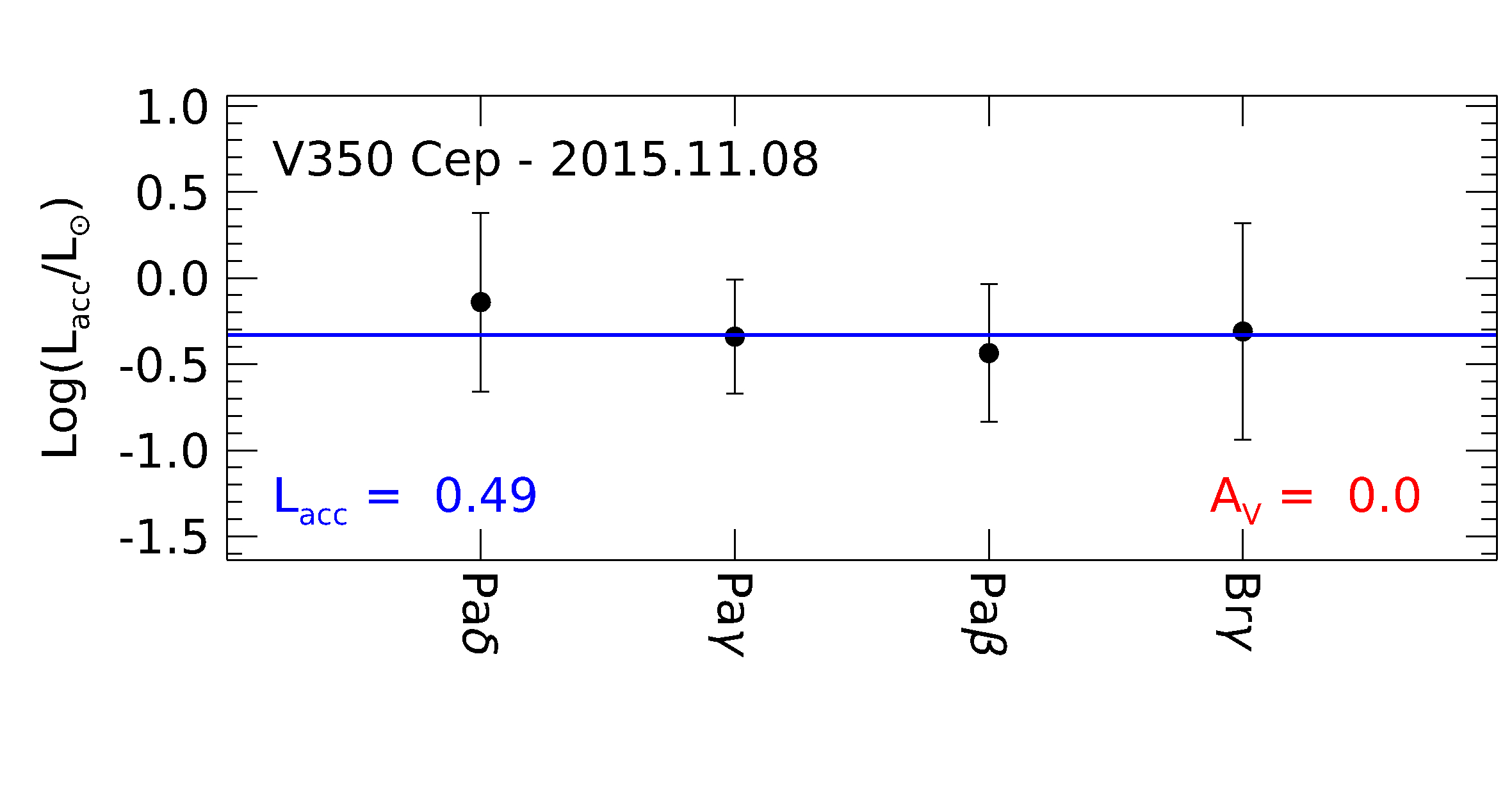}
\includegraphics[trim=0 0 0 0,width=1.0\columnwidth, angle=0]{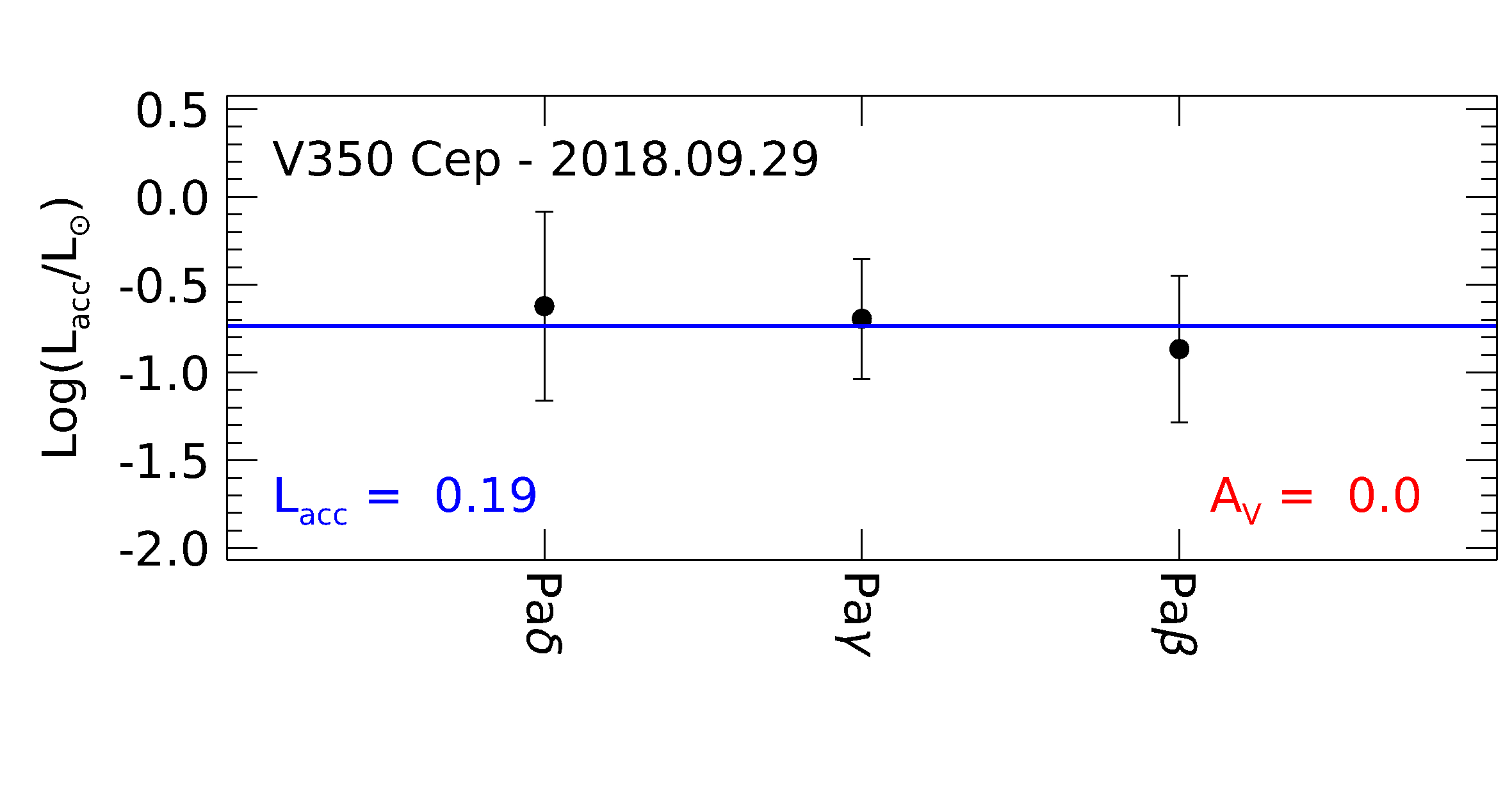}
\includegraphics[trim=0 0 0 0,width=1.0\columnwidth, angle=0]{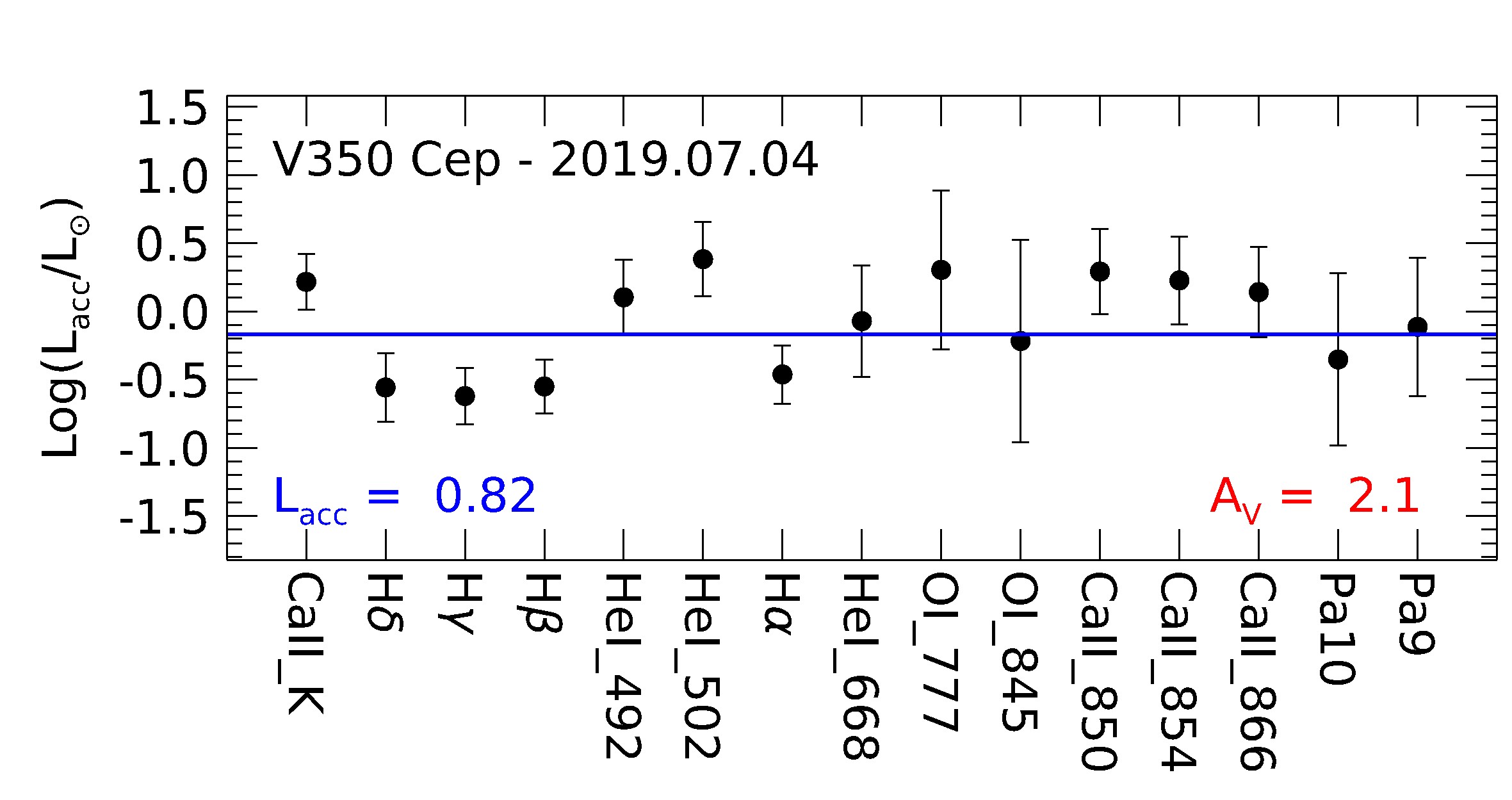}
\includegraphics[trim=0 0 0 0,width=1.0\columnwidth, angle=0]{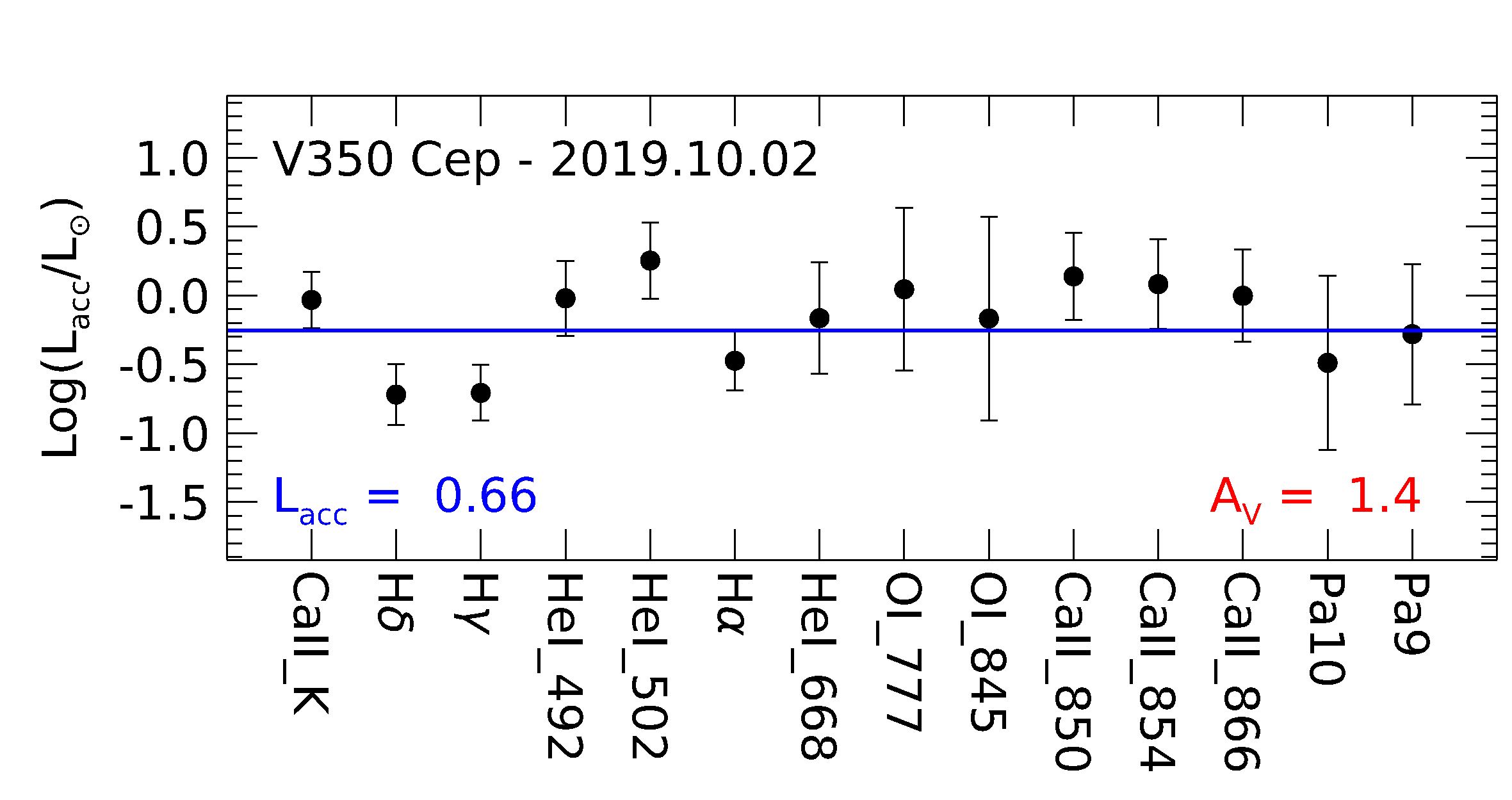}
\includegraphics[trim=0 0 0 0,width=1.0\columnwidth, angle=0]{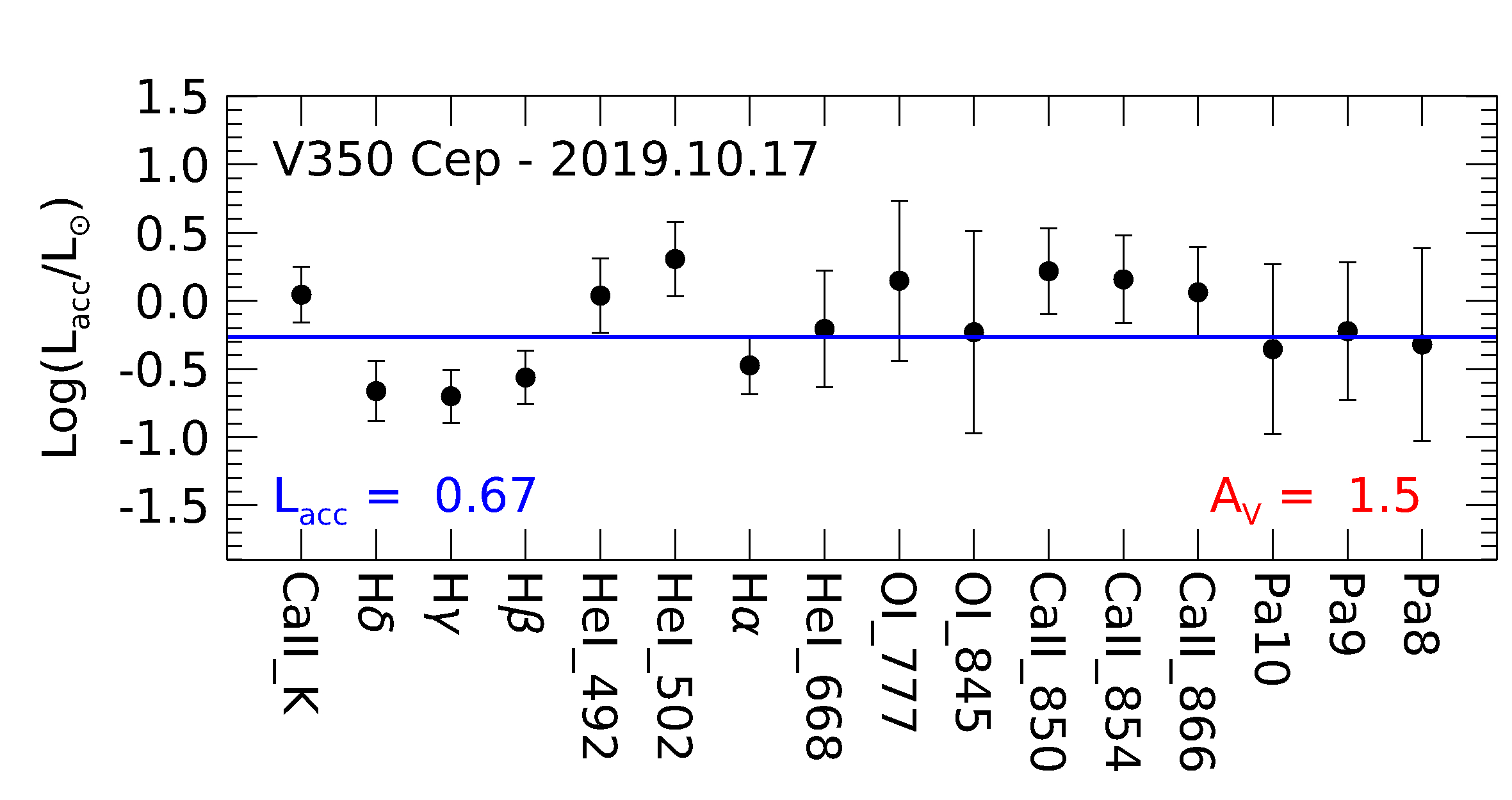}
\end{center}
\caption{As in Fig.\,\ref{fig:fig6}.
\label{fig:fig8}
}
\end{figure*}

We included in the fit only the lines that do not present any absorption component. We have verified that, rather than on the individual flux
errors, the main limitation of this method comes from the uncertainties on
the empirical relationships (Alcal{\'a} et al. 2017), which
give an error on \av\, of 0.5$-$1.5 mag, depending on how many lines are considered
in the fit. Further causes of uncertainty come from the assumptions on the extinction law and R$_V$. For example,
we have checked that if the extinction law by Weingartner \& Draine (2001) is applied, the fitted \av\, decreases by $\approx$\, 0.4 mag. Conversely, an increase in \av\, by $\sim$ 0.2-0.3 mag is produced for R$_V$\,=\,5.5. \\
In Figures\, \ref{fig:fig6}-\ref{fig:fig8} we show the results of our procedure, while in Table\,\ref{tab:tab5} we report the fitted extinction values and the number of lines used. 

\begin{figure*}
\begin{center}
\includegraphics[trim=10 5 5 5,width=2.\columnwidth, angle=0]{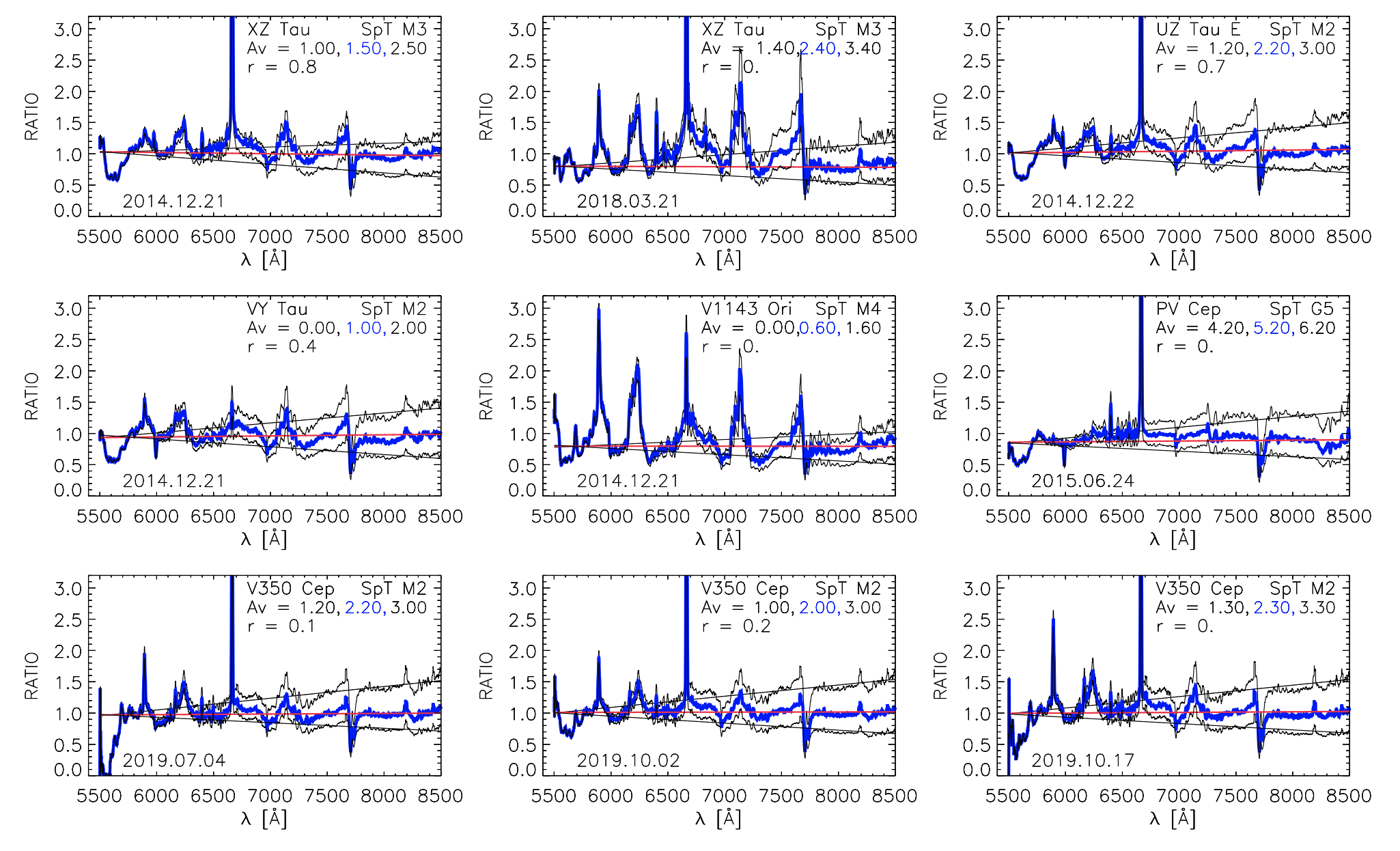}
\end{center}
\caption{Ratios between observed spectra and artificially reddened and veiling-corrected templates.
Ratios with minimum slopes (for which the extinction \av\, is estimated) are indicated in blue. For each panel, object name, date of observation, spectral type, and veiling are also reported.\label{fig:fig9}}
\end{figure*}

An independent way  to compute \av\, has been described by Alcal{\'a} et al. (2021, so-called 'continuum' method). 
The observed spectrum is iteratively divided by a template of the same spectral type (SpT), artificially reddened by varying  the value of \av. The best estimate of \av\, is that for which the ratio has a flat slope. Applying this method to strong accretors, one has to correct the spectrum for the excess continuum (veiling) induced by the accretion hot spot (UV) and disk (IR) emission. This has been estimated for the majority of our sources by applying the procedure described in Appendix\,\ref{sec:appendix_b}.\\
Following the prescriptions by Alcal{\'a} et al. (2014) and Fischer et al. (2011), we applied the continuum method in the spectral range between 550 and 800 nm, where the veiling excess is minimized and expected not to change more than the typical error of 0.3 associated with its determination (Appendix\,\ref{sec:appendix_b}).
We adopted as template spectra a grid of non-accreting (Class\,III) YSOs of the same SpT of the observed sources (Manara et al. 2013, 2017), reddened for \av\  between 0 and 15 mag in steps of 0.20 mag.  The results of the fitting procedure are provided in Figure\,\ref{fig:fig9} and the fitted \av\, are listed in Table\,\ref{tab:tab5}. Noticeably, in 6 out of the 9 optical spectra examined, the two \av\, determinations agree within 0.5 mag.


Finally, the case of V1647 Ori deserves a short discussion. This source is a well-known Class I/flat young source, still deeply embedded in the parental dusty envelope. 
As shown in several works (e.g. Nisini et al. 2016, Fiorellino et al. 2021), in embedded sources a relevant fraction of the optical emission originating close to the star is scattered in the cavity excavated inside the envelope. The effect is to enhance the emission at shorter wavelengths, that in turn simulates a lower extinction. For this reason, we consider not reliable the determination of \av\, (9.2 mag) derived in the optical range and will assume the extinction derived from the LUCI spectrum on the same date (13 mag, Sect.\,\ref{sec:sec6.1.2}).

\subsubsection{Extinction derived from near-IR spectra\label{sec:sec6.1.2}}

\begin{figure}
\begin{center}
\plotone{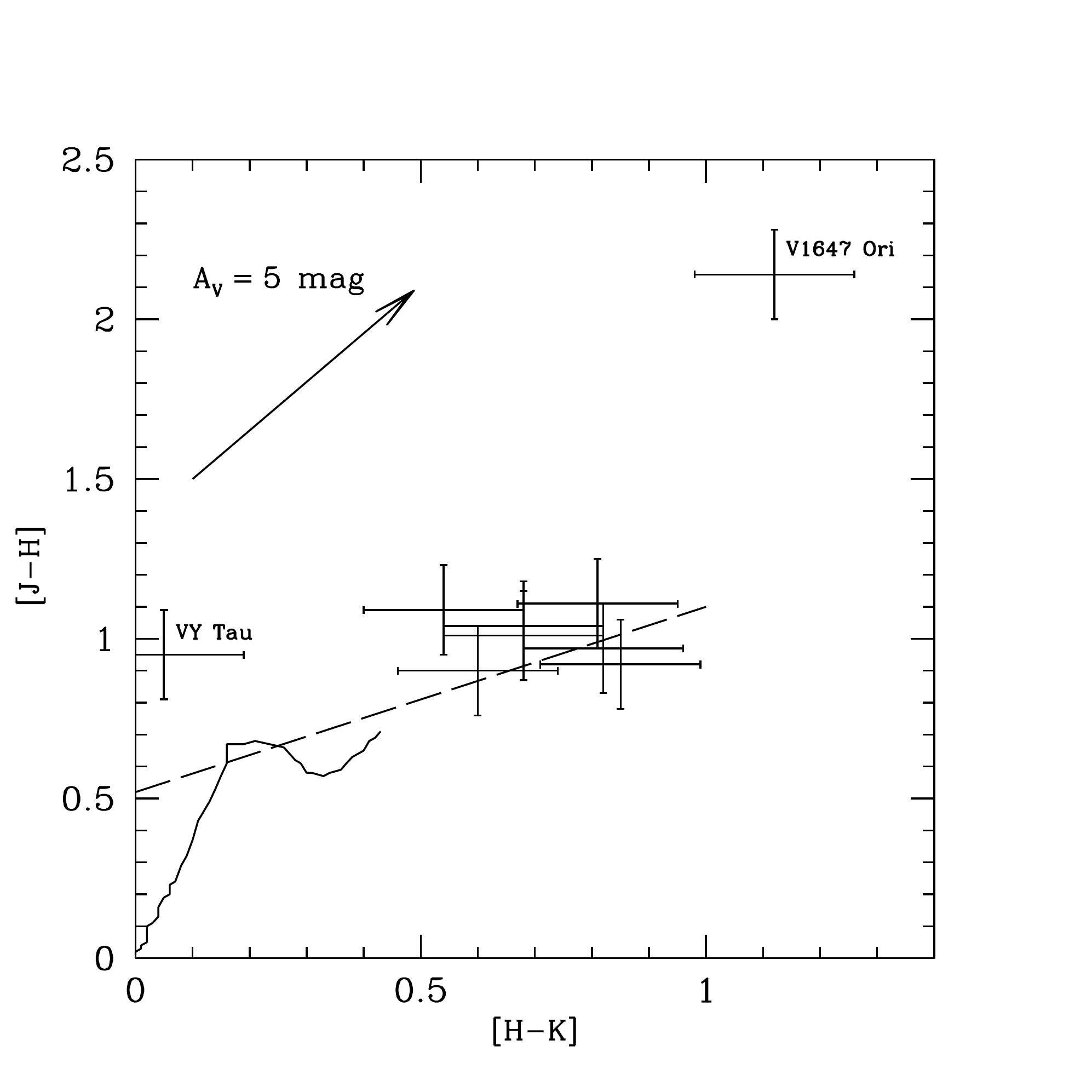}
\end{center}
\caption{[$J-H$] vs. [$H-K$] color-color diagram. The arrow shows the extinction vector for \av\,=5 mag, according to the reddening law by 
Cardelli et al. (1989). The dashed line marks the {\it locus} of the unreddened T Tauri stars (Meyer et al. 1997), 
while the continuous line represents Main Sequence stars (Luhman et al.\,2010). Data points of the objects discussed in the text are labelled.
\label{fig:fig10}}
\end{figure}

The \av\, estimate from the near-IR spectra was also obtained by applying two independent methods, briefly described in the following paragraph.

The first method is based on the [$J$-$H$] and [$H$-$K$] colors at the date of the observation. One photometric point is directly provided by the acquisition image (Table\,\ref{tab:tab4}), while the other two are computed from the flux-calibrated spectrum convolved  with the Johnson filter profiles\footnote{retrived at http://svo2.cab.inta-csic.es/theory/fps/, Spanish Virtual Observatory (SVO) Filter Profile Service (Rodrigo \& Solano 2020)}. We estimate that this procedure leads to an uncertainty of about $\pm$0.1 mag in each band.
From the magnitudes we have constructed the [$J$-$H$] vs. [$H$-$K$] color-color diagram shown in Figure\,\ref{fig:fig10}. Here, we indicate with an arrow the direction of the extinction vector, whose length corresponds to an extinction of 5 mag. We assume the reddening law by Cardelli et al. (1989) and  R$_V$\,=\,3.1. We estimate A$_V$ from the position of the observed colors and the unreddened Classical T Tauri locus (CTTs, Meyer et al. 1997) and list the values in Table\,\ref{tab:tab5}. Considering all the uncertainties involved in this procedure (absolute spectral calibration, propagation of magnitude errors, assumptions on extinction law and R$_V$), we conservatively evaluate a final error on \av\,  of 1 mag.

All sources have low or negligible A$_V$, with the exception of V1647 Ori  (A$_V$ $\sim$ 13 mag). In the case of VY Tau, which is located at the left edge of Figure\,\ref{fig:fig10}, we are not able to estimate a reliable value of 
\av, while for two observations  (iPTF15afq and V350 Cep) we cannot apply the method because the available spectra do not cover the three bands.

The second method relies on the detection of accretion line tracers, as done for the optical spectra, see Figures\,\ref{fig:fig6}-\ref{fig:fig8}. However, empirical relationships in the near-IR are available only for the Paschen lines (from Pa8 to P$\beta$) and for the  Br$\gamma$ (Alcal\'a et al. 2017), which are moreover less sensitive to extinction variations than the optical lines. For DR Tau, for which we have quasi-simultaneous MODS and LUCI spectra, the optical and near-IR lines have been considered all together in the fit. Conversely, as explained in the previous section, the lines of V1647 Ori observed with LUCI  have been fitted separately from those observed with MODS, although the two spectra were obtained on the same date. The fitted \av\, of 13 mag is in perfect agreement with that derived from the color-color diagram.

We were able to apply both methods to five spectra. The two \av\, determinations are in excellent agreement in UZ Tau E and V1647 Ori  (Table\,\ref{tab:tab5}), but differ between 0.9 and 3.3 mag in the other three sources (DR Tau, PV Cep and V350 Cep).
Since we have no reason to prefer one method over the other (also considering that the lines used in the fit of \lacc\, are few), we have assumed the entire range of \av\,determinations to compute the accretion parameters \lacc\, and \macc\, (see Sect. \ref{sec:sec6.2}).

\subsection{Accretion luminosity and mass accretion rate\label{sec:sec6.2}}

As explained in Sect.\,\ref{sec:sec6.1.1}, the accretion luminosity, \lacc\,, has been derived together with \av\, by fitting the individual \lacci\,. With reference to the fits shown in Figures
\,\ref{fig:fig6}-\ref{fig:fig8} we note that, in general, the \lacci\, derived from the individual lines agree within the errors. There are however lines whose accretion luminosity may be up to an order of magnitude higher than the average \lacc\,. In particular, we signal \caii\,K and \hei\, at 492 and 502 nm. However, thanks to the high number of lines used, \lacc\, does not change by more than 10\% whether or not these three lines are included in the fit.\\
For many sources, a further estimate of \lacc\, has been obtained by applying the relationships between \lacc\, and \lumi\, by fixing the \av\, to the value derived by the continuum method (in the optical) or by the color-color diagram (in the infrared), see Table\,\ref{tab:tab7}. This way, we have checked that \lacc\, does not change by more than 30\%.

The mass accretion rate, \macc\,, has been estimated using the relationship by Gullbring et al. (1998):

\begin{equation}
\dot{M}_{\mathrm{acc}} \approx 1.25 {\frac{L_{\mathrm{acc}} R_{\mathrm{*}}}{G M_{\mathrm{*}}}}                 
\end{equation}

where \rstar\, and \mstar\, are the stellar mass and radius, and G is the gravitational constant. We have adopted for \mstar\, the values given in the literature (Table\,\ref{tab:tab3}) and derived \rstar\, from the relationship: 

\begin{equation}
R_\mathrm{*} = {\frac{1}{2T_\mathrm{eff}^2}} \sqrt{{\frac{L_{\mathrm{*}}}{\pi\sigma}}}                  
\end{equation}

where $\sigma$ is the Stefan-Boltzmann constant, and \teff\, and \lstar\, are taken from Table\,\ref{tab:tab3}.

 The derived \macc\,values are listed in Table\,\ref{tab:tab7}.

It is worth noting that no remarkable differences exist between EXors and EXor candidates, either in \lacc\, or in \macc\,. Typical values are $-$2 \lapprox\, Log(\lacc/\lsun) \lapprox\, 1, and  $-$9 \lapprox\, Log(\macc/\msunyr) \lapprox\, $-$7. V1647 Ori is the object with significantly higher \lacc\,($\sim$ 50\,\lsun\,) and \macc\, ($\sim$ 2 10$^{-5}$ \msunyr\,), in line with its very young age and its classification as a Class I/flat source. This source underwent two outbursts during its recent history, one in 2003-2004 (Aspin et al. 2006) and another one in 2008-2011 (Aspin 2011). During the second outburst, its $r$-band photometry was $\sim$ 17.77 and \av\,=\,8 mag. In 2016, we measured $r$=18.68, namely about 1 mag fainter. Despite that, our determination of \macc\, is about one order of magnitude higher than the estimate of (4$\pm$2) 10$^{-6}$\msunyr\, measured by Aspin (2011). This disagreement is due to the higher \av\, (13 mag) that we have estimated from the LUCI spectrum. A considerable variability in \av\, is indeed exhibited by V1647 Ori. During quiescence ($r\sim$ 23), \av\, was around 19 mag (Aspin 2011), therefore it is plausible to assume that in an intermediate stage, as that we have probed, also the extinction was in between the two extreme values of 8 and 19 mag. It would therefore be relevant to measure \macc\, during the quiescent state (namely in the conditions of maximum \av\,) to disentangle the contributions of accretion and extinction in the large-amplitude variability exhibited by this source.

The sources for which we have more than one observation are
XZ Tau, UZ Tau E, NY Ori, DR Tau, ASASSN-13db, iPTF15afq, PV Cep,
and V350 Cep. Of these, all but NY Ori and ASASSN-13db present an amplitude variability \lapprox\,1 mag (Figure
\,\ref{fig:fig1}), in line with that observed in Classical T Tauri
stars (Costigan et al.  2014). These magnitude fluctuations
correspond to variations in \lacc\, and \macc\, of a factor  $\approx$ 1.5$-$3. In general, these results confirm the
accepted scenario in which accretion is a non-stationary
phenomenon, although they are insufficient to explain the spread of
more than two orders of magnitude found in the \macc\,
determinations among sources of the same mass (e.g. Antoniucci et al. 2014).

However, larger magnitude fluctuations imply significant variations in \lacc\, and \macc. In ASASSN-13b, in the 2 years between the outburst peak and the quiescence, \lacc\, changed from 0.2 to 0.03 \lsun\, and \macc\, from 3 10$^{-8}$ to 5 10$^{-10}$ \msunyr\,. Also, in NY Ori \lacc\, decreased by a factor 25 in about one year (2014 December - 2016 January). Although no photometric measurements are available for the indicated period, this change in \lacc\, is compatible with the variation of around 3 mag observed several times in this source (see Table\,\ref{tab:tab2}).

\subsection{Physical conditions of the \hi\, lines emitting region\label{sec:sec6.3}}
\begin{figure}
\plotone{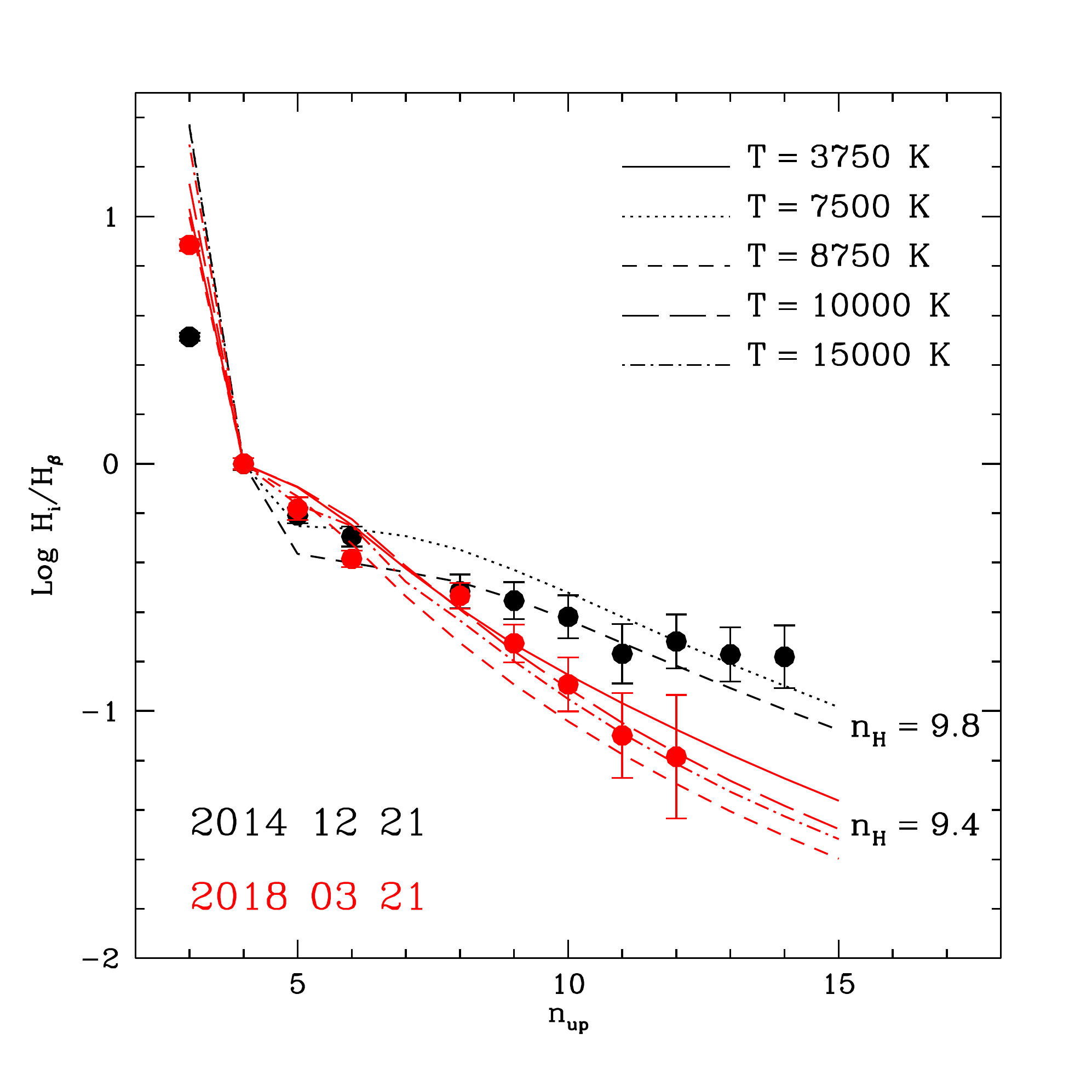}
\caption{Balmer decrement (H$_i$/H$\beta$ flux ratios vs. upper level of the transition) of XZ Tau in the two dates of observation (black: 2014/12/21, red: 2018/03/21). The hydrogen density fitted for the two sets of data is reported in logarithmic scale. Different line symbols represent different temperatures as indicated in the top-right corner of the figure. \label{fig:fig11}}
\end{figure}

 High-resolution spectroscopic observations indicate that both the high-velocity stellar winds and the gas channeled from the disk onto the stellar surface contribute to the hydrogen line profiles (e.g. Folha \& Emerson 2001, Antoniucci et al. 2017a, Moura et al. 2020). Although our low-resolution observations do not allow to kinematically separate these two gas components,  we can evaluate the average physical conditions of the hydrogen lines emitting regions using the observed flux ratios as temperature and density diagnostics.

Using their local line-excitation model, Kwan \& Fischer (2011) provided predictions of the Balmer decrements (ratios from H15 to H$\alpha$ with respect to H$\beta$), in the range of hydrogen density
8\,$\le$\, log[\nH(cm$^{-3}$)]\, $\le\,$12, and temperature between 3750 K and 15000 K. Similarly, Paschen decrements (ratios from 
Pa$\gamma$ to Pa12 with respect to Pa$\beta$), together with the ratio Br10/Br$\gamma$ are given by Edwards et al. (2013) for
 8\,$\le$\, log[\nH(cm$^{-3}$)]\, $\le\,$12.4 and T\,=\,5000$-$20000 K.
 
Fits through the intrinsic hydrogen line fluxes have been done by least squares method for sources where at least 4 lines are detected.
As an example, we show in Figure\,\ref{fig:fig11} the Balmer decrement measured in XZ Tau for the two observation dates (2014 and 2018). We plotted the H$_i$/H$\beta$ ratios up to H14/H$\beta$ and H12/H$\beta$ for the 
 2014 (black) and 2018 (red) observations, respectively. First, we note the different shape of the Balmer decrement in the two dates. Following the nomenclature of Antoniucci et al. 2017b, in 2014 the decrement shape was of 'type 4', typical of strong accretors, and changed in 2018 to the 'type 2' shape, which is the one most frequently seen in T Tauri stars. This shape can be explained by variation of a factor $\sim$ 2.5 in \nH, with ratios H$_i$/H$\beta$ increasing with n$_{up}$ in models with higher densities. Conversely,  no clear correlation is found between the temperature and the shape of the Balmer decrement, and in particular, all the temperatures considered in the Kwan \& Fischer model are compatible for the observation in 2018.

Table\,\ref{tab:tab8} summarizes the results for the objects of our sample, also reporting the brightness level of each source as derived from the photometric data and the number of lines involved in the fit.  
As pointed out for the case of XZ Tau, the temperature is not well constrained, apart from the case of UZ Tau E. 
It is therefore more interesting to focus on the results regarding the hydrogen density. In our previous study on the 2015 outburst of the classical EXor V1118 Ori, we highlighted the sudden increase in the hydrogen density of about 2 orders of magnitude, from log(\nH)=9.4 to log(\nH)=11.4,  passing from the quiescent phase to the peak of the outburst (Giannini et al. 2017).
In the objects of the sample presented here, we fit the hydrogen density in the range log(\nH)\, from 9.0 to 11.0, considering the uncertainties. Indeed, a certain level of correlation may be found between \nH\, and the activity level of the source. In Sect.\,\ref{sec:sec4} and \ref{sec:sec6.2}, we identified XZ Tau, PV Cep, iPTF15afq, ASASSN-13db, and NY Ori as sources that showed considerable photometric variability during the last years, often accompained by a significant \lacc\, and \macc\, variations. Specifically, in XZ Tau the $V$ band magnitude increased from $V\sim$ 12.8 in 2014 to $V\sim$ 14.2 in 2018. Our analysis shows that this resulted in a decrease of veiling from 0.8 to 0.0, of \macc\, by a factor of 5 and of \nH\, by a factor around 2.5-3.0.  PV Cep has been observed only during a period of high brightness. Close to the peak of the light-curve in 2015 we fit log(\nH)=11.0.  
In ASASSN-13db and in iPTF15afq, we are able to measure \nH\, only during the quiescent phase. It is interesting to note, however, that during periods of high brightness level (2014-2015 for ASASSN-13db and 2019 for iPTF15afq), the \hi\, lines are always seen in absorption, a signature typical of FUors or in general of sources with a high accretion rate (Connelley \& Reipurth 2018). In NY Ori the situation is more controversial. From our acquisition images we find  differences of about 2 magnitudes in the $r$-band between 2014 ($r \sim$ 12) and 2016 ($r \sim$ 14). All  \hi\, lines (from the Balmer and Paschen series) are seen in absorption in 2016, while the Paschen lines are in emission in 2014.  Given the spectral type of NY Ori (K4), a possibility is that during quiescence the observed \hi\, lines are in absorption because they originate in the photosphere, while during outburst
the accretion component in emission dominates. 

In the remaining sources, which are all observed during quiescence, log(\nH)\, is typically between 9.0 and 10.0. Only in V350 Cep we estimate log\,(\nH)\, between 10.6 and 10.8 in 2015. Unfortunatly, the photometric level at that  date is unknown.

\subsection{Mass loss variability and forbidden line emission\label{sec:sec6.4}}

Matter accretion from a disk necessarily imposes the removal of angular momentum through winds and jets. This makes the phenomena of mass accretion and ejection intimately correlated, so it is interesting to investigate whether they are also connected in terms of variability.

Mass loss rate determinations suffer from multiple assumptions on the geometry and physical conditions of the outflowing gas (e.g. Sperling et al. 2020). Therefore, rather than searching for correlations between the rates of accretion and ejection, we have preferred to look for flux variations of lines representative of the two phenomena. 

The \oi\,6300 \AA\, is the most prominent outflow emission feature. Indeed, it was detected in all our objects but VY Tau and V1143 Ori, namely the two sources with the lowest values of \macc\,.

As tracer of accretion we have selected the H$\alpha$ line, both because of its brightness and, more importantly, because the ratio H$\alpha$/\oi\,6300 \AA\,, being independent of extinction, is directly related to the intrinsic ratio between mass accretion and mass ejection rates.

Among the 5 sources showing a significant \macc\, variation, we have multi-epoch determinations of the \oi\,6300 \AA\, for XZ Tau, NY Ori and iPTF15afq. 
With reference to the line fluxes reported in Appendix\,\ref{sec:appendix_a}, 
we find that in XZ Tau F(\oi\,6300\,\AA\,) has remained fairly constant while F(H$\alpha$) decreased by a factor $\sim$ 2.4 from 2014 to 2018. This indicates that between 2014 and 2018 the variation of the mass accretion has been significantly larger than that of the  mass loss. Similarly, in NY Ori, F(H$\alpha$) and F(\oi\,6300\,\AA\,) decreased by a factor 7 and 3 in two years (2014$-$2016). In iPTF15afq we observe the same trend although the variation is low for both lines, due to the short temporal distance of just one year.

More striking is the trend in the classical EXor V1118 Ori. Indeed, comparing the H$\alpha$ and \oi\,6300 \AA\ fluxes during the 2015 outburst (Giannini et al. 2016, 2017), with those of quiescence (Lorenzetti et al. 2015) we find a variation of F(H$\alpha$) about ten times larger than the variation of F(\oi\,6300\,\AA\,).

\section{Conclusions\label{sec:sec7}}
We analysed optical and near-IR low resolution spectra taken at LBT between 2014 and 2019 of a sample of 11 variable young sources, composed of 5 well known eruptive variables (EXors) and 6 Pre-Main Sequence objects showing episodic variability likely attributable to intermittent accretion events (EXor candidates). Ten sources
have been observed more than once, therefore allowing us to investigate for correlations
between photometric and spectroscopic variability. At this stage, the collected observations represent the first flux-limited spectroscopic survey of candidate eruptive variables able to accurately determining 
fundamental quantities such as visual extinction; accretion parameters; physical parameters of the \hi\, lines emitting region, and signatures of mass outflowing. The main results of this survey can be 
summarized as follows:
\begin{itemize}
\item[-] The analysis of the light-curves indicates that six sources (UZ Tau E, VY Tau, V1143 Ori, DR Tau, V1647 Ori, and V350 Cep) have been observed while in quiescence and other two, XZ Tau, and PV Cep, during a high level of brightness. Remarkably,  ASASSN-13db and ipTF15afq  have been caught during outburst.
\item[-] All targets present lines that are tracers of accretion, like \hi\, recombination lines, both in the optical
and IR range. All IR spectra show prominent \caii\, and \hei\, 1.08\,$\mu$m lines. Metallic lines of many
species are also detected. Signatures of ejection activity (\oi\,6300\,\AA\,, \sii\,, \htwo\, 2.12$\mu$m) characterize about 80\% of the sources. No evident difference exists between the spectra of known and candidate EXors, all of them resembling spectra of accreting young T Tauri stars.

\item[-] For many sources the veiling excess was computed. We found that a significant variation of the veiling is associated with accretion events. During the 2015 outburst of ASASSN-13db the veiling at 710 nm increased from 0.1 to $>$ 3. 

\item[-] Since the visual extinction \av\, closely depends on the source activity level at the time of the observation, a major effort has been made to accurately evaluate this parameter. For both optical and near-IR spectra two independent methods (based on lines, continuum, and colors) were used to derive \av\ values.

\item[-] Empirical relationships between the accretion luminosity (\lacc\,) and the luminosity of selected lines allowed us to determine the average accretion luminosity for each source and date of observation.  
The mass accretion rate was also evaluated. We did not find any remarkable difference between the \lacc\, and \macc\, values of known EXors and EXor candidates of our sample, which are in the range $-$2 \lapprox\, Log(\lacc/\lsun) \lapprox\, 1 and $-$9 \lapprox\, Log(\macc/\msunyr) \lapprox\, $-$7, respectively.

\item[-] All sources observed more than once present significant \macc\, variability even if they have been observed during quiescence. In ASASSN-13db a decrease of \macc\, of two orders of magnitude is observed from the outburst peak to the quiescent phase. Also, the accretion luminosity of NY Ori decreased by a factor 25 in one year.

\item[-] Physical parameters (\nH\, and T) of the \hi\, emitting region have been evaluated from the observed hydrogen line ratios by using predictions provided in the literature. These ratios are much more effective in constraining the density (\nH\,) than the temperature, which remains poorly defined. Generally, a direct correlation is recognizable between density and accretion activity of the source, with significant density increases (up to two orders of magnitudes) associated with large brightness fluctuations. 

\item[-] Tracers of mass outflows, such as the \oi\,6300\,\AA\, line,  have been detected in most of the investigated sources. By
  comparing the \oi\, variations with those of a mass accretion tracer (H$\alpha$),  we conclude that mass accretion variations are larger than mass loss variations.
\end{itemize}

In the next future, other facilities appropriate for this research will be available in the southern hemisphere, which are expected both to
discover large numbers of eruptive variables (e.g. VRO-LSST), and to spectroscopically follow-up them in the
optical/near-IR band (e.g. SoXS at ESO-NTT). Their observations will probably deeply modify our knowledge of the eruptive accretion phenomenon. In such framework
we hope that our coherent database will represent a valuable reference study.

\section{Acknowledgements}
This work is based on observations made with the Large Binocular Telescope (LBT). The LBT is an international collaboration among institutions in the United States, Italy and Germany. LBT Corporation partners are: The University of Arizona on behalf of the Arizona university system; Istituto Nazionale di Astrofisica, Italy; LBT Beteiligungsgesellschaft, Germany, representing the Max-Planck Society, the Astrophysical Institute Potsdam, and Heidelberg University; The Ohio State University, and The Research Corporation, on behalf of The University of Notre Dame, University of Minnesota and University of Virginia. This  work  has  been  supported  by  the  project PRIN-INAF-MAIN-STREAM 2017  “Protoplanetary    disks    seen through  the  eyes  of  new-generation  instruments” and  by  the  European Union’s Horizon 2020 research and innovation programme un-der the Marie Sklodowska-Curie grant agreement No 823823 (DUST-BUSTERS).  We  also  acknowledge  the  support  by  INAF/Frontiera through  the  ‘Progetti  Premiali’  funding  scheme  of  the  Italian  Ministry  of  Education,  University,  and  Research  and  by  the  Deutsche Forschungs-Gemeinschaft (DFG, German Research Foundation) - Refno.  FOR  26341/1  TE  1024/1-1. This research also received financial support from the project PRIN-INAF 2019 "Spectroscopically Tracing the Disk Dispersal Evolution". ACG receveid fundings from the European Research Council (ERC) under the European Union's Horizon 2020 research and innovation programme (grant agreement No. 743029). A. Rossi acknowledges support from the INAF project Premiale Supporto Arizona \& Italia.

\begin{landscape}
\begin{deluxetable}{lllcccccc}
\tablewidth{553pt}
\tablecaption{\label{tab:tab1}  The investigated sample} 
\tabletypesize{\scriptsize}
\tablehead{Simbad      & RA (J2000)  & DEC (J2000)                          & D$^1$    & Location   & Binarity  & Jet/Outflow   & Other           & Ref\\
           Name        & (h:m:s)    &($^\circ$:$^\prime$:$^{\prime\prime}$) & (pc)  &            &           &               & designation     &     }     
\startdata                    
\cline{1-9}  
\multicolumn{9}{c}{\bf{EXors}}\\
\cline{1-9}   
XZ Tau                  & 04:31:40.07  & $+$18:13:57.2    & 150$^a$   & L1551          & Y(0\farcs30) &  HH152/Y           & Haro6-15$-$SVS169  & 1   \\
UZ Tau E                & 04:32:43.07  & $+$25:52:31.0    & 130.4     & B19            & Y(0\farcs34) &  \nodata/\nodata   & UZ Tau A           & 2   \\  
VY Tau                  & 04:39:17.41  & $+$22:47:53.4    & 153.4     & L1536          & Y(0\farcs66) &  Y(?)/\nodata      & SVS38              & 3,4 \\  
NY Ori                  & 05:35:36.01  & $-$05:12:25.3    & 403.5     & ONC            & \nodata      &  \nodata/\nodata   & Parenago 2119      & 5   \\  
V1143 Ori               & 05:38:03.90  & $-$04:16:42.8    & 395.3     & L1640          & \nodata      &  \nodata/\nodata   & Sugano's obj       & 5   \\   
\cline{1-9}                                                                                                                                    
\multicolumn{9}{c}{\bf{EXor candidates}}\\                                                                                                     
\cline{1-9}                            
DR Tau                  & 04:47:06.21  & $+$16:58:42.8    & 193.0     & L1558          & N            &  Y/\nodata         & \nodata             & 6 \\
\enddata
\tablecomments{\scriptsize{1: Distances taken from GAIA EDR3 (Gaia Collaboration (2016, 2020).
$^a$ assumed; References: [1] Coffey et al. 2004; [2] Jensen et al. 2007; [3] Leinert et al. 1993; 
[4] Banzatti et al. 2019; [5] Herbig 2008; [6] Kun et al. 2016; [7] Arce et al. 2010;  [8] Young et al. 2015; [9] Akeson et al. 2019; 
[10] Joncour et al. 2017; [11] Li et al. 2015; [12] Banzatti et al. 2014; [13] Principe et al. 2018.
}}
\end{deluxetable}
\end{landscape}

\begin{deluxetable}{ccccc}
\tablewidth{320pt}
\tablecaption{\label{tab:tab2}  Photometric variability} 
\tabletypesize{\scriptsize}
\tablehead{ID             & mag$_{min}-$mag$_{max}$ & band$^a$      &   Bursts                  & Ref\\
                          &  (mag)        &               &                           &    }     
\startdata                      
\cline{1-5}  
\multicolumn{5}{c}{\bf{EXors}}\\
\cline{1-5}  
XZ Tau                    &  16.6$-$10.4    &  V            &  many/last 2014$-$2016    & 1,2      \\
UZ Tau E                  &  18.5$-$12.5    &  V            &  many/last 2006           & 3,4      \\ 
VY Tau                    &  15.3$-$9.0     &  V            &  1960-1972/last 2014      & 2,3,5,6  \\ 
NY Ori                    &  $>$16$-$13.3   &  V            &  many/last 2001           & 7,8      \\ 
V1143 Ori                 &  16.7$-$13.5    &  V            &  many/last 1993           & 7,8      \\ 
\cline{1-5}                                      
\multicolumn{5}{c}{\bf{EXor candidates}}\\       
\cline{1-5}                                      
DR Tau                    & 16.0$-$10.5     &  V            &  1960$-$1980              & 3,9,10,11   \\
\enddata
\tablecomments{\scriptsize{References: [1] Coffey et al. 2004; [2] Dodin et al. 2016; [3] Herbig \& Bell 1988; [4] Jensen et al. 2007;
[5] Herbig 1990; [6] Stone 1983; [7] Herbig 2008; [8] Jurdana-\v{S}epi\'c et al. 2017; 
[9] Chavarria 1979; [10] Kenyon et al. 1994; [11] Grankin et al. 2007; [12] Holoien et al.2014; [13] Sicilia-Aguilar et al. 2017; 
[14] Brice{\~n}o et al. 2004; [15] Aspin et al. 2006; [16] Aspin et al. 2008; [17] Aspin \& Reipurth 2009; 
[18] Miller et al. 2015; [19] Hillenbrand 2019; [20] Cohen et al. 1981; 
[21] Caratti o Garatti et al. 2013; [22] Lorenzetti et al. 2011; [23] Jurdana-\v{S}epi\'c et al. 2018.
}} 
\end{deluxetable}

\begin{landscape}
\begin{deluxetable}{ccccccccccc}
\tablewidth{535pt}
\tablecaption{\label{tab:tab3}  Stellar parameters} 
\tabletypesize{\scriptsize}
\tablehead{Name           & A$_V$     & L$_{bol}$       &  SpT        & \lstar  & \teff   & \mstar          & \lacc        & \macc            & Class   & Ref \\
                          & (mag)     & (\lsun)         &             & (\lsun) & (K)     & (\msun)       & (\lsun)     & (10$^{-7}$\msunyr)          &     }     
\startdata                                                                                         
\cline{1-11}                                                                                        
\multicolumn{11}{c}{\bf{EXors}}\\                                                                   
\cline{1-11}                                                                                        
 XZ Tau               &  1.4$-$1.5     &  4.2$-$10.7     & M3         & 0.57    & 3410    & 0.37     &  0.347      &  1$-$10, 0.14  &    II      &  1,2,3,4  \\
 UZ Tau E             &  0.5$-$1.5     &  1.7            & M3         & 0.62    & 3410    & 0.3      &  0.025      &  1$-$3          &   II      &  1,4,5,6  \\
 VY Tau               &  0.5$-$0.85    &  0.72           & M1.5       & 0.39    & 3640    & 0.39     &  \nodata    &  \nodata        &   II      &  1,7,8    \\
 NY Ori               &  0.3           &  \nodata        & K4         & \nodata & 4430    & 1.0      &  \nodata    &  \nodata        &   \nodata &  9,10        \\  
 V1143 Ori            &  0             &  \nodata        & M2         & 0.20    & 3500    & 0.5      &  \nodata    &  \nodata        &   II      &  7          \\
\cline{1-11}                                                                                                                 
\multicolumn{11}{c}{\bf{EXor candidates}}\\                                                                                  
\cline{1-11}                                                                                                                 
 DR Tau               &  0.45$-$1.6   &  1.2            & K7          & 0.32    & 4070      & 0.75    &  0.74       &  4$-$9         &   II      &  1,4,11,12,13   \\
\enddata
\tablecomments{\scriptsize{References: [1] Herczeg \& Hillenbrand 2014; [2] Osorio et al. 2016; [3] Hartigan \& Kenyon 2003;  [4] Cs{\'e}p{\'a}ny, et al. 2017; 
[5] Yang et al. 2012; [6] Lorenzetti et al. 2007; [7] Sipos \& K{\'o}sp{\'a}l 2014; [8] Cabrit et al. 1990;  [9] Hillenbrand 1997; [10] Kounkel et al. 2019; [11] Muzerolle et al. 2003; 
[12] Lorenzetti et al. 2009; [13] Petrov et al. 2011; [14] Holoien et al. 2014; [15] Sicilia-Aguilar et al. 2017; [16] Aspin 2011; [17] Andrews et al. 2004; [18] Gramajo et al. 2014; [19] Moody \& Stahler 2017;
[20] Aspin et al. 2008; [21] Fischer et al. 2016; [22] Kun et al. 2011; [23] Hamidouche 2010; [24] Magakian \& Movsesian 2001; 
[25] Cohen et al. 1981; [26] The et al. 1994; [27] Caratti o Garatti et al. 2013; [28] Kun et al. 2009.
}}
\end{deluxetable}
\end{landscape}

\begin{landscape}
\begin{deluxetable}{cccccc|ccccc}
\tablewidth{0pt}
\tabletypesize{\scriptsize}
\tablecaption{\label{tab:tab4} Journal of observations.} 
\tablehead{
    Source           & \multicolumn{5}{c}{{MODS}}    & \multicolumn{5}{c}{LUCI} }                  
\startdata      
                        &  Date        & MJD     &T$_{int}$   (B/R) & Phot$^a$      & Filter       &  Date          & MJD        &T$_{int}$ (zJ/HK)& Phot$^a$     &  Filter\\         
                        &  (yy/mm/dd)  &         & (s)              & (mag)         &              &  (yy/mm/dd)    &            &(s)              & (mag)        &         \\    
\cline{1-11}                                                                        
XZ Tau               &  2014/12/21   & 57012    &  600/600          & 12.54         &  r           &  \nodata       & \nodata    & \nodata         & \nodata       & \nodata    \\
                     &  2018/03/21   & 58198    &  1200/1200        & 13.68         &  r           &  \nodata       & \nodata    & \nodata         & \nodata       & \nodata    \\
UZ Tau E             &  2014/12/22   & 57013    &  600/600          & 12.21         &  r           &  2016/01/25    & 57412      & 120/88          & 7.94          & [FeII]1.64 \\
VY Tau               &  2014/12/21   & 57012    &  600/600          & 12.88         &  r           &  2016/03/04    & 57451      & 160/80          & 9.43          & [FeII]1.64 \\ 
NY Ori               &  2014/12/21   & 57012    &  600/600          & 11.94         &  r           &  \nodata       & \nodata    & \nodata         & \nodata       & \nodata  \\
                     &  2016/01/31   & 57418    &  600/600          & 14.05         &  r           &  \nodata       & \nodata    & \nodata         & \nodata       & \nodata  \\
\enddata
\tablenotetext{a}{\scriptsize{Photometry obtained in the indicated filter during the same night as spectroscopy. 
Exceptions are flagged (1 = ASA-SN; 2 = ZTF, 3 = Campo Imperatore Telescope, Italy) with in parenthesis the number 
of days before (-) or after (+) the spectroscopic observation.}}
\end{deluxetable}
\end{landscape}

\begin{deluxetable}{c|c|cc|cc|c|c|cc}
\tablewidth{0pt}
\tabletypesize{\scriptsize}
\tablecaption{\label{tab:tab5} \av, and veiling derived from optical and near-infrared spectra.} 
\tablehead{
    Source  & \multicolumn{5}{c|}{{Optical}}    & \multicolumn{4}{c}{Near-infrared} }                  
\startdata      
& Date&\multicolumn{2}{c|}{Continuum}&\multicolumn{2}{c|}{Lines} &Date & Col-Col&\multicolumn{2}{c}{Lines}\\
&  & r$_{710}^a$  & \av                    & N$_{l}$ & \av &               &  \av   &   N$_{l}$  &  \av       \\
&  (yy/mm/dd)   &    &  (mag)     &&  (mag)       &  (yy/mm/dd)   &  (mag)         &  &(mag)        \\    
\cline{1-10}                                                                                  
XZ Tau      &  2014/12/21& 0.8  &  1.0$-$2.0    & 22     &  2.0$-$3.0&  \nodata  & \nodata  & \nodata    & \nodata  \\
            &  2018/03/21& 0.0  &  1.4$-$3.4    & 16     &  2.8$-$4.4&  \nodata  & \nodata  & \nodata    & \nodata  \\
UZ Tau E    &  2014/12/22& 0.7  &  1.6$-$3.0    & 25     &  2.0$-$3.0& 2016/01/25& 0.0$-$1.5& 3          & 0.0$-$1.5\\
VY Tau      &  2014/12/21& 0.4  &  0.0$-$2.0    &  9     &  1.6$-$4.0& 2016/03/04&  \nodata & 0          & \nodata  \\ 
NY Ori      &  2014/12/21& 0.2$^b$    & \nodata & 11     &  0.4$-$2.4&  \nodata  & \nodata  & \nodata    & \nodata  \\
\enddata
\tablenotetext{a}{\scriptsize{Typical error on r$_{710}$ is 0.3 (Appendix\,\ref{sec:appendix_b}). $^b$Derived by assuming the \av\, value fitted from the accretion lines (see Appendix\,\ref{sec:appendix_b} for more details). $^c$Optical and near-infrared lines have been combined.\\
}}
\end{deluxetable}

\begin{deluxetable}{cccc}
\tablewidth{0pt}
\tabletypesize{\scriptsize}
\tablecaption{\label{tab:tab7} Accretion luminosity and mass accretion rate.} 
\tablehead{
    Source        & Date (yy/mm/dd)  & \lacc/\lsun                     & log(\macc/\msunyr) }                  
\startdata                                                             
\cline{1-4}                                                                                                       
XZ Tau            &  2014/12/21     &  0.30\,\,$-$\,\,0.90                       & -7.16\,\,$-$\,\,-6.68       \\
                  &  2018/03/21     &  0.10\,\,$-$\,\,0.37                       &  -7.06\,\,$-$\,\,-7.63       \\
UZ Tau E          &  2014/12/22     &  0.40\,\,$-$\,\,0.55                       &  -6.92\,\,$-$\,\,-6.78       \\
                  &  2016/01/25     &  0.28\,\,$-$\,\,0.50                       &  -7.07\,\,$-$\,\,-6.83        \\
VY Tau            &  2014/12/21     & 0.004\,\,$-$\,\,0.04                       & -9.13\,\,$-$\,\,-8.16        \\
NY Ori            &  2014/12/21     & 0.60\,\,$-$\,\,3.60                        &  \nodata                    \\
                  &  2016/01/31     & 0.03\,\,$-$\,\,0.33                        &  \nodata                    \\
V1143 Ori         &  2014/12/21     & 0.006\,\,$-$\,\,0.01                       &  -9.23\,\,$-$\,\,-9.11       \\
\enddata               
\end{deluxetable}


\begin{deluxetable}{ccccccc}
\tablewidth{0pt}
\tabletypesize{\scriptsize}
\tablecaption{\label{tab:tab8} Hydrogen physical parameters derived from Balmer, Paschen, and Brackett line ratios.} 
\tablehead{
    Source        & Date (yy/mm/dd) & Bright. level & Series  & N$_{lines}$   & Log[n$_H$(cm$^{-3}$)]   &  T (K)   }                  
\startdata                                                                         
\cline{1-7}                                                                                                                   
XZ Tau            &  2014/12/21     &   H           &  Ba     &  11           &    9.8                  &  7500-8750         \\
                  &  2018/03/21     &   L           &  Ba     &   9           &    9.4                  &  all               \\
UZ Tau E          &  2014/12/22     &   L           &  Ba     &  11           &    9.8                  &  7500-8750         \\
                  &  2016/01/25     &   I           &  Pa+Br  &  4            &    10.6-10.8            &  5000-12500        \\
VY Tau            &  2014/12/21     &   L           &  Ba     &   7           &    9.2-9.4              &  all               \\
NY Ori            &  2014/12/21     &   H           &  Pa     &   4           &    9.0-10.4             &  5000-12500        \\
V1143 Ori         &  2014/12/21     &   I           &  Ba     &   11          &    9.2-9.4              &  all               \\
DR Tau            &  2014/10/31     &   L           &  Ba+Pa  &   9           &    10.0-10.8            &  5000-12500        \\
\enddata               
\tablenotetext{a}{Brightness level as retrieved from photometric data. H:high; I:intermediate; L:low.}
\end{deluxetable}

\appendix
\section{Appendix A}\label{sec:appendix_a}
Fluxes of the lines used in the analysis. For optical lines ($\lambda <$ 1 $\mu$m) we give the air wavelength, for near-IR lines ($\lambda >$ 1 $\mu$m) the vacuum wavelength.
\begin{deluxetable}{cccc}
\tablewidth{0pt}
\tabletypesize{\scriptsize}
\tablecaption{\label{tab:A_1} XZ Tau: observed lines.} 
\tablehead{&  & 2014/12/21 &  2018/03/21     }                
\startdata      
Line   & $\lambda_{air}$  & \multicolumn{2}{c}{F$\pm\Delta$F}              \\
       & (\AA)            & \multicolumn{2}{c}{(10$^{-14}$ erg s$^{-1}$ cm$^{-2}$)} \\
\cline{1-4}      
H15        & 3711.977   & 3.2$\pm$0.2      & \nodata       \\
H14        & 3721.938   & 3.5$\pm$0.2      & \nodata       \\
H13        & 3734.368   & 3.6$\pm$0.2      & \nodata       \\
H12        & 3750.151   & 4.1$\pm$0.2      & 0.18$\pm$0.02 \\
H11        & 3770.630   & 3.7$\pm$0.2      & 0.22$\pm$0.02 \\
H10        & 3797.898   & 5.3$\pm$0.2      & 0.36$\pm$0.02 \\
\enddata
\end{deluxetable}

\begin{deluxetable}{ccc|ccc}
\tablewidth{0pt}
\tabletypesize{\scriptsize}
\tablecaption{\label{tab:A_2} UZ Tau E: observed lines.} 
\tablehead{ \multicolumn{3}{c}{2014/12/22} & \multicolumn{3}{c}{2016/01/25} }                
\startdata      
Line & $\lambda_{air}$  & F$\pm\Delta$F                            &  Line    & $\lambda_{vac}$  & F$\pm\Delta$F        \\
     & (\AA)            & (10$^{-14}$ erg s$^{-1}$ cm$^{-2}$)      &          & (\AA)            & (10$^{-14}$ erg s$^{-1}$ cm$^{-2}$)\\
\cline{1-6}                                             
H15           & 3711.977   & 1.7$\pm$0.6                           & Pa$\delta$ & 10052.128        & 59.3$\pm$1.6 \\
H14           & 3721.938   & 2.5$\pm$0.6                           & He I       & 10833.2          & 68.8$\pm$1.0 \\
H13           & 3734.368   & 2.5$\pm$0.6                           & Pa$\gamma$ & 10941.090        & 92.0$\pm$0.9 \\
H12           & 3750.151   & 3.2$\pm$0.6                           & Pa$\beta$  & 12821.59         & 131$\pm$0.5  \\
H11           & 3770.630   & 3.7$\pm$0.6                           & Br12       & 16411.674        & 16.0$\pm$2.3 \\
H10           & 3797.898   & 5.0$\pm$0.6                           & Br11       & 16811.111        & 29.7$\pm$4.3 \\
\enddata
\end{deluxetable}

\begin{deluxetable}{ccc|ccc}
\tablewidth{0pt}
\tabletypesize{\scriptsize}
\tablecaption{\label{tab:A_3} VY Tau: observed lines.} 
\tablehead{ \multicolumn{3}{c}{2014/12/21} & \multicolumn{3}{c}{2016/03/04} }                
\startdata      
Line    & $\lambda_{air}$  & F$\pm\Delta$F                            &  Line    & $\lambda_{vac}$  & F$\pm\Delta$F        \\
        & (\AA)            & (10$^{-14}$ erg s$^{-1}$ cm$^{-2}$)      &          & (\AA)            & (10$^{-14}$ erg s$^{-1}$ cm$^{-2}$)\\
\cline{1-6}                                             
H10           & 3797.898   & 0.16$\pm$0.04                            & He I        & 10833.2          & 0.5$\pm$0.1    \\
H9            & 3835.384   & 0.19$\pm$0.08                            & CO 2-0      & 22930            & -9.2$\pm$0.7   \\
H8            & 3889.049   & 0.33$\pm$0.12                            & CO 3-1      & 23230            & -3.9$\pm$0.7   \\
CaII K        & 3933.660   & 1.94$\pm$0.12                            & \nodata     & \nodata          & \nodata        \\
CaII H/H7     & 3968.470   & 2.55$\pm$0.12                            & \nodata     & \nodata          & \nodata        \\
H$\delta$     & 4101.702   & 0.47$\pm$0.06                            & \nodata     & \nodata          & \nodata        \\
H$\gamma$     & 4340.464   & 0.97$\pm$0.08                            & \nodata     & \nodata          & \nodata        \\
H$\beta$      & 4861.325   & 2.83$\pm$0.13                            & \nodata     & \nodata          & \nodata        \\
H$\alpha$     & 6562.800   & 11.3$\pm$0.28                            & \nodata     & \nodata          & \nodata        \\
Pa9           & 9229.015   & 2.37$\pm$0.68                            & \nodata     & \nodata          & \nodata        \\
\enddata
\end{deluxetable}

\begin{deluxetable}{cccc}
\tablewidth{0pt}
\tabletypesize{\scriptsize}
\tablecaption{\label{tab:A_4} NY Ori: observed lines.} 
\tablehead{&  & 2014/12/21 &  2016/01/31     }                
\startdata      
Line  & $\lambda_{air}$  & \multicolumn{2}{c}{F$\pm\Delta$F}              \\
      & (\AA)            & \multicolumn{2}{c}{(10$^{-14}$ erg s$^{-1}$ cm$^{-2}$)} \\
\cline{1-4}      
H15          & 3711.977  & -1.7$\pm$0.4      & -0.09$\pm$0.02 \\
H14          & 3721.938  & -2.2$\pm$0.4      & -0.15$\pm$0.02 \\
H13          & 3734.368  & -2.4$\pm$0.6      & -0.23$\pm$0.02 \\
H12          & 3750.151  & -3.4$\pm$0.4      & -0.22$\pm$0.02 \\
H11          & 3770.630  & -3.0$\pm$0.4      & -0.23$\pm$0.02 \\
H10          & 3797.898  & 1.7(-3.6)$\pm$0.4 & -0.44$\pm$0.02 \\
\enddata
\end{deluxetable}

\begin{deluxetable}{ccc}
\tablewidth{0pt}
\tabletypesize{\scriptsize}
\tablecaption{\label{tab:A_5} V1143 Ori: observed lines.} 
\tablehead{\multicolumn{3}{c}{2014/12/21} }                   
\startdata      
\f{Line} & $\lambda_{air}$  & F$\pm\Delta$F              \\
         & (\AA)            & (10$^{-14}$ erg s$^{-1}$ cm$^{-2}$) \\
\cline{1-3}      
H14          & 3721.938   & 0.04$\pm$0.01     \\
H13          & 3734.368   & 0.05$\pm$0.01     \\
H12          & 3750.151   & 0.055$\pm$0.005   \\
H11          & 3770.630   & 0.067$\pm$0.005   \\
H10          & 3797.898   & 0.111$\pm$0.006   \\
H9           & 3835.384   & 0.161$\pm$0.006   \\
H8           & 3889.049   & 0.209$\pm$0.006   \\
CaII K       & 3933.660   & 0.371$\pm$0.005   \\
CaII H/H7    & 3968.470   & 0.540$\pm$0.005   \\
H$\delta$    & 4101.702   & 0.322$\pm$0.006   \\
H$\gamma$    & 4340.464   & 0.370$\pm$0.006   \\
HeI          & 4471.480   & 0.043$\pm$0.006   \\
H$\beta$     & 4861.325   & 0.486$\pm$0.005   \\
HeI          & 4921.931   & 0.058$\pm$0.006   \\
HeI          & 5015.678   & 0.056$\pm$0.007   \\
HeI          & 5875.621   & 0.11$\pm$0.01     \\
H$\alpha$    & 6562.800   & 1.55$\pm$0.01     \\ 
\enddata
\end{deluxetable}

\begin{deluxetable}{cccc|ccc}
\tablewidth{0pt}
\tabletypesize{\scriptsize}
\tablecaption{\label{tab:A_6} DR Tau: observed lines.} 
\tablehead{ & & 2014/10/31  & 2018/03/19& \multicolumn{3}{c}{2018/03/23}}              
\startdata      
Line &  $\lambda_{air}$ & \multicolumn{2}{c|}{F$\pm\Delta$F} &  Line  & $\lambda_{vac}$ & F$\pm\Delta$F     \\
     & (\AA)            & \multicolumn{2}{c|}{(10$^{-14}$ erg s$^{-1}$ cm$^{-2}$)}      &      &(\AA)  & (10$^{-14}$ erg s$^{-1}$ cm$^{-2}$)\\
\cline{1-7}                                             
H18           & 3691.554   & \nodata               & 1.04$\pm$0.06              & Pa$\delta$  & 10052.128   & 70.1$\pm$1.8 \\
H17           & 3697.151   & \nodata               & 1.35$\pm$0.06              & \f{[SII]}   & 10339.24    & 4.3$\pm$1.8 \\
H16           & 3703.852   & \nodata               & 2.57$\pm$0.08              & He I        & 10833.2     & (-35.7)12.1$\pm$0.1 \\ 
H15           & 3711.977   & 3.0$\pm$0.2           & 2.90$\pm$0.06              & Pa$\gamma$  & 10941.090   & 106.0$\pm$1.8 \\
H14           & 3721.938   & 4.3$\pm$0.3           & 4.1$\pm$0.1                & Pa$\beta$   & 12821.59    & 173.1$\pm$1.8\\
H13           & 3734.368   & 3.5$\pm$0.4           & 4.7$\pm$0.1                & Br14        & 15884.880   & 10.5$\pm$1.6 \\
\enddata
\end{deluxetable}

\begin{deluxetable}{ccccc|ccc}
\tablewidth{0pt}
\tabletypesize{\scriptsize}
\tablecaption{\label{tab:A_7} ASASSN-13db: observed lines.} 
\tablehead{ &  &  2014/12/21 & 2015/10/01 & 2017/02/21 & \multicolumn{3}{c}{2015/02/10} }                
\startdata      
Line & $\lambda_{air}$  & \multicolumn{3}{c|}{F$\pm\Delta$F}                         &  Line    & $\lambda_{vac}$  & F$\pm\Delta$F        \\
        & (\AA)         &\multicolumn{3}{c|}{(10$^{-14}$ erg s$^{-1}$ cm$^{-2}$)}    &          & (\AA)            & (10$^{-14}$ erg s$^{-1}$ cm$^{-2}$)\\
\cline{1-8}                                             
H9             & 3835.384   & \nodata              & \nodata              & 0.10$\pm$0.02       & He I        & 10833.2           & (-13.3)$\pm$0.1 \\ 
CaII K         & 3933.660   & (-1.9)1.2$\pm$0.2    & (-1.3)6.2$\pm$0.3    & (-0.01)0.21$\pm$0.01& Pa$\beta$   & 12821.59          & 5.7$\pm$1.5     \\
CaII H/H7      & 3968.470   & (-1.0)0.9$\pm$0.2    & (-1.5)3.6$\pm$0.3    & (-0.01)0.32$\pm$0.01& \nodata     & \nodata           & \nodata         \\
H$\delta$      & 4101.702   & \nodata              & \nodata              & 0.25$\pm$0.01       & \nodata     & \nodata           & \nodata         \\
H$\gamma$      & 4340.464   & \nodata              & \nodata              & 0.31$\pm$0.01       & \nodata     & \nodata           & \nodata         \\
H$\beta$       & 4861.325   & -2.6$\pm$0.2         & -3.0$\pm$0.3         & 0.47$\pm$0.01       & \nodata     & \nodata           & \nodata         \\
\enddata
\end{deluxetable}

\begin{deluxetable}{ccc}
\tablewidth{0pt}
\tabletypesize{\scriptsize}
\tablecaption{\label{tab:A_8} V1647 Ori: observed lines.} 
\tablehead{ \multicolumn{3}{c}{2016/01/26} }                
\startdata      
Line & $\lambda_{air/vac}$  & F$\pm\Delta$F                      \\
     & (\AA)            & (10$^{-14}$ erg s$^{-1}$ cm$^{-2}$)    \\
\cline{1-3}                                             
\f{[OI]}      & 6300.304   & 0.22$\pm$0.06                       \\ 
H$\alpha$     & 6562.800   & (-0.04)0.60$\pm$0.02                \\
OI            & 7773.055   & 0.03$\pm$0.04 				     	 \\ 
OI            & 8446.360   & 0.18$\pm$0.04                       \\
CaII          & 8498.020   & 1.09$\pm$0.04                       \\
CaII          & 8542.090   & 1.20$\pm$0.04                       \\
CaII          & 8662.140   & 1.09$\pm$0.04                       \\
Pa$\delta$    & 10052.128  & 2.8$\pm$0.3                         \\
He I          & 10833.2    & (-29.8)$\pm$1.0                     \\   
Pa$\gamma$    & 10941.090  & 10.3$\pm$1.0                        \\
Pa$\beta$     & 12821.59   & 73.2$\pm$2.0                        \\
\enddata
\end{deluxetable}

\begin{deluxetable}{cccc|cccc}
\tablewidth{0pt}
\tabletypesize{\scriptsize}

\tablecaption{\label{tab:A_9} iPTF15afq: observed lines.} 
\tablehead{& & 2016/01/27  &2017/02/15  & & & 2018/01/28 & 2019/12/02 }                
\startdata      
Line & $\lambda_{air}$  & \multicolumn{2}{c|}{F$\pm\Delta$F}                        &  Line    & $\lambda_{vac}$  & \multicolumn{2}{c}{F$\pm\Delta$F}        \\
     & (\AA)            & \multicolumn{2}{c|}{(10$^{-14}$ erg s$^{-1}$ cm$^{-2}$)}  &          & (\AA)            & \multicolumn{2}{c}{(10$^{-14}$ erg s$^{-1}$ cm$^{-2}$)}\\
\cline{1-8}                                             
\f{[OI]}      & 6300.304   & 0.35$\pm$0.01  & 0.21$\pm$0.01                         & Pa$\delta$ & 10052.128      & 0.46$\pm$0.02   & -0.6$\pm$0.1 \\
\f{[OI]}      & 6363.776   & 0.11$\pm$0.01  & 0.08$\pm$0.01                         & \f{[SII]}  & 10320.49       & 0.10$\pm$0.02   & \nodata \\
H$\alpha$     & 6562.800   & 1.46$\pm$0.01  & 1.14$\pm$0.01                         & \f{[SII]}  & 10336.41       & 0.16$\pm$0.02   & \nodata \\
\f{[NII]}     & 6583.46    & 0.03$\pm$0.01  & \nodata                               & He I       & 10833.2        & 0.62$\pm$0.02   & -2.2$\pm$0.1 \\
\f{[SII]}     & 6716.44    & 0.04$\pm$0.01  & 0.03$\pm$0.01                         & Pa$\gamma$ & 10941.090      & 0.67$\pm$0.02   & -1.6$\pm$0.2 \\
\f{[SII]}     & 6730.82    & 0.08$\pm$0.01  & 0.06$\pm$0.01                         & \f{[FeII]} & 12570.2068     & 0.26$\pm$0.02   & \nodata  \\
\enddata                                                                                                                           
\end{deluxetable}

\begin{deluxetable}{ccc|ccc}
\tablewidth{0pt}
\tabletypesize{\scriptsize}
\tablecaption{\label{tab:A_10} PV Cep: observed lines.} 
\tablehead{& & 2015/06/24    & & & 2016/12/08  }                
\startdata      
Line & $\lambda_{air}$  & F$\pm\Delta$F                         &  Line    & $\lambda_{vac}$  & F$\pm\Delta$F        \\
     & (\AA)            & (10$^{-14}$ erg s$^{-1}$ cm$^{-2}$)   &          & (\AA)            & (10$^{-14}$ erg s$^{-1}$ cm$^{-2}$)\\
\cline{1-6}                                             
H15        & 3711.977   & -0.05$\pm$0.02           & Pa8        & 9548.590       & 32$\pm$2 \\
H14        & 3721.938   & -0.13$\pm$0.02           & Pa$\delta$ & 10052.128      & 48$\pm$2 \\
H13        & 3734.368   & -0.10$\pm$0.02           & \f{[SII]}  & 10320.49       & 36$\pm$3 \\
H12        & 3750.151   & -0.10$\pm$0.02           & \f{[NI]}   & 10400.59       & 20$\pm$3 \\
H11        & 3770.630   & -0.10$\pm$0.02           & He I       & 10833.2        & -43$\pm$1 \\
H10        & 3797.898   & -0.15$\pm$0.02           & Pa$\gamma$ & 10941.090      & 61$\pm$3 \\
\enddata                                             
\end{deluxetable}

\begin{deluxetable}{cccc|ccccc}
\tablewidth{0pt}
\tabletypesize{\scriptsize}
\tablecaption{\label{tab:A_11}V350 Cep: observed lines.}          
\tablehead{&             & 2015/11/08    & 2018/09/29             &         &                 & 2019/07/04 & 2019/10/02 & 2019/10/17}                
\startdata      
 Line    & $\lambda_{vac}$  & \multicolumn{2}{c|}{F$\pm\Delta$F}   &    Line & $\lambda_{air}$  & \multicolumn{3}{c}{F$\pm\Delta$F}                         \\
 & (\AA) & \multicolumn{2}{c|}{(10$^{-14}$ erg s$^{-1}$ cm$^{-2}$)}&         & (\AA)            &  \multicolumn{3}{c}{(10$^{-14}$ erg s$^{-1}$ cm$^{-2}$)}  \\
\cline{1-9}                                             
Pa$\delta$ & 10052.128  & 2.8$\pm$0.2    & 1.1$\pm$0.1            &    CaII K     & 3933.660   & 1.48$\pm$0.08     & 2.17$\pm$0.03        & 2.24$\pm$0.02   \\
\f{[SII]}  & 10320.49   & 1.2$\pm$0.2    & \nodata                &    CaII H/H7  & 3968.470   & 0.85$\pm$0.07     & 0.67$\pm$0.03        & 1.03$\pm$0.02   \\
He I       & 10833.2    & -2.2$\pm$0.1   & -1.2$\pm$0.1           &    H$\delta$  & 4101.702   & 0.27$\pm$0.07     & 0.48$\pm$0.08        & 0.48$\pm$0.03   \\
Pa$\gamma$ & 10941.090  & 2.9$\pm$0.1    & 1.5$\pm$0.1            &    H$\gamma$  & 4340.464   & 0.32$\pm$0.05     & 0.65$\pm$0.08        & 0.58$\pm$0.03   \\   
Pa$\beta$  & 12821.59   & 4.1$\pm$0.1    & 1.6$\pm$0.1            &    H$\beta$   & 4861.325   & 0.78$\pm$0.04     &(-0.09)1.50$\pm$0.02  & 1.45$\pm$0.03   \\
Br12       & 16411.674  & 1.4$\pm$0.2    & \nodata                &    HeI        & 4921.931   & 0.39$\pm$0.05     & 0.61$\pm$0.02        & 0.63$\pm$0.03   \\
\enddata                                                                                                          
\end{deluxetable}

\section{Appendix B: Veiling determination}\label{sec:appendix_b}

\begin{figure}
\begin{center}
\includegraphics[trim=10 5 5 5,width=0.35\columnwidth, angle=0]{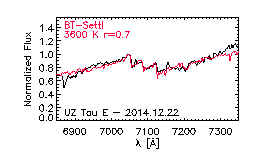}
\includegraphics[trim=10 5 5 5,width=0.35\columnwidth, angle=0]{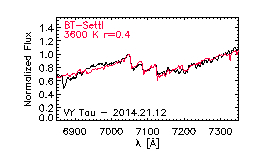}
\includegraphics[trim=10 5 5 5,width=0.35\columnwidth, angle=0]{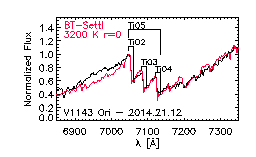}
\includegraphics[trim=10 5 5 5,width=0.35\columnwidth, angle=0]{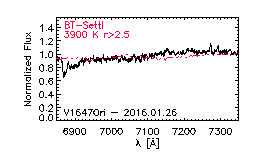}
\end{center}
\caption{Examples of BT-Settl models best matching the TiO bands. Spectra are normalized to the peak of the band. For each panel, object name, date of observation, temperature of the BT-Settl model, and veiling are reported. In the third panel TiO bands are indicated.\label{fig:fig13}}
\end{figure}

\begin{figure*}
\begin{center}
\includegraphics[trim=10 5 5 5,width=0.4\columnwidth, angle=0]{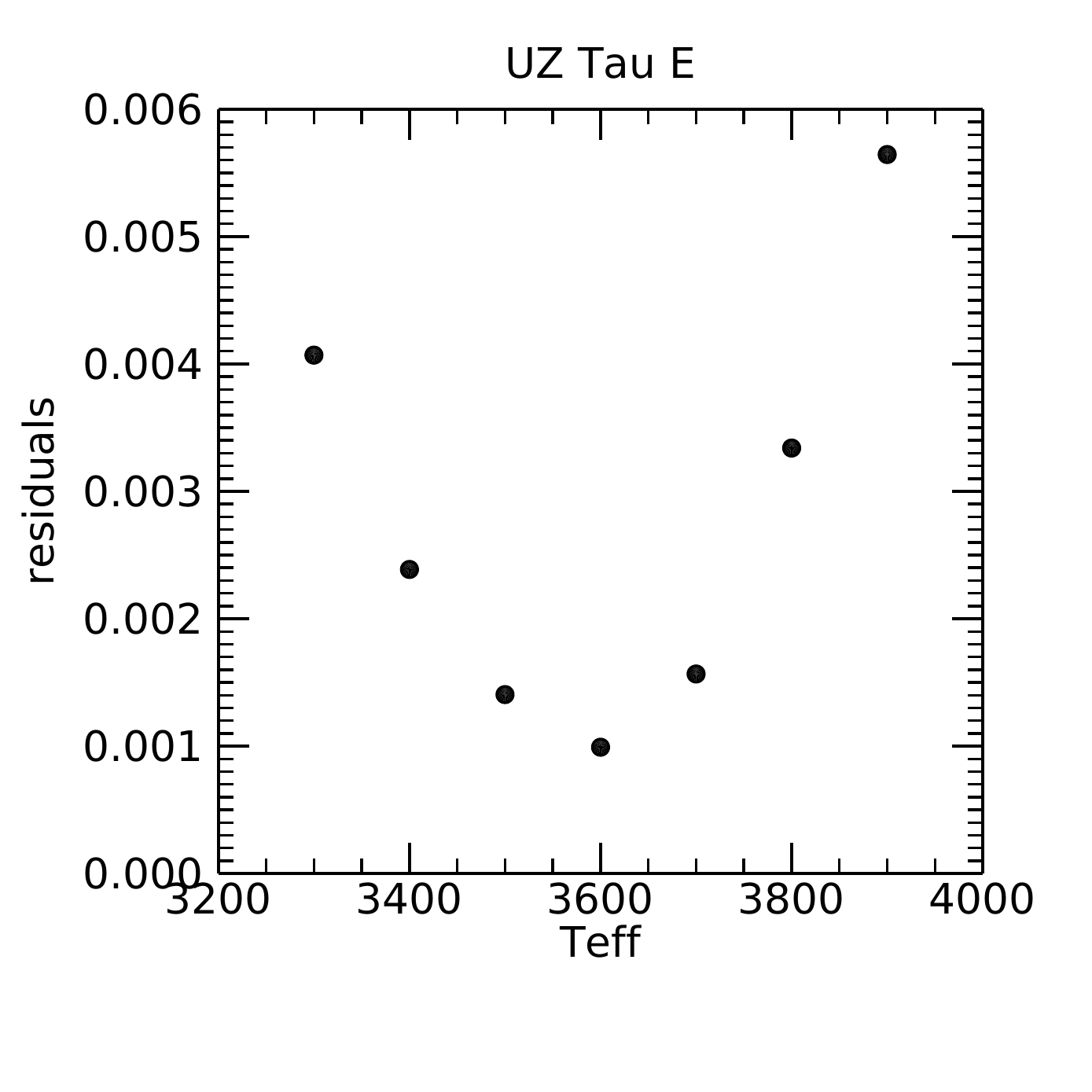}
\end{center}
\caption{\label{fig:fig14} Residuals on \teff\, fitting in UZ Tau E, assuming r=0.7. The BTSettl model that best matches the MODS spectrum corresponds to  \teff\,=\,3600 K, the same as that found with high-spectral resolution observations.}
\end{figure*}

In this Section we briefly discuss the strategy we adopted to estimate the veiling ($r$) in our sample. For a given source, we have firstly selected a number of BTSettl templates with temperature within 500 K of the \teff\, value given in the literature (Table\,\ref{tab:tab3}) and log$g$ between 3.5 and 4.5. A veiling (increasing in steps of 0.1) is then added to each template to find the value that best matches the spectral features observed in our sample of stars.

For M-type stars, which are the majority of our sources, we have fitted the portion of the continuum in the range 700-715 nm, where prominent TiO absorption bands are located (Fig.\,\ref{fig:fig13}). In all sources but two, our fit consistently provides a \teff\, in agreement with that of the literature, and determined $r$ with a typical error of 0.3, corresponding to a variation of  \teff\, less than 100 K.
The two exceptions are UZ Tau E and V1143 Ori (\teff\, of literature 3410 K and 3500 K, respectively). In the case of UZ Tau E we have accurate determinations of \teff\,=\,3600 K  and $r$\,=\,0.7 from high-resolution observations obtained with the HARPS-N instrument at the {\it Telescopio Nazionale Galileo} (Gangi et al. in preparation). Having assumed the veiling value, we have fitted 
\teff\, in the MODS spectrum, which is in perfect agreement with that derived from the HARPS-N spectrum (Figure\,\ref{fig:fig14}). In the case of V1143 Ori, we fit \teff\,=\,3200 K and $r$\,=\,0. Since for $r>$0 \teff\, is even lower, the literature value cannot be reproduced.\\ 
For the few sources that do not present the TiO bands because of their SpT (in particular NY Ori, DR Tau, and PV Cep), we applied an indirect method to evaluate the veiling at 710 nm. We firstly reddened the BTSettl template by assuming the \av\, derived from the 'line method' (Section\,\ref{sec:sec6.1.1}). By dividing the observed spectrum with the reddened template, we have derived the shape of the excess continuum flux. The ratio at 710 nm was then assumed as the veiling value, reported in Table\,\ref{tab:tab7}.

\begin{figure}
\begin{center}
\includegraphics[trim=10 5 5 5,width=0.35\columnwidth, angle=0]{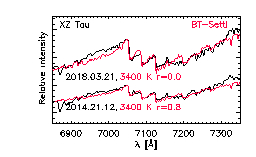}
\includegraphics[trim=10 5 5 5,width=0.35\columnwidth, angle=0]{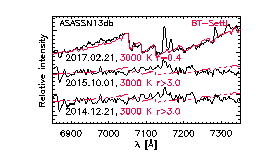}
\includegraphics[trim=10 5 5 5,width=0.35\columnwidth, angle=0]{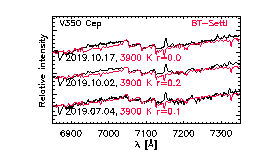}
\end{center}
\caption{TiO bands strength variability. Spectra are normalized to the peak of the band and vertically shifted for a better visualization. Date of observation,  temperature of the best matching BT-Settl model and veiling are also labelled. \label{fig:fig15}}
\end{figure}

We also looked for TiO spectral variations in the three sources of this sub-sample that were observed more than once (XZ Tau, V350 Cep and ASASSN13-db, see Fig. \ref{fig:fig15}). We have found that XZ Tau shows an increase in the depth of the TiO bands between 2014 and 2018 which corresponds to a decrease in $r$ from 0.8 to 0.0. 
TiO bands do not appear in the spectra of ASASSN13-db during the burst phases of 2014 and 2015 while they are present during the out-of-burst phase of 2017 with a corresponding veiling of 0.4. The absence of TiO bands in 2014 and 2015 indicates that during the outburst the veiling increased to values larger than 3.
Finally, the veiling measured in V350 Cep does not show variations that exceed the quoted uncertainty  of 0.3. 


\begin{thebibliography}{}
\bibitem[Akeson et al.(2019)]{2019ApJ...872..158A} Akeson, R.~L., Jensen, E.~L.~N., Carpenter, J., et al.\ 2019, \apj, 872, 158
\bibitem[Alcal{\'a} et al.(2021)] {2017A&A...600A..20A}Alcal{\'a}, J.~M., Gangi, M.~E., Biazzo, K. et al. \ 2021, \aap, in press
\bibitem[Alcal{\'a} et al.(2017)]{2017A&A...600A..20A} Alcal{\'a}, J.~M., Manara, C.~F., Natta, A., et al.\ 2017, \aap, 600, A20
\bibitem[Alcal{\'a} et al.(2014)]{2014A&A...561A...2A} Alcal{\'a}, J.~M., Natta, A., Manara, C.~F., et al.\ 2014, \aap, 561, A2
\bibitem[Andrews et al.(2004)]{2004ApJ...610L..45A} Andrews, S.~M., Rothberg, B., \& Simon, T.\ 2004, \apjl, 610, L45
\bibitem[Andreasyan et al.(2021)]{2021Ap.....64..187A} Andreasyan, H.~R., Magakian, T.~Y., Movsessian, T.~A., et al.\ 2021, Astrophysics, 64, 187
\bibitem[Antoniucci et al.(2013)]{2013prpl.conf2B055A} Antoniucci, S., Arkharov, A.~A., Di Paola, A., et al.\ 2013, Protostars and Planets VI Posters
\bibitem[Antoniucci et al.(2014)]{2014A&A...572A..62A} Antoniucci, S., Garc{\'\i}a L{\'o}pez, R., Nisini, B., et al.\ 2014, \aap, 572, A6
\bibitem[Antoniucci et al.(2017)]{2017A&A...606A..48A} Antoniucci, S., Nisini, B., Biazzo, K., et al.\ 2017a, \aap, 606, A48
\bibitem[Antoniucci et al.(2017)]{2017A&A...599A.105A} Antoniucci, S., Nisini, B., Giannini, T., et al.\ 2017b, \aap, 599, A105
\bibitem[Arce et al.(2010)]{2010ApJ...715.1170A} Arce, H.~G., Borkin, M.~A., Goodman, A.~A., et al.\ 2010, \apj, 715, 1170
\bibitem[Armitage(2016)]{2016ApJ...833L..15A} Armitage, P.~J.\ 2016, \apjl, 833, L15
\bibitem[Aspin(2011)]{2011AJ....142..135A} Aspin, C.\ 2011, \aj, 142, 135
\bibitem[Aspin et al.(2006)]{2006AJ....132.1298A} Aspin, C., Barbieri, C., Boschi, F., et al.\ 2006, \aj, 132, 1298 
\bibitem[Aspin et al.(2008)]{2008AJ....135..423A} Aspin, C., Beck, T.~L., \& Reipurth, B.\ 2008, \aj, 135, 423 
\bibitem[Aspin et al.(2009)]{2009AJ....137.2968A} Aspin, C., Greene, T.~P., \& Reipurth, B.\ 2009, \aj, 137, 2968
\bibitem[Aspin \& Reipurth(2009)]{2009AJ....138.1137A} Aspin, C., \& Reipurth, B.\ 2009, \aj, 138, 1137
\bibitem[Audard et al.(2014)]{2014prpl.conf..387A} Audard, M., {\'A}brah{\'a}m, P., Dunham, M.~M., et al.\ 2014, Protostars and Planets VI, 387 
\bibitem[Audard et al.(2010)]{2010A&A...511A..63A} Audard, M., Stringfellow, G.~S., G{\"u}del, M., et al.\ 2010, \aap, 511, A63. 
\bibitem[Banzatti et al.(2014)]{2014ApJ...780...26B} Banzatti, A., Meyer, M.~R., Manara, C.~F., et al.\ 2014, \apj, 780, 26
\bibitem[Banzatti et al.(2019)]{2019ApJ...870...76B} Banzatti, A., Pascucci, I., Edwards, S., et al.\ 2019, \apj, 870, 76
\bibitem[Biscaya et al.(1997)]{1997ApJ...491..359B} Biscaya, A.~M., Rieke, G.~H., Narayanan, G., et al.\ 1997, \apj, 491, 359
\bibitem[Bonnell \& Bastien(1992)]{1992ApJ...401L..31B} Bonnell, I. \& Bastien, P.\ 1992, \apjl, 401, L31
\bibitem[Brice{\~n}o et al.(2004)]{2004ApJ...606L.123B} Brice{\~n}o, C., Vivas, A.~K., Hern{\'a}ndez, J., et al.\ 2004, \apjl, 606, L123 
\bibitem[Cabrit et al.(1990)]{1990ApJ...354..687C} Cabrit, S., Edwards, S., Strom, S.~E., et al.\ 1990, \apj, 354, 687
\bibitem[Caratti o Garatti et al.(2013)]{2013A&A...554A..66C} Caratti o Garatti, A., Garcia Lopez, R., Weigelt, G., et al.\ 2013, \aap, 554, A66 
\bibitem[Cardelli et al.(1989)]{1989ApJ...345..245C} Cardelli, J.~A., Clayton, G.~C., \& Mathis, J.~S.\ 1989, \apj, 345, 245
\bibitem[Chavarria-K.(1979)]{1979A&A....79L..18C} Chavarria-K., C.\ 1979, \aap, 79, L18 
\bibitem[Cieza et al.(2018)]{2018MNRAS.474.4347C} Cieza, L.~A., Ru{\'\i}z-Rodr{\'\i}guez, D., Perez, S., et al.\ 2018, \mnras, 474, 4347
\bibitem[Coffey et al.(2004)]{2004A&A...419..593C} Coffey, D., Downes, T.~P., \& Ray, T.~P.\ 2004, \aap, 419, 593 
\bibitem[Cohen et al.(1981)]{1981ApJ...245..920C} Cohen, M., Kuhi, L.~V., Spinrad, H., \& Harlan, E.~A.\ 1981, \apj, 245, 920 
\bibitem[Connelley \& Reipurth(2018)]{2018ApJ...861..145C} Connelley, M.~S., \& Reipurth, B.\ 2018, \apj, 861, 145
\bibitem[Contreras Pe{\~n}a et al.(2017)]{2017MNRAS.465.3039C} Contreras Pe{\~n}a, C., Lucas, P.~W., Kurtev, R., et al.\ 2017, \mnras, 465, 3039.
\bibitem[Contreras Pe{\~n}a et al.(2019)]{2019MNRAS.486.4590C} Contreras Pe{\~n}a, C., Naylor, T., \& Morrell, S.\ 2019, \mnras, 486, 4590
\bibitem[Costigan et al.(2014)]{2014MNRAS.440.3444C} Costigan, G., Vink, J.~S., Scholz, A., et al.\ 2014, \mnras, 440, 3444
\bibitem[Cs{\'e}p{\'a}ny et al.(2017)]{2017A&A...603A..74C} Cs{\'e}p{\'a}ny, G., van den Ancker, M., {\'A}brah{\'a}m, P., et al.\ 2017, \aap, 603, A74
\bibitem[D'Angelo \& Spruit(2012)]{2012MNRAS.420..416D} D'Angelo, C.~R. \& Spruit, H.~C.\ 2012, \mnras, 420, 416. 
\bibitem[Dodin et al.(2016)]{2016AstL...42...29D} Dodin, A.~V., Emelyanov, N.~V., Zharova, A.~V., et al.\ 2016, Astronomy Letters, 42, 29 
\bibitem[Edwards et al.(2013)]{2013ApJ...778..148E} Edwards, S., Kwan, J., Fischer, W., et al.\ 2013, \apj, 778, 148
\bibitem[Fiorellino et al.(2021)]{2021arXiv210303863F} Fiorellino, E., Manara, C.~F., Nisini, B., et al.\ 2021, arXiv:2103.03863
\bibitem[Fischer et al.(2011)]{2011ApJ...730...73F} Fischer, W., Edwards, S., Hillenbrand, L., et al.\ 2011, \apj, 730, 73
\bibitem[Fischer et al.(2016)]{2016ApJ...827...96F} Fischer, W.~J., Padgett, D.~L., Stapelfeldt, K.~L., et al.\ 2016, \apj, 827, 96
\bibitem[Folha \& Emerson(2001)]{2001A&A...365...90F} Folha, D.~F.~M. \& Emerson, J.~P.\ 2001, \aap, 365, 90
\bibitem[Gaia Collaboration et al.(2020)]{2020A&A...595A...2G} Gaia Collaboration, Brown A.G.A, et al.\ 2020, A\&A in prep.
\bibitem[Gaia Collaboration et al.(2016)]{2016A&A...595A...1G} Gaia Collaboration, Prusti, T., de Bruijne, J.~H.~J., et al.\ 2016, \aap, 595, A1
\bibitem[Giannini et al.(2017)]{2017ApJ...839..112G} Giannini, T., Antoniucci, S., Lorenzetti, D., et al.\ 2017, \apj, 839, 112
\bibitem[Giannini et al.(2020)]{2020A&A...637A..83G} Giannini, T., Giunta, A., Lorenzetti, D., et al.\ 2020, \aap, 637, A83
\bibitem[Giannini et al.(2016)]{2016ApJ...819L...5G} Giannini, T., Lorenzetti, D., Antoniucci, S., et al.\ 2016, \apjl, 819, L5
\bibitem[Giannini et al.(2018)]{2018A&A...611A..54G} Giannini, T., Munari, U., Antoniucci, S., et al.\ 2018, \aap, 611, A54
\bibitem[Giannini et al.(2019)]{2019A&A...631A..44G} Giannini, T., Nisini, B., Antoniucci, S., et al.\ 2019, \aap, 631, A44
\bibitem[Gramajo et al.(2014)]{2014AJ....147..140G} Gramajo, L.~V., Rod{\'o}n, J.~A., \& G{\'o}mez, M.\ 2014, \aj, 147, 140
\bibitem[Grankin et al.(2007)]{2007IBVS.5752....1G} Grankin, K.~N., Artemenko, S.~A., \& Melnikov, S.~Y.\ 2007, Information Bulletin on Variable Stars, 5752, 1
\bibitem[Grankin et al.(2008)]{2008A&A...479..827G} Grankin, K.~N., Bouvier, J., Herbst, W., et al.\ 2008, \aap, 479, 827
\bibitem[Gullbring et al.(1998)]{1998ApJ...492..323G} Gullbring, E., Hartmann, L., Brice{\~n}o, C., et al.\ 1998, \apj, 492, 323
\bibitem[Guo et al.(2020)]{2020MNRAS.492..294G} Guo, Z., Lucas, P.~W., Contreras Pe{\~n}a, C., et al.\ 2020, \mnras, 492, 294
\bibitem[Hamidouche(2010)]{2010ApJ...722..204H} Hamidouche, M.\ 2010, \apj, 722, 204
\bibitem[Hartigan \& Kenyon(2003)]{2003ApJ...583..334H} Hartigan, P., \& Kenyon, S.~J.\ 2003, \apj, 583, 334
\bibitem[Hartmann \& Kenyon(1985)]{1985ApJ...299..462H} Hartmann, L. \& Kenyon, S.~J.\ 1985, \apj, 299, 462
\bibitem[Hartmann et al.(1993)]{1993prpl.conf..497H} Hartmann, L., Kenyon, S., \& Hartigan, P.\ 1993, Protostars and Planets III, 497
\bibitem[Herbig(1989)]{1989ESOC...33..233H} Herbig, G.~H.\ 1989, ESO Conference and Workshop Proceedings, 233
\bibitem[Herbig(1990)]{1990ApJ...360..639H} Herbig, G.~H.\ 1990, \apj, 360, 639 
\bibitem[Herbig(2008)]{2008AJ....135..637H} Herbig, G.~H.\ 2008, \aj, 135, 637 Herbig, G.~H. \& Bell, K.~R.\ 1988, Third catalog of emission-line stars of the Orion population., by G.H. gerbig and K.R. Bell.  Lick Observatory Bulletin \#1111, Santa Cruz: Lick Observatory, Jun 1988, 90 p.
\bibitem[Herbig \& Bell(1988)]{1988cels.book.....H} 
\bibitem[Herczeg \& Hillenbrand(2014)]{2014ApJ...786...97H} Herczeg, G.~J., \& Hillenbrand, L.~A.\ 2014, \apj, 786, 97
\bibitem[Hillenbrand(1997)]{1997AJ....113.1733H} Hillenbrand, L.~A.\ 1997, \aj, 113, 1733
\bibitem[Hillenbrand(2019)]{2019ATel13321....1H} Hillenbrand, L.~A.\ 2019, The Astronomer's Telegram 13321, 1
\bibitem[Hillenbrand et al.(2013)]{2013AJ....145...59H} Hillenbrand, L.~A., Miller, A.~A., Covey, K.~R., et al.\ 2013, \aj, 145, 59 
\bibitem[Holoien et al.(2014)]{2014ApJ...785L..35H} Holoien, T.~W.-S., Prieto, J.~L., Stanek, K.~Z., et al.\ 2014, \apjl, 785, L35
\bibitem[Jensen et al.(2007)]{2007AJ....134..241J} Jensen, E.~L.~N., Dhital, S., Stassun, K.~G., et al.\ 2007, \aj, 134, 241 
\bibitem[Joncour et al.(2017)]{2017A&A...599A..14J} Joncour, I., Duch{\^e}ne, G., \& Moraux, E.\ 2017, \aap, 599, A14
\bibitem[Jurdana-{\v S}epi{\'c} et al.(2017)]{2017A&A...602A..99J} Jurdana-{\v S}epi{\'c}, R., Munari, U., Antoniucci, S., et al.\ 2017, \aap, 602, A99 \
\bibitem[Jurdana-{\v{S}}epi{\'c} et al.(2018)]{2018A&A...614A...9J} Jurdana-{\v{S}}epi{\'c}, R., Munari, U., Antoniucci, S., et al.\ 2018, \aap, 614, A9
\bibitem[Kenyon et al.(1994)]{1994AJ....107.2153K} Kenyon, S.~J., Hartmann, L., Hewett, R., et al.\ 1994, \aj, 107, 2153 
\bibitem[K{\'o}sp{\'a}l et al.(2016)]{2016A&A...596A..52K} K{\'o}sp{\'a}l, {\'A}., {\'A}brah{\'a}m, P., Acosta-Pulido, J.~A., et al.\ 2016, \aap, 596, A52 
\bibitem[Kounkel et al.(2019)]{2019AJ....157..196K} Kounkel, M., Covey, K., Moe, M., et al.\ 2019, \aj, 157, 196 
\bibitem[Kuffmeier et al.(2018)]{2018MNRAS.475.2642K} Kuffmeier, M., Frimann, S., Jensen, S.~S., et al.\ 2018, \mnras, 475, 2642
\bibitem[Kun et al.(2009)]{2009ApJS..185..451K} Kun, M., Balog, Z., Kenyon, S.~J., et al.\ 2009, \apjs, 185, 451
\bibitem[Kun et al.(2011)]{2011MNRAS.413.2689K} Kun, M., Szegedi-Elek, E., Mo{\'o}r, A., et al.\ 2011, \mnras, 413, 2689
79
\bibitem[Kwan \& Fischer(2011)]{2011MNRAS.411.2383K} Kwan, J. \& Fischer, W.\ 2011, \mnras, 411, 2383

\bibitem[Leinert et al.(1993)]{1993A&A...278..129L} Leinert, C., Zinnecker, H., Weitzel, N., et al.\ 1993, \aap, 278, 129
\bibitem[Li et al.(2015)]{2015ApJS..219...20L} Li, H., Li, D., Qian, L., et al.\ 2015, \apjs, 219, 20
\bibitem[Lodato \& Clarke(2004)]{2004MNRAS.353..841L} Lodato, G. \& Clarke, C.~J.\ 2004, \mnras, 353, 841
\bibitem[Lorenzetti et al.(2015)]{2015ApJ...802...24L} Lorenzetti, D., Antoniucci, S., Giannini, T., et al.\ 2015, \apj, 802, 24 
\bibitem[Lorenzetti et al.(2007)]{2007ApJ...665.1182L} Lorenzetti, D., Giannini, T., Larionov, V.~M., et al.\ 2007, \apj, 665, 1182
\bibitem[Lorenzetti et al.(2011)]{2011ApJ...732...69L} Lorenzetti, D., Giannini, T., Larionov, V.~M., et al.\ 2011, \apj, 732, 69 
\bibitem[Lorenzetti et al.(2009)]{2009ApJ...693.1056L} Lorenzetti, D., Larionov, V.~M., Giannini, T., et al.\ 2009, \apj, 693, 1056
\bibitem[Luhman et al.(2010)]{2010ApJS..186..111L} Luhman, K.~L., Allen, P.~R., Espaillat, C., et al.\ 2010, \apjs, 186, 111 
\bibitem[MacFarlane et al.(2019)]{2019MNRAS.487.4465M} MacFarlane, B., Stamatellos, D., Johnstone, D., et al.\ 2019a, \mnras, 487, 4465
\bibitem[MacFarlane et al.(2019)]{2019MNRAS.487.5106M} MacFarlane, B., Stamatellos, D., Johnstone, D., et al.\ 2019b, \mnras, 487, 5106
\bibitem[Magakian \& Movsesian(2001)]{2001Ap.....44..419M} Magakian, T.~Y., \& Movsesian, T.~A.\ 2001, Astrophysics, 44, 419
\bibitem[Manara et al.(2013)]{2013A&A...558A.114M} Manara, C.~F., Beccari, G., Da Rio, N., et al.\ 2013, \aap, 558, A114
\bibitem[Manara et al.(2017)]{2017A&A...604A.127M} Manara, C.~F., Testi, L., Herczeg, G.~J., et al.\ 2017, \aap, 604, A127
\bibitem[Meng et al.(2019)]{2019ApJ...878....7M} Meng, H.~Y.~A., Rieke, G.~H., Kim, J.~S., et al.\ 2019, \apj, 878, 7
\bibitem[Meyer et al.(1997)]{1997AJ....114..288M} Meyer, M.~R., Calvet, N., \& Hillenbrand, L.~A.\ 1997, \aj, 114, 288
\bibitem[Miller et al.(2015)]{2015ATel.7428....1M} Miller, A.~A., Hillenbrand, L.~A., Bilgi, P., et al.\ 2015, The Astronomer's Telegram, 7428
\bibitem[Moody \& Stahler(2017)]{2017A&A...600A.133M} Moody, M.~S.~L. \& Stahler, S.~W.\ 2017, \aap, 600, A133. 
\bibitem[Morales-Calder{\'o}n et al.(2011)]{2011ApJ...733...50M} Morales-Calder{\'o}n, M., Stauffer, J.~R., Hillenbrand, L.~A., et al.\ 2011, \apj, 733, 50
\bibitem[Moura et al.(2020)]{2020MNRAS.494.3512M} Moura, T., Alencar, S.~H.~P., Sousa, A.~P., et al.\ 2020, \mnras, 494, 3512
\bibitem[Muzerolle et al.(2003)]{2003ApJ...592..266M} Muzerolle, J., Hillenbrand, L., Calvet, N., et al.\ 2003, \apj, 592, 266
\bibitem[Nisini et al.(2016)]{2016A&A...595A..76N} Nisini, B., Giannini, T., Antoniucci, S., et al.\ 2016, \aap, 595, A76
\bibitem[Osorio et al.(2016)]{2016ApJ...825L..10O} Osorio, M., Mac{\'\i}as, E., Anglada, G., et al.\ 2016, \apjl, 825, L10
\bibitem[Petrov et al.(2011)]{2011A&A...535A...6P} Petrov, P.~P., Gahm, G.~F., Stempels, H.~C., et al.\ 2011, \aap, 535, A6\bibitem[Pogge et al.(2010)]{2010SPIE.7735E..0AP} Pogge, R.~W., Atwood, B., Brewer, D.~F., et al.\ 2010, \procspie, 7735, 77350A
\bibitem[Principe et al.(2018)]{2018MNRAS.473..879P} Principe, D.~A., Cieza, L., Hales, A., et al.\ 2018, \mnras, 473, 879
\bibitem[Rodrigo \& Solano(2020)]{2020sea..confE.182R} Rodrigo, C. \& Solano, E.\ 2020, Contributions to the XIV.0 Scientific Meeting (virtual) of the Spanish Astronomical Society, 182
\bibitem[Schipani et al.(2018)]{2018SPIE10702E..0FS} Schipani, P., Campana, S., Claudi, R., et al.\ 2018, \procspie, 10702, 107020F
\bibitem[Seifert et al.(2003)]{2003SPIE.4841..962S} Seifert, W., Appenzeller, I., Baumeister, H., et al.\ 2003, \procspie, 4841, 962
\bibitem[Semkov et al.(2017)]{2017BlgAJ..27...75S} Semkov, E.~H., Ibryamov, S.~I., \& Peneva, S.~P.\ 2017, Bulgarian Astronomical Journal, 27, 75 
\bibitem[Shu et al.(1994)]{1994ApJ...429..797S} Shu, F.~H., Najita, J., Ruden, S.~P., \& Lizano, S.\ 1994, \apj, 429, 797 
\bibitem[Sicilia-Aguilar et al.(2012)]{2012A&A...544A..93S} Sicilia-Aguilar, A., K{\'o}sp{\'a}l, {\'A}., Setiawan, J., et al.\ 2012, \aap, 544, A93
\bibitem[Sicilia-Aguilar et al.(2017)]{2017A&A...607A.127S} Sicilia-Aguilar, A., Oprandi, A., Froebrich, D., et al.\ 2017, \aap, 607, A127 
\bibitem[Sipos et al.(2009)]{2009A&A...507..881S} Sipos, N., {\'A}brah{\'a}m, P., Acosta-Pulido, J., et al.\ 2009, \aap, 507, 881
\bibitem[Sipos \& K{\'o}sp{\'a}l(2014)]{2014IAUS..299..121S} Sipos, N., \& K{\'o}sp{\'a}l, {\'A}.\ 2014, Exploring the Formation and Evolution of Planetary Systems, 121
\bibitem[Sperling et al.(2020)]{2020A&A...642A.216S} Sperling, T., Eisl{\"o}ffel, J., Fischer, C., et al.\ 2020, \aap, 642, A216
\bibitem[Stock et al.(2020)]{2020A&A...643A.181S} Stock, C., Caratti o Garatti, A., McGinnis, P., et al.\ 2020, \aap, 643, A181
\bibitem[Stone(1983)]{1983IBVS.2380....1S} Stone, R.~P.~S.\ 1983, Information Bulletin on Variable Stars, 2380, 1 
\bibitem[The et al.(1994)]{1994A&AS..104..315T} The, P.~S., de Winter, D., \& Perez, M.~R.\ 1994, \aaps, 104, 315
\bibitem[Weingartner \& Draine(2001)]{2001ApJ...548..296W} Weingartner, J.~C. \& Draine, B.~T.\ 2001, \apj, 548, 296
\bibitem[Yang et al.(2012)]{2012ApJ...744..121Y} Yang, H., Herczeg, G.~J., Linsky, J.~L., et al.\ 2012, \apj, 744, 121
\bibitem[Young et al.(2015)]{2015AJ....150...40Y} Young, K.~E., Young, C.~H., Lai, S.-P., et al.\ 2015, \aj, 150, 40

\end{thebibliography}
\end{document}